\documentclass[12pt,a4paper]{article}
\usepackage{CJKutf8}
\usepackage{amsmath}
\usepackage{amssymb}
\usepackage[top=2cm, bottom=2cm, left=2cm, right=2cm]{geometry}
\usepackage{indentfirst}
\usepackage{hyperref}
\usepackage[numbers, sort&compress]{natbib}
\usepackage{color}

\allowdisplaybreaks[4]

\begin{document}
\begin{CJK}{UTF8}{gkai}

\begin{titlepage}

\begin{flushright}
USTC-ICTS-16-04
\end{flushright}

\vspace{10mm}
\begin{center}
{\Large\bf Wess-Zumino Model on Bosonic-Fermionic Noncommutative Superspace}
\vspace{16mm}

{\large Xu-Dong Wang\footnote{E-mail address: xudwang@ustc.edu.cn}

\vspace{6mm}
{\normalsize \em Interdisciplinary Center for Theoretical Study\\
 University of Science and Technology of China, Hefei, Anhui 230026, People's Republic of China}}

\end{center}

\vspace{10mm}
\centerline{{\bf{Abstract}}}
\vspace{6mm}

In our previous paper we construct a renormalizable Wess-Zumino action on BFNC superspace at the second order approximation of noncommutative parameters. The action contains about 200 terms which are necessary for renormalization. By removing chiral covariant derivatives and chiral coordinates we found that the BFNC  Wess-Zumino action can be transformed to a simpler form which have manifest 1/2 supersymmetry. Based on this discovery, we can extend the BFNC Wess-Zumino action to the all order of noncommutative parameters. At first we introduce global symmetries, then obtain divergent operators in the effective action by using dimensional analysis, the next step is to construct all possible BFNC parameters, at the end we combine the BFNC parameters with the divergent operators. We present the explicit action up to the fourth order of noncommutative parameters. Because the action contain all possible divergent operators, it is renormalizable to all order in perturbative theory.

\vskip 20pt
\noindent
{\bf PACS Number(s)}: 12.60.Jv, 11.30.Pb, 11.10.Nx, 11.10.Gh

\vskip 20pt
\noindent
{\bf Keywords}: Wess-Zumino Model, Bosonic-Fermionic Noncommutative Superspace, Renormalization

\end{titlepage}

\section{Introduction}

There are many studies on noncommutative field theory in recent years~\cite{Douglas:2001ba, Szabo:2001kg}. The introduction of supersymmetry on noncommutative spacetime is also very interesting. We can classify the noncommutative superspace by three kind. The first kind is Bosonic noncommutative superspace in which the algebraic relation is $\left[x^k,~x^l\right]=i~ \theta ^{k l}$. It can be obtained from string theory~\cite{Seiberg:1999vs}. Ref.~\cite{Ferrara:2003xy} constructed Wess-Zumino and Super Yang-Mills model on this superspace. The renormalization property of these models were studied in Ref.~\cite{Girotti:2000gc}. They proved that the Wess-Zumino model is renormalizable.

The second kind is non-anticommutative (NAC) superspace $\left\{\theta ^{\alpha },~\theta ^{\beta }\right\}=C^{\alpha  \beta }$. It can also be obtained from string theory~\cite{Ooguri:2003qp, Ooguri:2003tt}. In Ref.~\cite{Seiberg:2003yz} the author defined Wess-Zumino and Super Yang-Mills model on this superspace.
There are many work studying these models, for example Ref.~\cite{Grisaru:2003fd} calculate the 1PI effective action for NAC Wess-Zumino model, and found a new term $(-g^2/\epsilon) \int d^8z~\left(m^*\right)^2 U\left(D^2\Phi \right)^2$, where $U=\theta ^2~\bar{\theta }^2~C^2$ in the effective action. This term make the deformed action not renormalizable. By introducing two global $U(1)$ symmetries, Ref.~\cite{Britto:2003kg, Romagnoni:2003xt} prove that if one add this term to the deformed action, the new action is renormalizable to all order in perturbative theory.

The third kind is Bosonic-Fermionic noncommutative (BFNC) superspace, the algebraic relation is $\left[x^k,~\theta ^{\alpha }\right]=i ~\Lambda ^{k \alpha }$. In Ref.~\cite{de Boer:2003dn} the author found the origin of this structure in string theory. To study the field theory on BFNC superspace, in Ref.~\cite{Miao:2014oba, Miao:2014mia}, we constructed a deformed Wess-Zumino model. By calculating the 1PI effective action, we obtain three effective action $\Gamma\left(\Lambda ^2\right)=a_0 \,{\cal S}_{\Lambda }\left(\Lambda ^2\right)+a_1 \,\Gamma _{\rm 1st}\left(\Lambda ^2\right)+a_2 \, \Gamma _{\rm 2nd}\left(\Lambda ^2\right)+a_3 \, \Gamma _{\rm 3rd}\left(\Lambda ^2\right)$ and classified them by 1/2 supersymmetry invariant subset $\Gamma\left(\Lambda ^2\right)=\sum_{i=1}^{74} f_i$. By using these invariant subset, we obtain more general 1/2 supersymmetry invariant action ${\cal S}_{(3)}={\cal S}_{\rm WZ}+\int d^8z\left(\sum _{i=1}^{74} B_i\right)$ and prove that it is one-loop renormalizable. To extend the renormalizability to all order in perturbative theory, we introduce two global  $U(1)$ symmetries and obtain a modified action ${\cal S^{\prime}}_{(3)}={\cal S^{\prime}}_{\rm WZ}+\int d^8z\left(\sum _{i=1}^{74} c_i~ B_i\right)$, after adding the other divergent terms obtained by using dimensional analysis we obtain a BFNC Wess-Zumino model ${\cal S^{\prime}}_{(3)}+\int d^8z\left(\sum_{i=75}^{89}c_i ~B_i\right)$ at the second order approximation of noncommutative parameters. Because the action contain all of the divergent operators, it is renormalizable to all order in perturbative theory.

In our previous studies~\cite{Miao:2014oba, Miao:2014mia} the action contains only second order of noncommutative parameters. The associate property of BFNC star product need higher order of noncommutative parameters. Even at second order of noncommutative parameters, the BFNC Wess-Zumino model contain about 200 terms, this make the extension to higher oder approximation of noncommutative parameters very difficult. Inspired by recent studies on super Yang-Mills theory on NAC superspace~\cite{Grisaru:2005we} and BFNC superspace~\cite{Wang:2016c}, we found that by removing chiral covariant derivatives $D^2$ and chiral coordinates $\theta^2$, the BFNC  Wess-Zumino action can be transformed to more simpler form. With this discovery, we can extend the BFNC Wess-Zumino action to the all order of noncommutative parameters. To illustrate the idea, we will present the explicit action up to the fourth order of noncommutative parameters in this paper. 

The organization of this paper is as follows, in section~\ref{operators} we determine the divergent operators by using dimensional analysis method up to the fourth order of noncommutative parameters, in section~\ref{BFNC parameters} we construct all related BFNC parameters, in section~\ref{construct action} we construct 1/2 supersymmetric invariant action by combining the BFNC parameters with the divergent operators, we give our conclusion and outlook in section~\ref{conclusion}.

\section{Divergent operators}
\label{operators}

By using two global $U(1)$ symmetries, we can determine the divergent operators in the effective action. 
We list them as follows, at the order of $\Lambda^2$,  
\begin{itemize}
\label{}
\item
$N_{\Phi }=N_{D^2}$, $N_{\bar{D}}=0$, $N_{\partial \partial }=0$.
{\small
\begin{eqnarray}
\label{Lambda2-type1}
&&\Phi ^+,\quad 
\Phi ^+ \Phi ^+,\quad 
\Phi ^+ \Phi ^+ \Phi ^+,\quad 
\Phi ^+ \Phi ^+ \Phi ^+ \Phi ^+,\quad
\Phi ^+ \Phi ^+ \Phi ^+ \Phi ^+ \Phi ^+,\nonumber\\
&&D^2 \Phi,\quad
D^2 \Phi  \Phi ^+,\quad 
D^2 D^2 \Phi  \Phi ,\quad 
D^2 \Phi  \Phi ^+ \Phi ^+,\quad
D^2 D^2 \Phi  \Phi  \Phi ^+,\quad 
D^2 \Phi  \Phi ^+ \Phi ^+ \Phi ^+.
\end{eqnarray}
}
\item
$N_{\Phi }=N_{D^2}$, $N_{\bar{D}}=0$, $N_{\partial \partial }=1$.
{\small
\begin{eqnarray}
\label{Lambda2-type2}
&&\partial\partial \Phi ^+ \Phi ^+,\quad 
D^2 \partial\partial \Phi  \Phi ^+,\quad 
D^2 D^2 \partial\partial \Phi  \Phi ,\quad
\partial\partial\Phi ^+ \Phi ^+ \Phi ^+,\quad 
D^2 \partial\partial \Phi  \Phi ^+ \Phi ^+,\nonumber\\
&&\partial\partial\Phi ^+ \Phi ^+ \Phi ^+ \Phi ^+.
\end{eqnarray}
}
\item
$N_{\Phi }=N_{D^2}$, $N_{\bar{D}}=0$, $N_{\partial \partial }=2$.
{\small
\begin{eqnarray}
\label{Lambda2-type3}
&&\partial\partial\partial\partial \Phi ^+ \Phi ^+,\quad 
D^2 \partial\partial\partial\partial \Phi  \Phi ^+,\quad 
\partial\partial\partial\partial \Phi ^+ \Phi ^+ \Phi ^+.
\end{eqnarray}
}
\item
$N_{\Phi }=N_{D^2}$, $N_{\bar{D}}=2$, $N_{\partial \partial }=0$.
{\small
\begin{eqnarray}
\label{Lambda2-type4}
&&\bar{D}^2 \Phi ^+,\quad 
\bar{D}^2 \Phi ^+ \Phi ^+,\quad
\bar{D}^2 \Phi ^+ \Phi ^+ \Phi ^+,\quad
\bar{D}^2 \Phi ^+ \Phi ^+ \Phi ^+ \Phi ^+,\quad
\bar{D}^2 \Phi ^+ \Phi ^+ \Phi ^+ \Phi ^+ \Phi ^+,\nonumber\\
&&\bar{D}^2 \Phi ^+ \Phi ^+ \Phi ^+ \Phi ^+ \Phi ^+ \Phi ^+.
\end{eqnarray}
}
\item
$N_{\Phi }>N_{D^2}$, $N_{\bar{D}}=0$, $N_{\partial \partial }=0$.
{\small
\begin{eqnarray}
\label{Lambda2-type5}
&&D^2 \Phi  \Phi ,\quad 
D^2 \Phi  \Phi  \Phi ,\quad 
D^2 D^2 \Phi  \Phi  \Phi ,\quad 
D^2 \Phi  \Phi  \Phi ^+,\nonumber\\
&&D^2 D^2 \Phi  \Phi  \Phi  \Phi ,\quad
D^2 \Phi  \Phi  \Phi  \Phi ^+,\quad 
D^2 D^2 \Phi  \Phi  \Phi  \Phi ^+,\quad
D^2 \Phi  \Phi  \Phi ^+ \Phi ^+,\quad 
D^2 \Phi  \Phi  \Phi ^+ \Phi ^+ \Phi ^+,\nonumber\\
&&D^2 \Phi  \Phi  \Phi  \Phi ^+ \Phi ^+,\quad 
D^2 D^2 \Phi  \Phi  \Phi  \Phi  \Phi ^+, \quad
D^2 \Phi  \Phi  \Phi  \Phi ^+ \Phi ^+ \Phi ^+.
\end{eqnarray}
}
\item
The other operators,
{\small
\begin{eqnarray}
\label{Lambda2-type6}
&&D^2 \partial\partial\Phi  \Phi ,\quad
D^2 D^2 \partial\partial \Phi  \Phi  \Phi,\nonumber\\
&&D^2 \overline{D}^2 \Phi  \Phi ^+,\quad
\partial D  \bar{D} \Phi  \Phi ^+,\quad 
\partial\partial \Phi  \Phi ^+,\nonumber\\
&&D^2 \overline{D}^2 \Phi  \Phi ^+ \Phi ^+,\quad
\partial D  \bar{D}\Phi  \Phi ^+ \Phi ^+,\quad 
\partial\partial\Phi  \Phi ^+ \Phi ^+,\nonumber\\
&&D^2 D^2 \overline{D}^2 \Phi  \Phi  \Phi ^+,\quad
D^2 \partial D  \bar{D}\Phi  \Phi  \Phi ^+,\quad 
D^2 \partial\partial \Phi  \Phi  \Phi ^+,\nonumber\\
&&D^2 \overline{D}^2 \Phi  \Phi ^+ \Phi ^+ \Phi ^+,\quad
\partial D  \bar{D} \Phi  \Phi ^+ \Phi ^+ \Phi ^+, \quad 
\partial\partial \Phi  \Phi ^+ \Phi ^+ \Phi ^+,\nonumber\\
&&D^2 D^2 \overline{D}^2 \Phi  \Phi  \Phi ^+ \Phi ^+,\quad
D^2 \partial D  \bar{D} \Phi  \Phi  \Phi ^+ \Phi ^+,\quad
D^2 \partial\partial \Phi  \Phi  \Phi ^+ \Phi ^+, \quad\nonumber\\
&&D^2 \overline{D}^2 \Phi  \Phi ^+ \Phi ^+ \Phi ^+ \Phi ^+,\quad 
\partial D  \bar{D}\Phi  \Phi ^+ \Phi ^+ \Phi ^+ \Phi ^+,\quad 
\partial\partial\Phi  \Phi ^+ \Phi ^+ \Phi ^+ \Phi ^+.
\end{eqnarray}
}
\end{itemize}

At the order of $\Lambda^4$, 
\begin{itemize}
\label{}
\item
$N_{\Phi }=N_{D^2}$, $N_{\bar{D}}=0$, $N_{\partial \partial }=0$.
{\small
\begin{eqnarray}
\label{Lambda4-type1}
&&D^2 D^2 D^2 \Phi  \Phi  \Phi ,\nonumber\\
&&D^2 D^2 \Phi  \Phi  \Phi ^+ \Phi ^+,\quad 
D^2 D^2 D^2 \Phi  \Phi  \Phi  \Phi ^+,\quad 
D^2 D^2 D^2 D^2 \Phi  \Phi  \Phi  \Phi ,\nonumber\\
&&D^2 \Phi  \Phi ^+ \Phi ^+ \Phi ^+ \Phi ^+,\quad 
D^2 D^2 \Phi  \Phi  \Phi ^+ \Phi ^+ \Phi ^+,\quad 
D^2 D^2 D^2 \Phi  \Phi  \Phi  \Phi ^+ \Phi ^+,\nonumber\\
&&D^2 \Phi  \Phi ^+ \Phi ^+ \Phi ^+ \Phi ^+ \Phi ^+,\quad 
D^2 D^2 \Phi  \Phi  \Phi ^+ \Phi ^+ \Phi ^+ \Phi ^+,\nonumber\\
&&D^2 \Phi  \Phi ^+ \Phi ^+ \Phi ^+ \Phi ^+ \Phi ^+ \Phi ^+.
\end{eqnarray}
}
\item
$N_{\Phi }=N_{D^2}$, $N_{\bar{D}}=0$, $N_{\partial \partial }=1$.
{\small
\begin{eqnarray}
\label{Lambda4-type2}
&&D^2 D^2 \partial\partial \Phi  \Phi  \Phi ^+,\quad 
D^2 D^2 D^2 \partial\partial \Phi  \Phi  \Phi ,\nonumber\\
&&D^2 \partial\partial \Phi  \Phi ^+ \Phi ^+ \Phi ^+,\quad 
D^2 D^2 \partial\partial \Phi  \Phi  \Phi ^+ \Phi ^+,\quad 
D^2 D^2 D^2 \partial\partial \Phi  \Phi  \Phi  \Phi ^+,\nonumber\\
&&\partial\partial \Phi ^+ \Phi ^+ \Phi ^+ \Phi ^+ \Phi ^+,\quad 
D^2 \partial\partial \Phi  \Phi ^+ \Phi ^+ \Phi ^+ \Phi ^+,\quad 
D^2 D^2 \partial\partial \Phi  \Phi  \Phi ^+ \Phi ^+ \Phi ^+,\nonumber\\
&&\partial\partial \Phi ^+ \Phi ^+ \Phi ^+ \Phi ^+ \Phi ^+ \Phi ^+,\quad 
D^2 \partial\partial \Phi  \Phi ^+ \Phi ^+ \Phi ^+ \Phi ^+ \Phi ^+,\nonumber\\
&&\partial\partial \Phi ^+ \Phi ^+ \Phi ^+ \Phi ^+ \Phi ^+ \Phi ^+ \Phi ^+.
\end{eqnarray}
}
\item
$N_{\Phi }=N_{D^2}$, $N_{\bar{D}}=0$, $N_{\partial \partial }=2$.
{\small
\begin{eqnarray}
\label{Lambda4-type3}
&&D^2 D^2 \partial\partial \partial\partial \Phi  \Phi ,\nonumber\\
&&D^2 \partial\partial \partial\partial \Phi  \Phi ^+ \Phi ^+,\quad 
D^2 D^2 \partial\partial \partial\partial \Phi  \Phi  \Phi ^+,\quad 
D^2 D^2 D^2 \partial\partial \partial\partial \Phi  \Phi  \Phi ,\nonumber\\
&&\partial\partial \partial\partial \Phi ^+ \Phi ^+ \Phi ^+ \Phi ^+,\quad 
D^2 \partial\partial \partial\partial \Phi  \Phi ^+ \Phi ^+ \Phi ^+,\quad 
D^2 D^2 \partial\partial \partial\partial \Phi  \Phi  \Phi ^+ \Phi ^+,\nonumber\\
&&\partial\partial \partial\partial \Phi ^+ \Phi ^+ \Phi ^+ \Phi ^+ \Phi ^+,\quad 
D^2 \partial\partial \partial\partial \Phi  \Phi ^+ \Phi ^+ \Phi ^+ \Phi ^+,\nonumber\\
&&\partial\partial \partial\partial \Phi ^+ \Phi ^+ \Phi ^+ \Phi ^+ \Phi ^+ \Phi ^+.
\end{eqnarray}
}
\item
$N_{\Phi }>N_{D^2}$, $N_{\bar{D}}=0$, $N_{\partial \partial }=0$.
{\small
\begin{eqnarray}
\label{Lambda4-type4}
&&D^2 D^2 D^2 \Phi  \Phi  \Phi  \Phi ,\nonumber\\
&&D^2 D^2 \Phi  \Phi  \Phi  \Phi ^+ \Phi ^+,\quad 
D^2 D^2 D^2 \Phi  \Phi  \Phi  \Phi  \Phi ,\quad 
D^2 D^2 D^2 \Phi  \Phi  \Phi  \Phi  \Phi ^+,\quad 
D^2 D^2 D^2 D^2 \Phi  \Phi  \Phi  \Phi  \Phi ,\nonumber\\
&&D^2 \Phi  \Phi  \Phi ^+ \Phi ^+ \Phi ^+ \Phi ^+,\quad 
D^2 D^2 \Phi  \Phi  \Phi  \Phi ^+ \Phi ^+ \Phi ^+,\quad 
D^2 D^2 \Phi  \Phi  \Phi  \Phi  \Phi ^+ \Phi ^+,\nonumber\\
&&D^2 D^2 D^2 \Phi  \Phi  \Phi  \Phi  \Phi ^+ \Phi ^+,\quad 
D^2 D^2 D^2 \Phi  \Phi  \Phi  \Phi  \Phi  \Phi ^+,\quad 
D^2 D^2 D^2 D^2 \Phi  \Phi  \Phi  \Phi  \Phi  \Phi ,\nonumber\\
&&D^2 \Phi  \Phi  \Phi ^+ \Phi ^+ \Phi ^+ \Phi ^+ \Phi ^+,\quad 
D^2 \Phi  \Phi  \Phi  \Phi ^+ \Phi ^+ \Phi ^+ \Phi ^+,\quad 
D^2 D^2 \Phi  \Phi  \Phi  \Phi ^+ \Phi ^+ \Phi ^+ \Phi ^+,\nonumber\\
&&D^2 D^2 \Phi  \Phi  \Phi  \Phi  \Phi ^+ \Phi ^+ \Phi ^+,\quad 
D^2 D^2 D^2 \Phi  \Phi  \Phi  \Phi  \Phi  \Phi ^+ \Phi ^+,\nonumber\\
&&D^2 \Phi  \Phi  \Phi ^+ \Phi ^+ \Phi ^+ \Phi ^+ \Phi ^+ \Phi ^+,\quad 
D^2 \Phi  \Phi  \Phi  \Phi ^+ \Phi ^+ \Phi ^+ \Phi ^+ \Phi ^+,\quad
D^2 D^2 \Phi  \Phi  \Phi  \Phi  \Phi ^+ \Phi ^+ \Phi ^+ \Phi ^+, \nonumber\\
&&D^2 \Phi  \Phi  \Phi  \Phi ^+ \Phi ^+ \Phi ^+ \Phi ^+ \Phi ^+ \Phi ^+.
\end{eqnarray}
}
\item
The other operators,
{\small
\begin{eqnarray}
\label{Lambda4-type5}
&&D^2 D^2 D^2 \partial\partial \Phi  \Phi  \Phi  \Phi ,\nonumber\\
&&D^2 D^2 D^2 \overline{D}^2 \Phi  \Phi  \Phi  \Phi ^+,\quad 
D^2 D^2 \partial D \overline{D} \Phi  \Phi  \Phi  \Phi ^+,
D^2 D^2 \partial\partial \Phi  \Phi  \Phi  \Phi ^+,\nonumber\\
&&D^2 D^2 \overline{D}^2 \Phi  \Phi  \Phi ^+ \Phi ^+ \Phi ^+,\quad 
D^2 \partial D \overline{D} \Phi  \Phi  \Phi ^+ \Phi ^+ \Phi ^+,\quad 
D^2 \partial\partial \Phi  \Phi  \Phi ^+ \Phi ^+ \Phi ^+,\nonumber\\
&&D^2 D^2 D^2 \overline{D}^2 \Phi  \Phi  \Phi  \Phi ^+ \Phi ^+,\quad 
D^2 D^2 \partial D \overline{D} \Phi  \Phi  \Phi  \Phi ^+ \Phi ^+,\quad
D^2 D^2 \partial\partial \Phi  \Phi  \Phi  \Phi ^+ \Phi ^+,\nonumber\\
&&D^2 D^2 D^2 D^2 \overline{D}^2 \Phi  \Phi  \Phi  \Phi  \Phi ^+,\quad 
D^2 D^2 D^2 \partial D \overline{D} \Phi  \Phi  \Phi  \Phi  \Phi ^+,\quad 
D^2 D^2 D^2 \partial\partial \Phi  \Phi  \Phi  \Phi  \Phi ^+,\nonumber\\
&&D^2 \overline{D}^2 \Phi  \Phi ^+ \Phi ^+ \Phi ^+ \Phi ^+ \Phi ^+,\quad 
\partial D \overline{D} \Phi  \Phi ^+ \Phi ^+ \Phi ^+ \Phi ^+ \Phi ^+,\quad 
\partial\partial \Phi  \Phi ^+ \Phi ^+ \Phi ^+ \Phi ^+ \Phi ^+,\nonumber\\
&&D^2 D^2 \overline{D}^2 \Phi  \Phi  \Phi ^+ \Phi ^+ \Phi ^+ \Phi ^+,\quad 
D^2 \partial D \overline{D} \Phi  \Phi  \Phi ^+ \Phi ^+ \Phi ^+ \Phi ^+,\quad 
D^2 \partial\partial \Phi  \Phi  \Phi ^+ \Phi ^+ \Phi ^+ \Phi ^+,\nonumber\\
&&D^2 D^2 D^2 \overline{D}^2 \Phi  \Phi  \Phi  \Phi ^+ \Phi ^+ \Phi ^+,\quad 
D^2 D^2 \partial D \overline{D} \Phi  \Phi  \Phi  \Phi ^+ \Phi ^+ \Phi ^+,\quad 
D^2 D^2 \partial\partial \Phi  \Phi  \Phi  \Phi ^+ \Phi ^+ \Phi ^+,\nonumber\\
&&D^2 \overline{D}^2 \Phi  \Phi ^+ \Phi ^+ \Phi ^+ \Phi ^+ \Phi ^+ \Phi ^+,\quad 
\partial D \overline{D} \Phi  \Phi ^+ \Phi ^+ \Phi ^+ \Phi ^+ \Phi ^+ \Phi ^+,\quad 
\partial\partial \Phi  \Phi ^+ \Phi ^+ \Phi ^+ \Phi ^+ \Phi ^+ \Phi ^+,\nonumber\\
&&D^2 D^2 \overline{D}^2 \Phi  \Phi  \Phi ^+ \Phi ^+ \Phi ^+ \Phi ^+ \Phi ^+,\quad 
D^2 \partial D \overline{D} \Phi  \Phi  \Phi ^+ \Phi ^+ \Phi ^+ \Phi ^+ \Phi ^+,\quad 
D^2 \partial\partial \Phi  \Phi  \Phi ^+ \Phi ^+ \Phi ^+ \Phi ^+ \Phi ^+,\nonumber\\
&&D^2 \overline{D}^2 \Phi  \Phi ^+ \Phi ^+ \Phi ^+ \Phi ^+ \Phi ^+ \Phi ^+ \Phi ^+,\quad 
\partial D \overline{D} \Phi  \Phi ^+ \Phi ^+ \Phi ^+ \Phi ^+ \Phi ^+ \Phi ^+ \Phi ^+,\quad 
\partial\partial \Phi  \Phi ^+ \Phi ^+ \Phi ^+ \Phi ^+ \Phi ^+ \Phi ^+ \Phi ^+.\nonumber\\
\end{eqnarray}
}
\end{itemize}

\section{BFNC parameters}
\label{BFNC parameters}

In order to combine with possible divergent operators, we need BFNC parameters with the following indices,
\begin{equation}
\label{index}
k~l,\quad 
\alpha~ \beta ,\quad 
\dot{\alpha }~\dot{\beta },\quad 
k ~\dot{\alpha }~\beta ,\quad 
k~l~\alpha ~\beta ,\quad 
k~l~n~o,
\end{equation}
they can only be constructed by using the following elements,
\begin{eqnarray}
\label{all_symbol}
& & \epsilon _{\alpha  \beta }, \quad \epsilon ^{\alpha  \beta },\quad \epsilon _{\dot{\alpha }\dot{\beta }},\quad \epsilon ^{\dot{\alpha}\dot{\beta }},\quad \eta _{k l}, \quad \eta ^{k l},\nonumber \\ 
& &  \epsilon ^{{klmn}}, \quad  (\sigma ^{k l} )^{\alpha  \beta },\quad  (\bar{\sigma }^k )^{\dot{\alpha }\beta },\quad  \Lambda^k{}_{\alpha}.
\end{eqnarray}

By using the algebraic relations of covariant derivatives $D$ and $\bar{D}$, we found some BFNC parameters have certain symmetries. For example, if the BFNC parameter contains only $\alpha ~\beta$ and $\dot{\alpha }~\dot{\beta }$, it must be antisymmetric with respect to these indices.
On the other hand, we can obtain some BFNC parameters from others. 
For example, if we combine parameters with indices $k~l~\alpha~ \beta $ and  parameters with indices $\epsilon _{\alpha \beta }$ or  $\eta _{{kl}}$, we can obtain parameters with indices $k~l$ or $\alpha ~\beta$. The parameters with indices $k~ l$ can also be obtained from parameters with indices $k~l~n~o$.
Because the indices $\dot{\alpha }~\beta$ can be projected on $\left(\bar{\sigma }^k\right)^{\dot{\alpha }\beta }$, if we find all of the parameters with indices $k~l$, then combine them with $\left(\bar{\sigma }^k\right)^{\dot{\alpha }\beta }$ by using $\eta _{{kl}}$, we will obtain all parameters with indices $k ~\dot{\alpha }~\beta$
Because the indices $k~l~\alpha ~\beta $ can be projected on $\epsilon^{\alpha \beta }$ and $\left(\sigma ^{{kl}}\right)^{\alpha \beta }$, to construct parameters with indices $k~l~\alpha ~\beta $, we can combine $\epsilon^{\alpha \beta }$ and parameters with indices  $k~l$, or combine $\left(\sigma ^{{kl}}\right)^{\alpha \beta }$ and parameters with indices  $k~l$ or $k~l~n~o$.
We do not need to consider parameters with more Bosonic indices, because combining $\left(\sigma ^{{kl}}\right)^{\alpha \beta }$ and these parameters equivalent to combining $\left(\sigma ^{{kl}}\right)^{\alpha \beta }$ and parameters with indices  $k~l$ or $k~l~n~o$.

Based on above consideration, the task to construct parameters with indices in Eq.~(\ref{index}) is equivalent to constructing all parameters with indices $k~l~n~o$.
We give the details in Appendix.~\ref{construct BFNC parameters} and only list the BFNC parameters at order $\Lambda ^0$, $\Lambda ^2$ and $\Lambda ^4$.
\begin{itemize}
\item
At order $\Lambda ^0$, parameters with indices $kl$ and $klno$,
{\small
 \begin{eqnarray}
\label{}
\eta^{kl};\qquad \qquad 
\eta^{kl}~\eta^{no}, \quad 
\epsilon ^{klno}.
\end{eqnarray}
}
\item
At order $\Lambda ^2$, parameters without any indices,
{\small
\begin{equation}
\label{}
\Lambda ^2,\quad 
\sigma \Lambda \Lambda.
\end{equation}
}
with indices $\alpha~\beta$,
{\small
\begin{equation}
\label{}
\Lambda ^2~ \epsilon ^{\alpha  \beta },\quad 
\sigma \Lambda \Lambda  ~\epsilon ^{\alpha  \beta };\qquad\qquad
\Lambda ^2 ~\epsilon ^{\dot{\alpha } \dot{\beta }},\quad 
\sigma \Lambda \Lambda  ~\epsilon ^{\dot{\alpha } \dot{\beta }}.
\end{equation}
}
with indices $k~l$,
{\small
 \begin{eqnarray}
\label{Lambda2kl}
&&\Lambda ^2~\eta ^{kl}, \quad 
\sigma \Lambda \Lambda~\eta ^{kl},\quad 
\Lambda ^{kl}, \quad 
\left(\eta \sigma \Lambda \Lambda ^{k}\right){}^{l}.
\end{eqnarray}
}
with indices $k~\dot{\alpha }~\beta$,
{\small
\begin{eqnarray}
\label{}
& &\Lambda ^2 ~(\bar{\sigma }^k)^{\dot{\alpha } \beta },\quad 
\sigma \Lambda \Lambda~  (\bar{\sigma }^k)^{\dot{\alpha } \beta },\quad 
(\bar{\sigma }^n)^{\dot{\alpha } \beta }~\Lambda ^k{}_n , \nonumber\\
& &\eta _{n l} ~(\bar{\sigma }^l)^{\dot{\alpha } \beta }~ (\eta \sigma \Lambda \Lambda ^n)^k,\quad
\eta _{n l} ~(\bar{\sigma }^l)^{\dot{\alpha } \beta }~ (\eta \sigma \Lambda \Lambda ^k)^n;
\end{eqnarray}
}
with indices $k~l~n~o$,
{\small
\begin{eqnarray}
\label{Lambda2klno}
&&\Lambda ^2~\eta ^{kl}~\eta ^{no}, \quad 
\sigma \Lambda \Lambda~\eta ^{kl}~\eta ^{no} ,\quad
\Lambda ^2~\epsilon ^{klno},\nonumber\\
&&\eta ^{kl}~\Lambda ^{no}, \quad 
\eta ^{kl}~\left(\eta \sigma \Lambda \Lambda ^{n}\right){}^{o}, \quad 
\epsilon ^{klnp}~\Lambda _{p}{}^{o}, \quad 
\left(\sigma \Lambda \Lambda ^{kl}\right){}^{no}.
\end{eqnarray}
}
with indices  $k~ l~ \alpha ~ \beta$,
{\small
\begin{eqnarray}
\label{Lambda2klab}
&&\Lambda ^2~\eta ^{kl}~\epsilon ^{\alpha  \beta},\quad 
\sigma \Lambda \Lambda~ \eta ^{kl}~\epsilon ^{\alpha  \beta},\quad 
\Lambda ^2~\left(\sigma ^{kl}\right){}^{\alpha  \beta},\quad 
\sigma \Lambda \Lambda ~\left(\sigma ^{kl}\right){}^{\alpha  \beta},\nonumber\\
&&\left(\eta \sigma \Lambda \Lambda ^{k}\right){}^{l}~\epsilon ^{\alpha  \beta},\quad 
\Lambda ^{kl}~\epsilon ^{\alpha  \beta},\quad 
\Lambda ^{k\alpha}~\Lambda ^{l\beta },\quad
\epsilon ^{klno}~\Lambda _{n}{}^{\alpha }~\Lambda _{o}{}^{\beta},\nonumber\\
&&\left(\sigma ^{kn}\right){}^{\alpha  \beta}~\Lambda _{n}{}^{l},\quad 
\left(\sigma \Lambda ^{kn}\right){}^{l\alpha}~\Lambda _{n}{}^{\beta},\quad 
\left(\eta \sigma \Lambda ^{k}\right){}^{\alpha}~\Lambda ^{l\beta},\quad
\eta ^{kl}~\left(\eta \sigma \Lambda ^{n}\right){}^{\alpha }~\Lambda _{n}{}^{\beta}.
\end{eqnarray}
}
\item
At order $\Lambda ^4$, parameters without any indices,
{\small
\begin{eqnarray}
\label{}
\Lambda ^2~\Lambda ^2,\quad 
\Lambda ^2~\sigma \Lambda \Lambda,\quad
\sigma \Lambda \Lambda  ~\sigma \Lambda \Lambda 
\end{eqnarray}
}
with indices $\alpha~\beta$,
{\small
\begin{eqnarray}
\label{}
&&\Lambda ^2~\Lambda ^2~\epsilon ^{\alpha  \beta },\quad 
\Lambda ^2~\sigma \Lambda \Lambda~ \epsilon ^{\alpha  \beta },\quad 
\sigma \Lambda \Lambda  ~\sigma \Lambda \Lambda ~\epsilon ^{\alpha  \beta };\nonumber\\
&&\Lambda ^2~\Lambda ^2~\epsilon ^{\dot{\alpha } \dot{\beta }},\quad 
\Lambda ^2~\sigma \Lambda \Lambda ~\epsilon ^{\dot{\alpha } \dot{\beta }},\quad 
\sigma \Lambda \Lambda ~ \sigma \Lambda \Lambda ~\epsilon ^{\dot{\alpha } \dot{\beta }}.
\end{eqnarray}
}
with indices $k~ l$,
{\small
\begin{eqnarray}
\label{Lambda4kl}
&&\Lambda ^2~\Lambda ^2~\eta ^{kl},\quad 
\Lambda ^2~\sigma \Lambda \Lambda~\eta ^{kl} ,\quad 
\sigma \Lambda \Lambda ~ \sigma \Lambda \Lambda ~\eta ^{kl},\quad
\Lambda ^2~\Lambda ^{kl},\quad 
\sigma \Lambda \Lambda ~\Lambda ^{kl},\nonumber\\
&&\Lambda ^2~\left(\eta \sigma \Lambda \Lambda ^{k}\right){}^{l},\quad
\sigma \Lambda \Lambda~\left(\eta \sigma \Lambda \Lambda ^{k}\right){}^{l},\quad 
\left(\eta \sigma \Lambda \Lambda ^{n}\right){}^{k}~\Lambda _{n}{}^{l}.
\end{eqnarray}
}
with indices $k~\dot{\alpha }~\beta$,
{\small 
\begin{eqnarray}
\label{}
&&\Lambda ^2~ \Lambda ^2 ~\left(\bar{\sigma }^{k}\right)^{\dot{\alpha } \beta },\quad 
\Lambda ^2~ \sigma \Lambda \Lambda  ~\left(\bar{\sigma }^{k}\right)^{\dot{\alpha } \beta },\quad 
\sigma \Lambda \Lambda  ~\sigma \Lambda \Lambda ~ \left(\bar{\sigma }^{k}\right)^{\dot{\alpha } \beta },\nonumber\\
&&\Lambda ^2~ \left(\bar{\sigma }^{n}\right)^{\dot{\alpha } \beta } ~\Lambda ^{k}{}_{n},\quad 
\sigma \Lambda \Lambda~  \left(\bar{\sigma }^{n}\right)^{\dot{\alpha } \beta } ~\Lambda ^{k}{}_{n},\nonumber\\
&&\Lambda ^2~ \eta _{{n_1} {n_2}} ~\left(\bar{\sigma }^{{n_2}}\right)^{\dot{\alpha } \beta }~ \left(\eta \sigma \Lambda \Lambda ^{k}\right)^{{n_1}},\quad 
\Lambda ^2 ~\eta _{{n_1} {n_2}} ~\left(\bar{\sigma }^{{n_2}}\right)^{\dot{\alpha } \beta } ~\left(\eta \sigma \Lambda \Lambda ^{{n_1}}\right)^{k},\nonumber\\
&&\sigma \Lambda \Lambda  ~\eta _{{n_1} {n_2}}~ \left(\bar{\sigma }^{{n_2}}\right)^{\dot{\alpha } \beta } ~\left(\eta \sigma \Lambda \Lambda ^{k}\right)^{{n_1}},\quad 
\sigma \Lambda \Lambda ~ \eta _{{n_1} {n_2}} ~\left(\bar{\sigma }^{{n_2}}\right)^{\dot{\alpha } \beta } ~\left(\eta \sigma \Lambda \Lambda ^{{n_1}}\right)^{k},\nonumber\\
&&\eta _{{n_1} {n_2}} ~\left(\bar{\sigma }^{{n_2}}\right)^{\dot{\alpha } \beta } ~\left(\eta \sigma \Lambda \Lambda ^{{n_3}}\right)^{{n_1}} \Lambda ^{k}{}_{{n_3}},\quad
\Lambda _{{n_1} {n_2}} ~\left(\bar{\sigma }^{{n_2}}\right)^{\dot{\alpha } \beta } ~\left(\eta \sigma \Lambda \Lambda ^{{n_1}}\right)^{k}.
\end{eqnarray}
}
with indices $k~ l~ n~ o$, 
{\small
\begin{eqnarray}
\label{Lambda4klno}
&&\Lambda ^2~\Lambda ^2~\eta ^{kl}~\eta ^{no},\quad 
\Lambda ^2~\sigma \Lambda \Lambda~\eta ^{kl}~\eta ^{no} ,\quad 
\sigma \Lambda \Lambda  ~\sigma \Lambda \Lambda~ \eta ^{kl}~\eta ^{no},\quad 
\Lambda ^2~\Lambda ^2~\epsilon ^{klno},\nonumber\\
&&\Lambda ^2~\eta ^{kl}~\Lambda ^{no},\quad
\sigma \Lambda \Lambda ~\eta ^{kl}~\Lambda ^{no},\quad 
\Lambda ^2~\eta ^{kl}~\left(\eta \sigma \Lambda \Lambda ^{n}\right){}^{o},\quad 
\sigma \Lambda \Lambda~\eta ^{kl}~\left(\eta \sigma \Lambda \Lambda ^{n}\right){}^{o},\nonumber\\
&&\Lambda ^2~\epsilon ^{klnp}~\Lambda _{p}{}^{o},\quad
\Lambda ^2~\left(\sigma \Lambda \Lambda ^{kl}\right){}^{no},\quad 
\sigma \Lambda \Lambda~\left(\sigma \Lambda \Lambda ^{kl}\right){}^{no},\nonumber\\
&&\Lambda ^{kl}~\Lambda ^{no}, \quad
\left(\eta \sigma \Lambda \Lambda ^{k}\right){}^{l}~\Lambda ^{no},\quad
\left(\sigma \Lambda \Lambda ^{kp}\right){}^{ln}~\Lambda _{p}{}^{o},\quad
\eta ^{kl}~\left(\eta \sigma \Lambda \Lambda ^{p}\right){}^{n}~\Lambda _{p}{}^{o},\nonumber\\
&&\epsilon ^{klpq}~\Lambda _{p}{}^{n}~\Lambda _{q}{}^{o}.
\end{eqnarray}
}
with indices $k~ l ~\alpha ~ \beta$,
{\small
\begin{eqnarray}
\label{Lambda4klab}
&&\Lambda ^2~ \Lambda ^2~\eta ^{kl} ~\epsilon ^{\alpha \beta} ,\quad 
\Lambda ^2 ~\sigma \Lambda \Lambda~\eta ^{kl} ~\epsilon ^{\alpha \beta}  ,\quad 
\sigma \Lambda \Lambda ~ \sigma \Lambda \Lambda~\eta ^{kl} ~\epsilon ^{\alpha \beta}\nonumber\\
&&\Lambda ^2 ~\Lambda ^2 ~\left(\sigma ^{kl}\right){}^{\alpha \beta},\quad 
\Lambda ^2 ~\sigma \Lambda \Lambda  ~\left(\sigma ^{kl}\right){}^{\alpha \beta},\quad 
\sigma \Lambda \Lambda  ~\sigma \Lambda \Lambda  \left(\sigma ^{kl}\right){}^{\alpha \beta},\nonumber\\
&&\Lambda ^2 ~\left(\eta \sigma \Lambda \Lambda ^{k}\right){}^{l}~\epsilon ^{\alpha \beta} ,\quad 
\sigma \Lambda \Lambda ~ \left(\eta \sigma \Lambda \Lambda ^{k}\right){}^{l}~\epsilon ^{\alpha \beta},\quad
\Lambda ^2 ~\Lambda ^{kl}~\epsilon ^{\alpha \beta},\quad 
\sigma \Lambda \Lambda ~\Lambda ^{kl}~\epsilon ^{\alpha \beta} ,\nonumber\\
&&\Lambda ^2 ~\Lambda ^{k \alpha} ~\Lambda ^{l \beta},\quad 
\sigma \Lambda \Lambda ~ \Lambda ^{k \alpha} ~\Lambda ^{l \beta},\quad
\Lambda ^2 ~\epsilon ^{kl n o}~\Lambda _{n}{}^{\beta} ~\Lambda _{o}{}^{\alpha},\nonumber\\
&&\Lambda ^2 ~\left(\sigma ^{k n}\right){}^{\alpha \beta}~ \Lambda ^{l}{}_{n},\quad 
\sigma \Lambda \Lambda ~ \left(\sigma ^{k n}\right){}^{\alpha \beta} ~\Lambda ^{l}{}_{n},\quad
\Lambda ^2 ~\left(\sigma \Lambda ^{k n}\right){}^{l \alpha }~ \Lambda _{n}^{\beta},\quad
\sigma \Lambda \Lambda  ~\left(\sigma \Lambda ^{k n}\right){}^{l \alpha} ~\Lambda _{n}^{\beta},\nonumber\\
&&\Lambda ^2 ~\left(\eta \sigma \Lambda ^{k}\right){}^{\alpha}~ \Lambda ^{l \beta},\quad
\sigma \Lambda \Lambda ~ \left(\eta \sigma \Lambda ^{k}\right){}^{\alpha} ~\Lambda ^{l \beta},\quad
\Lambda ^2~ \eta ^{kl} ~\left(\eta \sigma \Lambda ^{n}\right){}^{\alpha} ~\Lambda _{n}^{\beta},\quad
\sigma \Lambda \Lambda ~ \eta ^{kl} ~\left(\eta \sigma \Lambda ^{n}\right){}^{\alpha} ~\Lambda _{n}^{\beta},\nonumber\\
&&\left(\eta \sigma \Lambda \Lambda ^{n}\right){}^{k} ~\Lambda ^{l}{}_{n}~\epsilon ^{\alpha \beta} ,\quad 
\Lambda ^{k}{}_{n}~ \Lambda ^{l}{}_{o}~\left(\sigma ^{no}\right){}^{\alpha \beta},\quad
\Lambda ^{kl}~ \left(\eta \sigma \Lambda ^{n}\right){}^{\alpha}~ \Lambda _{n}^{\beta},\quad
\epsilon ^{k n o p} ~\Lambda ^{l}{}_{n}~ \Lambda _{o}^{\alpha} ~\Lambda _{p}^{\beta},\nonumber\\
&&\left(\sigma \Lambda \Lambda ^{no}\right){}^{kl} ~\Lambda _{n}^{\alpha}~ \Lambda _{o}^{\beta},\quad
\left(\eta \sigma \Lambda \Lambda ^{n}\right){}^{k} ~\Lambda ^{l \alpha} ~\Lambda _{n}^{\beta},\quad
\left(\sigma \Lambda ^{no}\right){}^{k \alpha} ~\Lambda ^{l}{}_{n} ~\Lambda _{o}^{\beta},\quad
\left(\eta \sigma \Lambda ^{n}\right){}^{\alpha}~ \Lambda ^{k \beta} ~\Lambda ^{l}{}_{n}.
\end{eqnarray}
}
\end{itemize}

\section{Construct 1/2 supersymmetric invariant action}
\label{construct action}

We can construct 1/2 supersymmetric invariant action by using the BFNC parameters and divergent operators as follows,
\begin{enumerate}
\label{}
\item 
We can remove $\theta ^2$ in  $\int d^4x~ d^4\theta  ~\theta ^4$ by integrating by part of $D^2$ .

\item
We do not apply $D_{\alpha }$, $\overline{D}_{\dot{\alpha }}$ on the same superfield, because we will obtain $\left(\sigma ^k\right){}_{\alpha  \dot{\beta }}~\partial _k$ by using $D$ algebraic relations.

\item
Take into account the possible simplification when combining BFNC parameters  with $\partial _k\partial _l$ or $D_{\alpha }D_{\beta }$.

\item
For operators with the same number of superfields and containing $D^2\overline{D}^2$, $\partial D \overline{D}$ or $\partial \partial$, we only keep the $\overline{D}^2$. 

\item
We define new parameters by using the BFNC parameters. At the order of $\Lambda ^2$, we define $X_i$, $X_i{}^{{k_1} {k_2}}$, $X_i{}^{{k_1} {k_2} {k_3} {k_4}}$, $\overline{X}_i{}^{{k_1} {k_2}, {k_3} {k_4}}$, $X_i{}^{{k_1} \dot{\alpha }\beta }$, $X_i{}^{{k_1} {k_2} \alpha  \beta }$, $\overline{X}_i{}^{{k_1} {k_2} \alpha  \beta }$, see Eq.~(\ref{X0})-(\ref{X4-4}).
At the order of $\Lambda ^4$, we define 
$Y_i$, $Y_i{}^{{k_1} {k_2}}$, $\overline{Y}_i{}^{{k_1} {k_2}}$, $Y_i{}^{{k_1} {k_2} {k_3} {k_4}}$, $\overline{Y}_i{}^{{k_1},{k_2} {k_3} {k_4}}$, $\widetilde{Y}_i{}^{{k_1} {k_2},{k_3} {k_4}}$, $\hat{Y}_i{}^{{k_1} {k_2},{k_3} {k_4}}$, $\widetilde{\overline{Y}}_i{}^{{k_1},{k_2} {k_3},{k_4}}$, $Y_i{}^{{k_1} \dot{\alpha }\beta }$, $Y_i{}^{{k_1} {k_2} \alpha  \beta }$, $\overline{Y}_i{}^{{k_1} {k_2} \alpha  \beta }$, $\widetilde{Y}_i{}^{{k_1} {k_2} \alpha  \beta }$, see Eq.~(\ref{Y0})-(\ref{Ylast}).

\end{enumerate}

At the order of $\Lambda ^2$, by using Eq.~(\ref{Lambda2-type1}) we can construct the following action,
{\small
\begin{eqnarray}
\label{L2-1}
&&\int d^4x~d^4\theta~\Bigg\{
X_1~\theta ^4 ~\Phi ^+
+X_2~\theta ^4 \left(\Phi ^+\right)^2
+X_3 ~\theta ^4 \left(\Phi ^+\right)^3
+X_4 ~\theta ^4 \left(\Phi ^+\right)^4
+X_5 ~\theta ^4  \left(\Phi ^+\right)^5
+X_6~\overline{\theta }^2 ~\Phi 
\nonumber\\&&
+X_7~\overline{\theta }^2~ \Phi ~\Phi ^+
+X_8~\overline{\theta }^2 \left(D^2 \Phi \right) \Phi 
+X_9~\overline{\theta }^2~ \Phi  \left(\Phi ^+\right)^2
+X_{10}~\overline{\theta }^2 \left(D^2 \Phi \right) \Phi ~ \Phi ^+
+X_{11}~\overline{\theta }^2~ \Phi  \left(\Phi ^+\right)^3
\Bigg\}.\nonumber\\
\end{eqnarray}
}
by using Eq.~(\ref{Lambda2-type2}) we obtain,
{\small
\begin{eqnarray}
\label{L2-2}
&&\int d^4x~d^4\theta~\Bigg\{
X_{12}{}^{{k_1} {k_2}}~\theta ^4 \left(\partial _{{k_1}}\partial _{{k_2}}\Phi ^+\right) \Phi ^+
+X_{13}{}^{{k_1} {k_2}}~\overline{\theta }^2~ \Phi  \left(\partial _{{k_1}}\partial _{{k_2}}\Phi ^+\right)
+X_{14}{}^{{k_1} {k_2}}~\overline{\theta }^2 \left(\partial _{{k_1}}\partial _{{k_2}}D^2 \Phi \right) \Phi 
\nonumber\\&&
+X_{15}{}^{{k_1} {k_2}}~\theta ^4 \left(\partial _{{k_1}}\partial _{{k_2}}\Phi ^+\right) \left(\Phi ^+\right)^2
+X_{16}{}^{{k_1} {k_2}}~\overline{\theta }^2 ~\Phi  \left(\partial _{{k_1}}\Phi ^+\right)\left(\partial _{{k_2}}\Phi ^+\right)
+X_{17}{}^{{k_1} {k_2}}~\overline{\theta }^2~ \Phi  \left(\partial _{{k_1}}\partial _{{k_2}}\Phi ^+\right) \Phi ^+
\nonumber\\&&
+X_{18}{}^{{k_1} {k_2}}~\theta ^4 \left(\partial _{{k_1}}\partial _{{k_2}}\Phi ^+\right) \left(\Phi ^+\right)^3
\Bigg\}.
\end{eqnarray}
}
by using Eq.~(\ref{Lambda2-type3}) we obtain,
{\small
\begin{eqnarray}
\label{L2-3}
&&\int d^4x~d^4\theta~\Bigg\{
X_{19}{}^{{k_1} {k_2} {k_3} {k_4}}~\theta ^4 \left(\partial _{{k_1}}\partial _{{k_2}}\partial _{{k_3}}\partial _{{k_4}}\Phi ^+ \right) \Phi ^+
+X_{20}{}^{{k_1} {k_2} {k_3} {k_4}}~\overline{\theta }^2~\Phi  \left(\partial _{{k_1}}\partial _{{k_2}}\partial _{{k_3}}\partial _{{k_4}}\Phi ^+\right) 
\nonumber\\&&
+X_{21}{}^{{k_1} {k_2} {k_3} {k_4}}~\theta ^4 \left(\partial _{{k_1}}\partial _{{k_2}}\partial _{{k_3}}\partial _{{k_4}}\Phi ^+\right)  \left(\Phi ^+\right)^2
+\overline{X}_{22}{}^{{k_1} {k_2}, {k_3} {k_4}}~\theta ^4 \left(\partial _{{k_1}}\partial _{{k_2}}\Phi ^+\right) \left(\partial _{{k_3}}\partial _{{k_4}}\Phi ^+\right) \Phi ^+
\Bigg\}.
\end{eqnarray}
}
by using Eq.~(\ref{Lambda2-type4}) we obtain,
{\small
\begin{eqnarray}
\label{L2-4}
&&\int d^4x~d^4\theta~\Bigg\{
X_{23}~\theta ^4 \left(\bar{D}^2 \Phi ^+\right)
+X_{24}~\theta ^4 \left(\bar{D}^2 \Phi ^+\right) \Phi ^+
+X_{24}~\epsilon ^{\dot{\alpha } \dot{\beta }} ~\theta ^4 \left(\bar{D}_{\dot{\alpha }} \Phi ^+\right) \left(\bar{D}_{\dot{\beta }} \Phi ^+\right)
\nonumber\\&&
+\frac{1}{2} ~X_{25}~\theta ^4 \left(\bar{D}^2 \Phi ^+\right) \left(\Phi ^+\right)^2
+X_{25}~\epsilon ^{\dot{\alpha } \dot{\beta }} ~\theta ^4 \left(\bar{D}_{\dot{\alpha }} \Phi ^+\right) \left(\bar{D}_{\dot{\beta }} \Phi ^+\right) \Phi ^+
+\frac{1}{3} ~X_{26}~ \theta ^4 \left(\bar{D}^2 \Phi ^+\right) \left(\Phi ^+\right)^3
\nonumber\\&&
+X_{26}~ \epsilon ^{\dot{\alpha } \dot{\beta }} ~\theta ^4 \left(\bar{D}_{\dot{\alpha }} \Phi ^+\right) \left(\bar{D}_{\dot{\beta }} \Phi ^+\right)\left(\Phi ^+\right)^2
+\frac{1}{4} ~X_{27}~ \theta ^4 \left(\bar{D}^2 \Phi ^+\right) \left(\Phi ^+\right)^4
+X_{27}~ \epsilon ^{\dot{\alpha } \dot{\beta }} ~\theta ^4 \left(\bar{D}_{\dot{\alpha }} \Phi ^+\right) \left(\bar{D}_{\dot{\beta }} \Phi ^+\right) \left(\Phi ^+\right)^3
\nonumber\\&&
+\frac{1}{5} ~X_{28}~ \theta ^4 \left(\bar{D}^2 \Phi ^+\right) \left(\Phi ^+\right)^5
+X_{28}~\epsilon ^{\dot{\alpha } \dot{\beta }} ~\theta ^4 \left(\bar{D}_{\dot{\alpha }} \Phi ^+\right) \left(\bar{D}_{\dot{\beta }} \Phi ^+\right)\left(\Phi ^+\right)^4
\Bigg\}.
\end{eqnarray}
}
by using Eq.~(\ref{Lambda2-type5}) we obtain,
{\small
\begin{eqnarray}
\label{L2-5}
&&\int d^4x~d^4\theta~\Bigg\{
X_{29}~\overline{\theta }^2~\Phi^2
+X_{30}~\overline{\theta }^2~\Phi^3
+X_{31}~\overline{\theta }^2~\left(D^2 \Phi \right) \Phi^2
+X_{32}~\overline{\theta }^2~\left(\Phi\right)^2 ~\Phi ^+
+X_{33}~\overline{\theta }^2~\left(D^2 \Phi \right) \Phi^3
\nonumber\\&&
+X_{34}~\overline{\theta }^2~\Phi^3 ~\Phi ^+
+X_{35}~\overline{\theta }^2~\left(D^2 \Phi \right) \Phi^2 ~\Phi ^+
+X_{36}~\overline{\theta }^2~\Phi^2 \left(\Phi ^+\right)^2
+X_{37}~\overline{\theta }^2~\Phi^2 \left(\Phi ^+\right)^3
\nonumber\\&&
+X_{38}~\overline{\theta }^2~\Phi^3 \left(\Phi ^+\right)^2
+X_{39}~\overline{\theta }^2~\left(D^2 \Phi \right) \Phi^3 ~\Phi ^+
+X_{40}~\overline{\theta }^2~\Phi^3 \left(\Phi ^+\right)^3
\Bigg\}.
\end{eqnarray}
}
by using Eq.~(\ref{Lambda2-type6}) we obtain,
{\small
\begin{eqnarray}
\label{L2-6}
&&\int d^4x~d^4\theta~\Bigg\{
X_{41}{}^{{k_1} {k_2}}~\overline{\theta }^2\left( \partial _{{k_1}}\partial _{{k_2}}\Phi  \right)\Phi
+X_{42}{}^{{k_1} {k_2}}~\overline{\theta }^2\left(\partial _{{k_1}}\partial _{{k_2}}D^2 \Phi \right) \Phi^2
+X_{43}{}^{{k_1} {k_2}}~\overline{\theta }^2 \left(D^2 \Phi \right) \left(\partial _{{k_1}}\partial _{{k_2}}\Phi \right) \Phi
\nonumber\\&&
+X_{44}{}^{{k_1} {k_2} \alpha  \beta }~\overline{\theta }^2 \left(\partial _{{k_1}}\partial _{{k_2}}D_{\alpha } \Phi \right) \left(D_{\beta } \Phi \right) \Phi
+\overline{X}_{45}{}^{{k_1} {k_2} \alpha  \beta }~\overline{\theta }^2 \left(\partial _{{k_1}}D_{\alpha } \Phi \right) \left(\partial _{{k_2}}D_{\beta } \Phi \right) \Phi
+X_{46}~\overline{\theta }^2 ~\Phi  \left(\overline{D}^2 \Phi ^+\right)
\nonumber\\&&
+X_{47}~\overline{\theta }^2 ~\Phi  \left(\overline{D}^2 \Phi ^+\right) \Phi ^+
+X_{48}~\epsilon ^{\dot{\alpha } \dot{\beta }}~\overline{\theta }^2 ~\Phi  \left(\overline{D}_{\dot{\alpha }} \Phi ^+\right) \left(\overline{D}_{\dot{\beta }} \Phi ^+\right)
+X_{49}~\overline{\theta }^2 \left(D^2 \Phi \right) \Phi  \left(\overline{D}^2 \Phi ^+\right)
\nonumber\\&&
+X_{50}~ \epsilon ^{\alpha  \beta }~\overline{\theta }^2 \left(D_{\alpha } \Phi \right) \left(D_{\beta } \Phi \right) \left(\overline{D}^2 \Phi ^+\right)
+X_{51}{}^{{k_1} \dot{\alpha }\beta }~\overline{\theta }^2 \left(\partial _{{k_1}}D_{\beta } \Phi \right) \Phi  \left(\overline{D}_{\dot{\alpha }} \Phi ^+\right)
\nonumber\\&&
+X_{52}{}^{{k_1} \dot{\alpha }\beta }~\overline{\theta }^2 \left(D_{\beta } \Phi \right) \left(\partial _{{k_1}}\Phi \right) \left(\overline{D}_{\dot{\alpha }} \Phi ^+\right)
+X_{53}{}^{{k_1} {k_2}}~\overline{\theta }^2 \left(\partial _{{k_1}}\partial _{{k_2}}\Phi \right) \Phi  ~\Phi ^+
+X_{54}{}^{{k_1} {k_2}}~\overline{\theta }^2 \left(\partial _{{k_1}}\Phi \right) \left(\partial _{{k_2}}\Phi \right) \Phi ^+
\nonumber\\&&
+X_{55}~\overline{\theta }^2 ~\Phi  \left(\overline{D}^2 \Phi ^+\right) \left(\Phi ^+\right)^2
+X_{56}~\epsilon ^{\dot{\alpha } \dot{\beta }}~\overline{\theta }^2 ~\Phi  \left(\overline{D}_{\dot{\alpha }} \Phi ^+\right) \left(\overline{D}_{\dot{\beta }} \Phi ^+\right) \Phi ^+
+X_{57}~\overline{\theta }^2 \left(D^2 \Phi \right) \Phi  \left(\overline{D}^2 \Phi ^+\right) \Phi ^+
\nonumber\\&&
+X_{58}~\epsilon ^{\dot{\alpha } \dot{\beta }}~\overline{\theta }^2 \left(D^2 \Phi \right) \Phi  \left(\overline{D}_{\dot{\alpha }} \Phi ^+\right) \left(\overline{D}_{\dot{\beta }} \Phi ^+\right)
+X_{59}~\epsilon ^{\alpha  \beta }~\overline{\theta }^2 \left(D_{\alpha } \Phi \right) \left(D_{\beta } \Phi \right) \left(\overline{D}^2 \Phi ^+\right) \Phi ^+
\nonumber\\&&
+X_{60}{}^{{k_1} \dot{\alpha }\beta }~\overline{\theta }^2 \left(\partial _{{k_1}}D_{\beta } \Phi \right) \Phi  \left(\overline{D}_{\dot{\alpha }} \Phi ^+\right) \Phi ^+
+X_{61}{}^{{k_1} \dot{\alpha }\beta }~\overline{\theta }^2 \left(D_{\beta } \Phi \right) \left(\partial _{{k_1}}\Phi \right) \left(\overline{D}_{\dot{\alpha }} \Phi ^+\right) \Phi ^+
\nonumber\\&&
+X_{62}{}^{{k_1} \dot{\alpha }\beta }~\overline{\theta }^2 \left(D_{\beta } \Phi \right) \Phi  \left(\partial _{{k_1}}\overline{D}_{\dot{\alpha }} \Phi ^+\right) \Phi ^+
+X_{63}{}^{{k_1} {k_2}}~\overline{\theta }^2 \left(\partial _{{k_1}}\partial _{{k_2}}\Phi \right) \Phi ~ \left(\Phi ^+\right)^2
\nonumber\\&&
+X_{64}{}^{{k_1} {k_2}}~\overline{\theta }^2 \left(\partial _{{k_1}}\Phi \right) \left(\partial _{{k_2}}\Phi \right) \left(\Phi ^+\right)^2
+X_{65}{}^{{k_1} {k_2}}~\overline{\theta }^2 ~\Phi^2  \left(\partial _{{k_1}}\partial _{{k_2}}\Phi ^+\right) \Phi ^+
\nonumber\\&&
+X_{66}~\overline{\theta }^2 ~\Phi  \left(\overline{D}^2 \Phi ^+\right) \left(\Phi ^+\right)^3
+X_{67}~\epsilon ^{\dot{\alpha } \dot{\beta }}~\overline{\theta }^2 ~\Phi  \left(\overline{D}_{\dot{\alpha }} \Phi ^+\right) \left(\overline{D}_{\dot{\beta }} \Phi ^+\right) \left(\Phi ^+\right)^2
\Bigg\}.
\end{eqnarray}
}
At the order of $\Lambda ^4$, we can construct the action corresponding to the operators in Eq.~(\ref{Lambda4-type1})-(\ref{Lambda4-type5}), see Eq.~(\ref{L4-1})-(\ref{L4-5}) in the Appendix.

\section{Conclusion and outlook}
\label{conclusion}

In summary, at the order of $\Lambda ^2$, we construct actions in Eq.~(\ref{L2-1})-(\ref{L2-6}) for divergent operators in Eq.~(\ref{Lambda2-type1})-(\ref{Lambda2-type6}). 
Except for the action in Eq.(\ref{L2-4}), all of the parameters are independent. 
After removing $D^2$, we found $\int d^4x d^4\theta ~ \theta ^4 \to  \int d^4x d^4\theta ~ \overline{\theta }^2$ for some terms in Eq.~(\ref{L2-1})-(\ref{L2-4}), and all terms in Eq.~(\ref{L2-5})-(\ref{L2-6}).

Because the deformed action Eq.~(\ref{S_NC}) have 1/2 supersymmetry, its effective action should also have 1/2 supersymmetry. 
The action in Eq.~(\ref{L2-1})-(\ref{L2-6}) that writen as a superspace integral $\int d^4x d^4\theta ~ \overline{\theta }^2$ have manifest 1/2 supersymmetry, because there is no $\theta ^2$ that do not commute with the supersymmetry generator $Q_{\alpha}$. The 1/2 supersymmetry invariance of the action with superspace integral $\int d^4x d^4\theta ~ \theta ^4$ in Eq.~(\ref{L2-1})-(\ref{L2-4}) can also be verified easily.
At the order of $\Lambda ^2$, we can add the action in Eq.~(\ref{L2-1})-(\ref{L2-6}) to the deformed action in Eq.~(\ref{S_NC}) and define a new action, which is renormalizable to all order in perturbative theory, because it contains all of the divergent operators and have 1/2 supersymmetry.

If we add the $\Lambda ^2$ order action in Eq.~(\ref{L2-1})-(\ref{L2-6}) and the $\Lambda ^4$ oder action in Eq.~(\ref{L4-1})-(\ref{L4-5}) to the deformed action in Eq.~(\ref{S_NC}), we obtain a renormalizable BFNC Wess-Zumino action at the order of $\Lambda ^4$. It is straightforward to extend the construction to the order of $\Lambda ^8$, which is the highest order of noncommutative parameters, because there are 8 different Fermionic noncommutative parameters $\Lambda ^{k \alpha }$.

We found there are many free parameters for BFNC Wess-Zumino model and expect the introduction of supergauge symmetry~\cite{Wang:2016c} will reduce the number of free parameters.

\section*{Acknowledgments}
The author would like to thank Professors Jian-Xin Lu, Yan-Gang Miao, Zhi-Guang Xiao for helpful discussions. This work was supported by a key grant from the NSF of China with Grant No: 11235010.

\appendix
\setcounter{equation}{0}
\renewcommand\theequation{A\arabic{equation}}

\section{The deformed Wess-Zumino model on BFNC superspace}

The Wess-Zumino model is defined on superspace as follows,
\begin{eqnarray}
\label{WZ_superfield_form}
{\cal S}_{\rm WZ}&=&\int d^8z\left\{\Phi ^+\Phi 
-\frac{m}{8}~\Phi  \left(\frac{D^2}{\square }\Phi \right)
-\frac{m^*}{8}~\Phi ^+ \left(\frac{\bar{D}^2}{\square }\Phi ^+\right)\right.
\nonumber\\&&
\hspace{15mm}\left.
-\frac{g}{12}~\Phi ~ \Phi \left(\frac{D^2}{\square }\Phi \right)
-\frac{g^*}{12}~\Phi ^+ ~\Phi ^+ \left(\frac{\bar{D}^2}{\square }\Phi ^+\right)\right\}.
\end{eqnarray}

BFNC star product is defined as follows,
\begin{equation}
\label{star product}
{\bf F} \star {\bf G}\equiv\mu \left\{ \exp \left[\frac{i}{2}~\Lambda ^{k \alpha }\left(\frac{\partial }{\partial y^k}\otimes \frac{\partial }{\partial \theta ^{\alpha }}-\frac{\partial }{\partial \theta ^{\alpha }}\otimes \frac{\partial }{\partial y^k}\right)\right]\triangleright({\bf F}\otimes {\bf G}) \right\},
\end{equation}

Taylor expansion,
\begin{eqnarray}
\label{star_product_expansion}
{\bf F}\star {\bf G}&=&{\bf F} {\bf G}-\frac{i}{2}~\Lambda ^{k \alpha }\left(\partial _{\alpha }{\bf F}\right)\left(\partial _k{\bf G}\right)+(-1)^{|{\bf F}|}~\frac{i}{2}~\Lambda ^{k \alpha }\left(\partial _k{\bf F}\right)\left(\partial _{\alpha }{\bf G}\right) \nonumber \\
&&+\frac{1}{8}~\Lambda ^{k \alpha }~\Lambda ^{l \beta }\left(\partial _k\partial _l{\bf F}\right)\left(\partial _{\alpha }\partial _{\beta }{\bf G}\right)+\frac{1}{8}~\Lambda ^{k \alpha }~\Lambda ^{l \beta }\left(\partial _{\alpha }\partial _{\beta }{\bf F}\right)\left(\partial _k\partial _l{\bf G}\right) \nonumber \\
&&+(-1)^{|{\bf F}|}~\frac{1}{4}~\Lambda ^{k \alpha }~\Lambda ^{l \beta }\left(\partial _{\beta }\partial _k{\bf F}\right)\left(\partial _{\alpha }\partial _l{\bf G}\right)\nonumber \\
&&-\frac{i}{16}~\Lambda ^{k \alpha }~\Lambda ^{l \beta }~\Lambda ^{m \zeta }\left(\partial _{\alpha }\partial _l\partial _m{\bf F}\right)\left(\partial _{\beta }\partial _{\zeta }\partial _k{\bf G}\right)\nonumber \\
&&+(-1)^{|{\bf F}|}~\frac{i}{16}~\Lambda ^{k \alpha }~\Lambda ^{l \beta }~\Lambda ^{m \zeta }\left(\partial _{\alpha }\partial _{\beta }\partial _m{\bf F}\right)\left(\partial _{\zeta }\partial _k\partial _l{\bf G}\right)\nonumber \\
&&-\frac{1}{64}~\Lambda ^{k \alpha }~\Lambda ^{l \beta }~\Lambda ^{m \zeta }~\Lambda ^{n \iota }\left(\partial _{\alpha }\partial _{\zeta }\partial _l\partial _n{\bf F}\right)\left(\partial _{\beta }\partial _{\iota }\partial _k\partial _m{\bf G}\right),
\end{eqnarray}
where $\partial_k\equiv \frac{\partial }{\partial y^k}$, $\partial _{\alpha }\equiv \frac{\partial }{\partial \theta ^{\alpha }}$.

By replacing the ordinary products in  Eq.~(\ref{WZ_superfield_form}) by star products Eq.~(\ref{star_product_expansion}), we obtain,
\begin{eqnarray}
\label{S_NC}
{\cal S}_{{\rm NC}}
&=&\int d^8z \Bigg\{\Phi ^+\Phi 
-\frac{m}{8}~\Phi  \left(\frac{D^2}{\square }\Phi \right)
-\frac{m^*}{8}~\Phi ^+ \left(\frac{\bar{D}^2}{\square }\Phi ^+\right)\nonumber\\
&&~~~~-\frac{g}{12}~\Phi  ~\Phi \left(\frac{D^2}{\square }\Phi \right)
-\frac{g^*}{12}~\Phi ^+~ \Phi ^+ \left(\frac{\bar{D}^2}{\square }\Phi ^+\right)\nonumber\\
&&~~~~-\frac{g}{32}~\Lambda ^{k l} ~\theta ^4 ~\Phi  \left(D^2 \Phi \right) \left(\partial _l\partial _k D^2 \Phi \right)
-\frac{g}{32}~\Lambda ^{k l}~ \theta ^4 ~\Phi  \left(\partial _k D^2 \Phi \right) \left(\partial _l D^2 \Phi \right)\nonumber\\
&&~~~~-\frac{g^*}{6} ~\Lambda ^{k l} ~\theta ^4 ~\Phi ^+ \left(\square \Phi ^+ \right) \left(\partial _k\partial _l\Phi ^+ \right) 
-\frac{g^*}{3} ~\left(\sigma \Lambda \Lambda ^{k l}\right)^{n o} ~\theta ^4 ~\Phi ^+ \left(\partial _k\partial _n\Phi ^+\right)  \left(\partial _l\partial _o\Phi ^+\right)\nonumber\\
&&~~~~+\frac{g^*}{6} ~\eta ^{k l} ~\Lambda ^{n o} ~\theta ^4 ~\Phi ^+ \left(\partial _k\partial _n\Phi ^+ \right) \left(\partial _l\partial _o\Phi ^+ \right) 
-\frac{g}{16} ~\epsilon ^{\alpha  \beta }~ \Lambda ^{k l}~ \theta ^4~ \left(\partial _k D_{\alpha } \Phi \right) \left(\partial _l D_{\beta } \Phi \right) \left(D^2 \Phi \right)\nonumber\\
&&~~~~-\frac{g}{16}~\epsilon ^{\alpha  \beta } ~\epsilon ^{\zeta  \iota }~ \Lambda ^k{}_{\beta } ~\Lambda ^l{}_{\iota }~ \theta ^4 \left(\partial _k D_{\alpha } \Phi \right) \left(\partial _l D_{\zeta } \Phi \right) \left(D^2 \Phi \right)\nonumber\\
&&~~~~-\frac{g}{3072}~\Lambda ^{k l} ~\Lambda ^{n o} ~\theta ^4 \left(D^2 \Phi \right)\left( \partial _l\partial _k D^2 \Phi \right) \left(\partial _o\partial _n D^2 \Phi \right)\Bigg\},
\end{eqnarray}

\section{Determine divergent operators}

We introduce  two global $U(1)$ symmetries.
\begin{table}[t]
\centering
\begin{tabular}{|c|c|c|c|c|c|c|c|}
\hline
  &  ${\rm dim}$& $U(1)_R$ & $U(1)_{\Phi}$ &  & ${\rm dim}$& $U(1)_R$ & $U(1)_{\Phi}$\\
\hline
$m$ & 1 & 0 & -2 & $m^*$ & 1 & 0 & 2\\
\hline
$g$&0&-1&-3&$g^*$&0&1&3\\
\hline
$ (\Lambda ^k{}_{\alpha } )^2$&-3&2&0&V&-5&2&0\\
\hline
$d^4\theta$ &2&0&0&$\theta ^4$&-2&0&0\\
\hline
$\Phi$ &1&1&1&$\Phi ^+$&1&-1&-1\\
\hline
$D_{\alpha}$&$\frac{1}{2}$&-1&0&$\bar{D}_{\dot{\alpha }}$&$\frac{1}{2}$&1&0\\
\hline
${D}^2$&1&-2&0&$\bar{D}^2$&1&2&0\\
\hline
$\partial _k$&1&0&0&$d^4 x$&-4&0&0\\
\hline
\end{tabular}
\end{table}

We obtain the following constraints on the divergent operators,
\begin{eqnarray}
\label{effective_action}
\Gamma &=&\int d^4x~ \lambda~  {\mathcal O}, \quad
\lambda \sim \Lambda_{UV} ^d~g^{x-R}~{g^*}^x\left(\frac{m}{\Lambda_{UV} }\right)^y\left(\frac{{m^*}}{\Lambda_{UV} }\right)^{y+\frac{S-3R}{2}},\nonumber\\
 {\mathcal O}&=&d^4\theta~ (D^2)^{\gamma }~  (\bar{D}^2)^{\delta } ~(\partial D  \bar{D})^{\eta }~(\partial\partial)^{\zeta }~V^{\rho }~\Phi ^{\alpha } ~(\Phi ^+)^{\beta },\quad
V\equiv(\Lambda ^k{}_{\alpha } )^2~\theta ^4, \nonumber\\
d&=&2-\alpha -\beta -\gamma -\delta -2~ \zeta -2 ~\eta +5~ \rho, \quad 
R=-\alpha +\beta +2~ \gamma -2~ \delta -2 ~\rho, \quad 
S=-\alpha +\beta,\nonumber\\
P&=&d+\frac{3}{2}R-\frac{1}{2}S-2~ y=2-2~ y-2 ~\alpha +2 ~\gamma -4~ \delta -2 ~\zeta -2 ~\eta +2~ \rho \geq 0,\nonumber\\
\gamma &\leq&  \alpha -\eta, \quad 
\delta \leq \beta -\eta,\nonumber\\
\gamma &\geq& 0,\quad 
\delta \geq 0,\quad 
\alpha \geq 0,\quad 
\eta \geq 0,\quad 
\rho \geq 0,\quad 
\zeta \geq 0,\quad 
\beta \geq 0,\nonumber\\
\rho &=&2,\nonumber\\
y &\geq& 0,\quad 
x\geq 0,\quad 
x+\alpha -\beta -2 ~\gamma +2~ \delta +2~ \rho \geq 0,\quad 
y+\alpha -\beta -3~ \gamma +3 ~\delta +3~ \rho \geq 0.
\end{eqnarray}

\section{Convention and identities}

We define the following symbols,
\begin{align}
\label{}
\Lambda ^{k l}&\equiv\epsilon ^{\alpha  \beta }~\Lambda ^k{}_{\beta }~\Lambda ^l{}_{\alpha }, &
\Lambda ^2&\equiv\eta _{k l}~\Lambda ^{k l},\nonumber\\
\sigma \Lambda \Lambda &\equiv\eta _{k n}~\eta _{l o}~\left(\sigma ^{k l}\right)^{\alpha  \beta } ~\Lambda ^n{}_{\alpha } ~\Lambda ^o{}_{\beta },&
\left(\sigma \Lambda ^{k l}\right)^{n \alpha }&\equiv\left(\sigma ^{k l}\right)^{\beta  \alpha }~ \Lambda ^n{}_{\beta },\nonumber\\
\left(\eta \sigma \Lambda ^k\right)^{\alpha }&\equiv\eta _{l n} ~\left(\sigma ^{n k}\right)^{\beta  \alpha } ~\Lambda ^l{}_{\beta },&
\left(\eta \sigma \Lambda \Lambda ^k\right)^l&\equiv\eta _{n o} ~\left(\sigma ^{o k}\right)^{\alpha  \beta }~ \Lambda ^n{}_{\alpha }~ \Lambda ^l{}_{\beta },\nonumber\\
\left(\sigma \Lambda \Lambda ^{k l}\right)^{n o}&\equiv\left(\sigma ^{k l}\right)^{\alpha  \beta } ~\Lambda ^n{}_{\alpha } ~\Lambda ^o{}_{\beta }.
\end{align}

Combine $\sigma$,
\begin{eqnarray}
\label{}
\sigma ^l{}_{\beta  \dot{\gamma }} ~\left(\bar{\sigma }^k\right)^{\dot{\alpha } \beta }&=&-\delta _{\dot{\alpha } \dot{\gamma }} ~\eta ^{k l}+2 \left(\bar{\sigma }^{k l}\right)^{\dot{\alpha }}{}_{\dot{\gamma }},\nonumber\\
\sigma ^k{}_{\alpha  \dot{\beta }} ~\left(\bar{\sigma }^l\right)^{\dot{\beta } \gamma }&=&-\delta _{\alpha  \gamma } ~\eta ^{k l}+2 \left(\sigma ^{k l}\right)_{\alpha }^{\gamma }.
\end{eqnarray}

Reduce the number of $\sigma$,
\begin{eqnarray}
\label{}
\sigma ^k{}_{\gamma  \dot{\zeta }} ~\left(\bar{\sigma }^{l m}\right)^{\dot{\zeta }}{}_{\dot{\alpha }}&=&\frac{1}{2} ~\eta ^{k m}~ \sigma ^l{}_{\gamma  \dot{\alpha }}-\frac{1}{2}~ \eta ^{k l} ~\sigma ^m{}_{\gamma  \dot{\alpha }}-\frac{i}{2} ~\eta _{n o} ~\epsilon ^{o k l m}~ \sigma ^n{}_{\gamma  \dot{\alpha }},\nonumber\\
\left(\bar{\sigma }^m\right)^{\dot{\beta } \gamma } ~\left(\bar{\sigma }^{k l}\right)^{\dot{\alpha }}{}_{\dot{\beta }}&=&-\frac{1}{2} ~\eta ^{l m} ~\left(\bar{\sigma }^k\right)^{\dot{\alpha } \gamma }+\frac{1}{2} ~\eta ^{k m} ~\left(\bar{\sigma }^l\right)^{\dot{\alpha } \gamma }+\frac{i}{2} ~\eta _{n o} ~\epsilon ^{o k l m} ~\left(\bar{\sigma }^n\right)^{\dot{\alpha } \gamma },\nonumber\\
\left(\bar{\sigma }^m\right)^{\dot{\alpha } \beta } ~\left(\sigma ^{k l}\right)_{\beta }^{\gamma }&=&\frac{1}{2} ~\eta ^{l m}~ \left(\bar{\sigma }^k\right)^{\dot{\alpha } \gamma }-\frac{1}{2}~ \eta ^{k m} ~\left(\bar{\sigma }^l\right)^{\dot{\alpha } \gamma }+\frac{i}{2}~\eta _{n o} ~\epsilon ^{o k l m} ~\left(\bar{\sigma }^n\right)^{\dot{\alpha } \gamma },\nonumber\\
\sigma ^m{}_{\beta  \dot{\gamma }}~ \left(\sigma ^{k l}\right)_{\alpha }^{\beta }&=&-\frac{1}{2}~ \eta ^{l m} ~\sigma ^k{}_{\alpha  \dot{\gamma }}+\frac{1}{2}~ \eta ^{k m} ~\sigma ^l{}_{\alpha  \dot{\gamma }}-\frac{i}{2}~\eta _{n o}~ \epsilon ^{o k l m} ~\sigma ^n{}_{\alpha  \dot{\gamma }},\nonumber\\
\left(\bar{\sigma }^{k l}\right)^{\dot{\alpha }}{}_{\dot{\beta }} ~\left(\bar{\sigma }^{m n}\right)^{\dot{\beta }}{}_{\dot{\alpha }}&=&\frac{i}{2}~\epsilon ^{k l m n}+\frac{1}{2}~ \eta ^{k n} ~\eta ^{l m}-\frac{1}{2} ~\eta ^{k m}~ \eta ^{l n},\nonumber\\
\left(\sigma ^{k l}\right)_{\alpha }^{\zeta }~ \left(\sigma ^{m n}\right)_{\zeta }^{\alpha }&=&-\frac{i}{2}~ \epsilon ^{k l m n}+\frac{1}{2}~ \eta ^{k n} ~\eta ^{l m}-\frac{1}{2} ~\eta ^{k m} ~\eta ^{l n}.
\end{eqnarray}

We can reduce the number of $\sigma$ from 2 to 1 for the following combinations,
\begin{eqnarray}
\label{}
&&\left(\sigma ^{k l}\right)^{\alpha  \beta } ~\sigma \Lambda \Lambda ^{n o}{}_{k l},\quad 
\eta _{k l} ~\left(\sigma ^{l n}\right)^{\alpha  \beta } ~\left(\eta \sigma \Lambda \Lambda ^o\right)^k,\quad 
\eta _{k l} ~\left(\sigma \Lambda ^{n o}\right)^{l \alpha } ~\left(\eta \sigma \Lambda ^k\right)^{\beta }, \nonumber\\
&& \eta _{l n} ~\left(\eta \sigma \Lambda \Lambda ^n\right)^q ~\left(\sigma \Lambda \Lambda ^{o r}\right)^{l s},\quad
\eta _{k l} ~\eta _{n o} ~\left(\sigma \Lambda ^{l q}\right)^{o \beta }~ \left(\sigma \Lambda ^{n p}\right)^{k \alpha },\nonumber\\
&&\left(\eta \sigma \Lambda ^q\right)^{\beta } ~\left(\eta \sigma \Lambda ^p\right)^{\alpha },\quad 
\left(\eta \sigma \Lambda \Lambda ^o\right)^p ~\left(\eta \sigma \Lambda \Lambda ^s\right)^q,\quad 
\sigma \Lambda \Lambda ^{k l}{}_{n o} ~\sigma \Lambda \Lambda ^{n o}{}_{k l},\nonumber\\
&&\eta _{k l} ~\eta _{n o} ~\left(\eta \sigma \Lambda \Lambda ^l\right)^o ~\left(\eta \sigma \Lambda \Lambda ^n\right)^k,\quad
\eta _{k l} ~\eta _{n o} ~\left(\sigma \Lambda \Lambda ^{l o}\right)^{n p} ~\left(\sigma \Lambda \Lambda ^{q r}\right)^{k s},\nonumber\\
&&\eta _{k l} ~\eta _{n o}~ \left(\sigma \Lambda \Lambda ^{l p}\right)^{o q} ~\left(\sigma \Lambda \Lambda ^{n r}\right)^{k s},\quad
\Lambda _{k l} ~\left(\sigma \Lambda \Lambda ^{k p}\right)^{q r} ~\left(\eta \sigma \Lambda \Lambda ^l\right)^s, \nonumber\\ 
&&\Lambda _{k l} ~\Lambda _{n o} ~\left(\sigma \Lambda \Lambda ^{k n}\right)^{l_1 l_2} ~\left(\sigma \Lambda \Lambda ^{l o}\right)^{l_3 l_4}, \quad
\Lambda _{l k_1} ~\left(\sigma ^{k_1 k_2}\right){}^{\alpha  \beta } ~\left(\eta \sigma \Lambda \Lambda ^l\right)^o,\nonumber\\
&&\Lambda _{k_1 k_3} ~\eta _{k_2 l_3}~\left(\sigma ^{k_1 k_2}\right){}^{\alpha  \beta } ~\left(\sigma \Lambda \Lambda ^{k_3 k_4}\right){}^{l_3 l_4}.
\end{eqnarray}

Other identities,
\begin{eqnarray}
\label{}
\left(\sigma \Lambda \Lambda ^{k_3 k_4}\right){}^{l_3 l_4} ~\Lambda ^{l_5 l_6}&=&-\left(\sigma \Lambda \Lambda ^{k_3 k_4}\right){}^{l_6 l_4} ~\Lambda ^{l_3 l_5}-\left(\sigma \Lambda \Lambda ^{k_3 k_4}\right){}^{l_5 l_4} ~\Lambda ^{l_3 l_6},\nonumber\\
\Lambda ^{k l} ~\Lambda _{k l}&=&-\frac{1}{2} ~\Lambda ^2 ~\Lambda ^2,\nonumber\\
\left(\eta \sigma \Lambda \Lambda ^k\right)^l ~\Lambda _{k l}&=&-\frac{1}{2} ~\Lambda ^2~ \sigma \Lambda \Lambda,\nonumber\\
\Lambda ^{k n} ~\Lambda ^{l o} ~\eta _{k l}&=&-\frac{1}{2} ~\Lambda ^2 ~\Lambda ^{n o},\nonumber\\
\left(\eta \sigma \Lambda \Lambda ^{{n_1}}\right)^{{k_3}} ~\Lambda ^{{k_4}}{}_{{n_1}} \left(\partial _{{k_3}}\partial _{{k_4}} X \right)&=&-\frac{1}{2} ~\sigma \Lambda \Lambda  ~\Lambda ^{{k_3} {k_4}} \left(\partial _{{k_3}}\partial _{{k_4}} X \right),\nonumber\\
\Lambda ^{{k_1} {k_2}} ~\left(\eta \sigma \Lambda \Lambda ^{{k_3}}\right)^{{k_4}} \left( \partial _{{k_1}}\partial _{{k_2}}\partial _{{k_4}} X \right) &=&0,\nonumber\\
\Lambda ^{{k_1} {k_3}} ~\Lambda ^{{k_2} {k_4}}\left(\partial _{{k_1}}\partial _{{k_2}}X\right)&=& \frac{1}{2}~\Lambda ^{{k_1} {k_2}} ~\Lambda ^{{k_3} {k_4}}\left(\partial _{{k_1}}\partial _{{k_2}}X\right).
\end{eqnarray}

\section{BFNC parameters}
\label{construct BFNC parameters}

We do not allow combination of $\epsilon$ and $\sigma$, because we can use identity to cancel $\epsilon$. For example, if we consider $\sigma \Lambda \Lambda~\epsilon ^{k_1k_2k_3k_4}$, then we have,
\begin{eqnarray}
\label{}
\sigma \Lambda \Lambda~\epsilon ^{k_1k_2k_3k_4}&=&-2 i \left(\sigma \Lambda \Lambda ^{k_1 k_2}\right){}^{k_3 k_4}+2 i \left(\sigma \Lambda \Lambda ^{k_1 k_3}\right){}^{k_2 k_4}-2 i \left(\sigma \Lambda \Lambda ^{k_2 k_3}\right){}^{k_1 k_4},\nonumber\\
&&-2 i \left(\sigma \Lambda \Lambda ^{k_1 k_4}\right){}^{k_2 k_3}+2 i \left(\sigma \Lambda \Lambda ^{k_2 k_4}\right){}^{k_1 k_3}-2 i \left(\sigma \Lambda \Lambda ^{k_3 k_4}\right){}^{k_1 k_2}.\nonumber
\end{eqnarray}

We also do not consider combination that contain two $\epsilon$, because we can transform $\epsilon$ to $\delta$ by using $\epsilon ^{k_1k_2k_3k_4}~\epsilon _{l_1l_2l_3l_4}=\delta ^{k_1}{}_{\left[l_1\right.}~\delta ^{k_2}{}_{l_2}~\delta ^{k_3}{}_{l_3}~\delta ^{k_4}{}_{\left.l_4\right]}$.

We will not use $\eta ^{k l}$ when constructing parameters, but we can combine $\eta ^{k l}$ and parameters with indices $n~o$ to obtain parameters with indices $k~l~n~o$.

Further more, we demand the number of $\sigma$ can not be reduced in the combinations, we also require the combinations not producing $\Lambda ^2$ or $\sigma \Lambda \Lambda$.

\subsection{$\Lambda ^2$}

Some of the parameters with indices $k~l$ and $k~l~n~o$ can be obtained by using $\Lambda ^2$ and $\sigma \Lambda \Lambda$. The parameters with indices $k~l$ are,
\begin{eqnarray}
\label{Lambda_0_2}
\Lambda ^2~\eta ^{kl}, \quad \sigma \Lambda \Lambda~\eta ^{kl}
\end{eqnarray}
with indices $k~l~n~o$ are,
\begin{eqnarray}
\label{Lambda_0_4}
\Lambda ^2~\eta ^{kl}~\eta ^{no}, \quad 
\sigma \Lambda \Lambda~\eta ^{kl}~\eta ^{no}, \quad 
\Lambda ^2~\epsilon ^{klno}.
\end{eqnarray}

To construct new parameters with indices $k~l$ and $k~l~n~o$, we consider the following combinations,
\begin{enumerate}
\label{}
\item 
To construct parameters with indices $k~l~n~o$ by using $\epsilon ^{k_1 k_2 k_3 k_4} ~\Lambda ^{l_1 \beta _1} ~\Lambda ^{l_2 \beta _2}$, we can contract it with $\epsilon _{\beta _1\beta _2}$ and obtain $\epsilon ^{k_1k_2k_3k_4}~\Lambda ^{l_1l_2}$, where we define $\Lambda ^{l_1l_2}=\Lambda ^{l_1\beta _1}~\Lambda ^{l_2\beta _2}~\epsilon _{\beta _1\beta _2}$. There are 6 indices in $\epsilon ^{k_1k_2k_3k_4}~\Lambda ^{l_1l_2}$, and have to contract with a $\eta$. We can not contract $k_i$, because the result is 0. If we contract  $l_i$, the result is $\Lambda ^2$, which is defined as $\Lambda ^2=\eta _{k l}~\Lambda ^{k l}$. If we obtain $\Lambda ^2$, then the parameters are contained in Eq.~(\ref{Lambda_0_2}), (\ref{Lambda_0_4}). As a result we can only contract $k_i$ and $l_i$ in $\epsilon ^{k_1k_2k_3k_4}~\Lambda ^{l_1l_2}$. Considering the symmetries of $k_i$ and $l_i$, the only possible contraction is $k_1$ and $l_1$, the result is $\epsilon ^{n_1k_2k_3k_4}~\Lambda _{n_1}{}^{l_2}$. We can only obtain parameters with indices $k~l~n~o$ by using $\epsilon ^{k_1 k_2 k_3 k_4} ~\Lambda ^{l_1 \beta _1}~ \Lambda ^{l_2 \beta _2}$. If we want to construct parameters with indices $k~l$, then we must contract $\epsilon ^{n_1k_2k_3k_4}~\Lambda _{n_1}{}^{l_2}$ with a $\eta$, due to the antisymmetric indices of $\epsilon$, the result is 0.

\item 
Starting from $\Lambda ^{l_1 \beta _1} ~\Lambda ^{l_2 \beta _2}$, if we contract it with $\epsilon _{\beta _1\beta _2}$, we will obtain $\Lambda ^{l_1l_2}$. If we combine $\Lambda ^{l_1l_2}$ and $\eta ^{l_3l_4}$, the result is $\Lambda ^{l_1l_2}~\eta ^{l_3l_4}$.

\item 
For $\left(\sigma ^{k_1 k_2}\right){}^{\alpha _1 \alpha _2} ~\Lambda ^{l_1 \beta _1} ~\Lambda ^{l_2 \beta _2}$, if we contract it with $\epsilon _{\beta _1\beta _2}$, the result have symmetric indices $\alpha _1\alpha _2$. If we contract $\alpha _1\alpha _2$ with $\epsilon _{\alpha _1\alpha _2}$, the result is 0. 
So we can only contract $\left(\sigma ^{k_1 k_2}\right){}^{\alpha _1 \alpha _2} ~\Lambda ^{l_1 \beta _1} ~\Lambda ^{l_2 \beta _2}$ with $\epsilon _{\alpha _1\beta _1}~\epsilon _{\alpha _2\beta _2}$ and obtain $\left(\sigma \Lambda \Lambda ^{k_1 k_2}\right){}^{l_1l_2}$. We define $\left(\sigma \Lambda \Lambda ^{k_1 k_2}\right){}^{l_1l_2}=\left(\sigma ^{k_1k_2}\right){}^{\alpha  \beta }~\Lambda ^{l_1}{}_{\alpha }~\Lambda ^{l_2}{}_{\beta }$ (where $k_1$, $k_2$ are antisymmetric, $l_1$, $l_2$ are also antisymmetric). We also define $\Lambda ^l{}_{\alpha }=\epsilon _{\alpha  \beta }~\Lambda ^{l \beta }$. $\left(\sigma \Lambda \Lambda ^{k_1 k_2}\right){}^{l_1l_2}$ contain indices $k~l~n~o$, we can obtain parameters with indices $k~l$ by contracting it with $\eta _{k_1 l_1}$, the result is $\left(\eta \sigma \Lambda \Lambda ^{k_2}\right){}^{l_2}$. We define $\left(\eta \sigma \Lambda \Lambda ^{k_2}\right){}^{l_2}=\eta _{k_1l_1}~\left(\sigma \Lambda \Lambda ^{k_1 k_2}\right){}^{l_1l_2}$. We can construct parameters with indices $k~l~n~o$ by using $\left(\eta \sigma \Lambda \Lambda ^{k_2}\right){}^{l_2}$, that is $\left(\eta \sigma \Lambda \Lambda ^{k_2}\right){}^{l_2}~\eta ^{l_3l_4}$.

\item 
Let us consider $\left(\sigma ^{k_1 k_2}\right){}^{\alpha _1 \alpha _2} ~\left(\sigma ^{k_3 k_4}\right){}^{\alpha _3 \alpha _4} ~\Lambda ^{l_1 \beta _1} ~\Lambda ^{l_2 \beta _2}$. We have to contract $\alpha _1~\alpha _2~\alpha _3~\alpha _4~\beta _1~\beta _2$ by using $\epsilon$. The indices $\alpha _i$ can not be contracted with $\epsilon$, otherwise the number of $\sigma$ will be reduced from 2 to 1. We have already considered the combination with one $\sigma$. If we contract $\beta _i$ with $\epsilon$, then we must also contract $\alpha _i$ with $\epsilon$. Anyway, we have to contract  $\alpha _i$. So this combination do not give new parameters.

\item 
The combinations with more $\sigma$ such as $\left(\sigma ^{k_1 k_2}\right){}^{\alpha _1\alpha _2}~\left(\sigma ^{k_3 k_4}\right){}^{\alpha _3\alpha _4}~\left(\sigma ^{k_5 k_6}\right){}^{\alpha _5\alpha _6}~\Lambda ^{l_1\beta _1}~\Lambda ^{l_2\beta _3}$ will not give new parameters.

\end{enumerate}
At the end we obtain all of the parameters with indices $k~l$ and $k~l~n~o$, see Eq.~(\ref{Lambda2kl}), (\ref{Lambda2klno}).\\

In the following, we construct parameters with indices $k~ l ~\alpha~ \beta$ from parameters with indices $k~l$ or $k~l~n~o$.
\begin{enumerate}
\label{}
\item 
Combine $\epsilon ^{\alpha_1  \alpha_2 }$ and  parameters with indices $k~ l$, we obtain 
$\left(\eta \sigma \Lambda \Lambda ^{k_1}\right)^{l_1}~\epsilon ^{\alpha_1  \alpha_2 }$, 
$\Lambda ^{l_1 l_2}~\epsilon ^{\alpha_1  \alpha_2 }$, 
$\Lambda ^2~\eta ^{k_1 k_2}~\epsilon ^{\alpha_1  \alpha_2 }$ and 
$\sigma \Lambda \Lambda~\eta ^{k_1 k_2}~\epsilon ^{\alpha_1  \alpha_2  }$.

\item 
Combine $\left(\sigma ^{k_1 k_2}\right)^{\alpha_1  \alpha_2 }$ with parameters without indices, we obtain $\Lambda ^2~\left(\sigma ^{k_1 k_2}\right)^{\alpha_1  \alpha_2  }$ and $\sigma \Lambda \Lambda ~\left(\sigma ^{k_1 k_2}\right)^{\alpha_1  \alpha_2 }$.

\item 
Combine $\left(\sigma ^{k_1 k_2}\right)^{\alpha_1  \alpha_2  }$ and parameters with indices $k~l$, we have 3 possible forms,
\begin{enumerate}
\label{}
\item 
$\Lambda ^2~\left(\sigma ^{k_1 k_2}\right){}^{\alpha_1  \alpha_2  }~\eta ^{l_1 l_2}$ and  $\sigma \Lambda \Lambda ~\left(\sigma ^{k_1 k_2}\right){}^{\alpha_1  \alpha_2  }~\eta ^{l_1 l_2}$ do not give new parameters.

\item
From $\left(\sigma ^{k_1 k_2}\right){}^{\alpha_1  \alpha_2  }~\Lambda ^{l_1 l_2}$ we can obtain $\left(\sigma ^{n_1 k_2}\right){}^{\alpha_1  \alpha_2 }~\Lambda _{n_1}{}^{l_2}$.

\item 
We can reduce the number of  $\sigma$ in $\left(\sigma ^{k_1 k_2}\right){}^{\alpha_1  \alpha_2 }~\left(\eta \sigma \Lambda \Lambda ^{k_3}\right){}^{l_3}$ by using identities.
\end{enumerate}

\item 
Now consider combining $\left(\sigma ^{k_1 k_2}\right)^{\alpha_1  \alpha_2  }$ with parameters with indices $k~ l~ n ~o$:
\begin{enumerate}
\label{}
\item 
We can cancel $\epsilon ^{l_1 l_2 l_3 l_4}$ in $\Lambda ^2~\left(\sigma ^{k_1 k_2}\right){}^{\alpha_1  \alpha_2 } ~\epsilon ^{l_1 l_2 l_3 l_4}$.

\item 
$\Lambda ^2~\left(\sigma ^{k_1 k_2}\right){}^{\alpha_1  \alpha_2  } ~\eta ^{l_1 l_2}~ \eta ^{l_3 l_4}$ and $\sigma \Lambda \Lambda~\left(\sigma ^{k_1 k_2}\right){}^{\alpha_1  \alpha_2  } ~\eta ^{l_1 l_2} ~\eta ^{l_3 l_4} $ do not give new parameters.

\item 
The $ \epsilon ^{n_1l_1l_2l_3}$ in $\left(\sigma ^{k_1 k_2}\right){}^{\alpha_1  \alpha_2  } ~\epsilon ^{n_1l_1l_2l_3}~\Lambda _{n_1}{}^{l_4}$ can be removed.

\item 
For $\left(\sigma ^{k_1 k_2}\right){}^{\alpha_1  \alpha_2  }~ \left(\sigma \Lambda \Lambda ^{k_3 k_4}\right){}^{l_3 l_4}$, if we contract $k_1$, $k_2$ with $k_3$, $k_4$, then we can reduce the number of $\sigma$. If we contract $k_1$, $k_2$ with $l_3$, $l_4$, we can also reduce the number of $\sigma$ by using identities.

\item 
We can reduce the number of $\sigma$ in $\left(\sigma ^{k_1 k_2}\right){}^{\alpha_1  \alpha_2  } ~\eta ^{l_1 l_2} ~\left(\eta \sigma \Lambda \Lambda ^{k_3}\right){}^{l_3}$.

\item 
$\left(\sigma ^{k_1 k_2}\right){}^{\alpha_1  \alpha_2 } ~\eta ^{l_1 l_2} ~\Lambda ^{l_3 l_4}$ do not give new parameters.
\end{enumerate}
We conclude that no new parameters will be produced by combing $\left(\sigma ^{k_1 k_2}\right)^{\alpha_1  \alpha_2  }$ with parameters with indices $k~ l ~n~ o$.
\end{enumerate}

We found that we only need to construct parameters with indices $k ~l ~\alpha ~\beta$ by using at most one $\left(\sigma ^{k_1 k_2}\right){}^{{\alpha_1} {\alpha_2}}$, at most one $\epsilon ^{k_1k_2k_3k_4}$, one $\Lambda ^{l_1 {\beta_1}}$ and one $\Lambda ^{l_2 {\beta_2}}$. We list them as follows,
\begin{enumerate}
\label{}
\item 
Without $\epsilon ^{k_1k_2k_3k_4}$ and $\left(\sigma ^{k_1 k_2}\right){}^{\alpha _1\alpha _2}$ we obtain $\Lambda ^{l_1 {\beta_1}}~\Lambda ^{l_2 {\beta_2}}$.

\item 
Form $\epsilon ^{k_1k_2k_3k_4}~\Lambda ^{l_1\beta _1}~\Lambda ^{l_2\beta _2}$ we obtain $\epsilon ^{k_1k_2n_1n_2}~\Lambda _{n_1}{}^{\beta_1 }~\Lambda _{n_2}{}^{\beta_2 }$.

\item 
Contract $\left(\sigma ^{k_1 k_2}\right){}^{{\alpha_ 1} {\alpha_ 2}} ~\Lambda ^{l_1 {\beta_ 1}} ~\Lambda ^{l_2 {\beta_ 2}}$ with $\epsilon _{\beta _1\beta _2}$, we obtain $\left(\sigma ^{k_1 k_2}\right){}^{\alpha _1\alpha _2}~\Lambda ^{l_1l_2}$. Contract it with $\eta _{k_1 l_1}$, the result is $\left(\sigma ^{n_1 k_2}\right){}^{\alpha _1\alpha _2}~\Lambda _{n_1}{}^{l_2}$.

\item  
Contract $\left(\sigma ^{k_1 k_2}\right){}^{{\alpha_ 1} {\alpha_ 2}} ~\Lambda ^{l_1 {\beta_ 1}} ~\Lambda ^{l_2 {\beta_ 2}}$ with $\epsilon _{\alpha _1\beta _1}$, the result is $\left(\sigma \Lambda ^{k_1 k_2}\right){}^{l_1\alpha _2}~\Lambda ^{l_2\beta _2}$, then we have 2 possibilities: 
\begin{enumerate}
\label{}
\item 
Contract $\left(\sigma \Lambda ^{k_1 k_2}\right){}^{l_1\alpha _2}~\Lambda ^{l_2\beta _2}$ with $\eta _{k_1 l_1}$, we obtain $\left(\eta \sigma \Lambda ^{k_1}\right){}^{\alpha _2}~\Lambda ^{l_2\beta _2}$. Contract it with $\eta _{{k_1} {l_2}}$, the result is $\left(\eta \sigma \Lambda ^{{n_1}}\right)^{{\alpha_{1}}}~\Lambda _{{n_1}}{}^{{\beta_{2}}}$. By combining it with $\eta ^{{k_1} {k_2}}$ we obtain $\eta ^{k_1 k_2}~\left(\eta \sigma \Lambda ^{n_1}\right){}^{\alpha _1}~\Lambda _{n_1}{}^{\beta _2}$.

\item 
Contract $\left(\sigma \Lambda ^{k_1 k_2}\right){}^{l_1\alpha _2}~\Lambda ^{l_2\beta _2}$ with $\eta _{k_2 l_2}$, the result is $\left(\sigma \Lambda ^{n_1 k_2}\right){}^{l_1\alpha _2}~\Lambda _{n_1}{}^{\beta _2}$.
\end{enumerate}
\end{enumerate}
At the end we obtain all possible parameters with indices  $k~ l ~\alpha~  \beta$, see Eq.~(\ref{Lambda2klab}).

\subsection{$\Lambda ^4$}

In the following we will construct parameters with indices $k~l$ and $k~l~n~o$.
\begin{enumerate}
\label{}
\item 
From $\Lambda ^{l_1\beta _1}~\Lambda ^{l_2\beta _2}~\Lambda ^{l_3\beta _3}~\Lambda ^{l_4\beta _4}$ we obtain $\Lambda ^{l_1l_2}~\Lambda ^{l_3l_4}$.

\item 
From $\epsilon ^{k_1k_2k_3k_4}~\Lambda ^{l_1\beta _1}~\Lambda ^{l_2\beta _2}~\Lambda ^{l_3\beta _3}~\Lambda ^{l_4\beta _4}$ we have $\epsilon ^{k_1k_2k_3k_4}~\Lambda ^{l_1l_2}~\Lambda ^{l_3l_4}$. By contracting it with $\eta _{k_1 l_1}~\eta _{k_2 l_3}$, we obtain $\epsilon ^{n_1n_2k_3k_4}~\Lambda _{n_1}{}^{l_2}~\Lambda _{n_2}{}^{l_4}$.

\item 
Contract $\left(\sigma ^{k_1 k_2}\right){}^{\alpha _1\alpha _2}~\Lambda ^{l_1\beta _1}~\Lambda ^{l_2\beta _2}~\Lambda ^{l_3\beta _3}~\Lambda ^{l_4\beta _4}$ with $\epsilon _{\alpha _1\beta _1}~\epsilon _{\alpha _2\beta _2}~\epsilon _{\beta _3\beta _4}$, the result is $\left(\sigma \Lambda \Lambda ^{k_1 k_2}\right){}^{l_1l_2}~\Lambda ^{l_3l_4}$. If we contract $l_i$, we will obtain $\Lambda ^2$. So we need to consider the following two cases,
\begin{enumerate}
\label{}
\item 
Contract $\left(\sigma \Lambda \Lambda ^{k_1 k_2}\right){}^{l_1l_2}~\Lambda ^{l_3l_4}$ with $\eta _{k_1 l_1}$ we have $\left(\eta \sigma \Lambda \Lambda ^{k_1}\right){}^{l_1}~\Lambda ^{l_3l_4}$.
Then contract it with $\eta _{k_1 l_3}$, the result is $\left(\eta \sigma \Lambda \Lambda ^{n_1}\right){}^{l_1}~\Lambda _{n_1}{}^{l_4}$. The other contractions will generate $\Lambda ^2$ or $\sigma \Lambda \Lambda$.

\item 
Contract $\left(\sigma \Lambda \Lambda ^{k_1 k_2}\right){}^{l_1l_2}~\Lambda ^{l_3l_4}$ with $\eta _{k_1 l_3}$ we obtain $\left(\sigma \Lambda \Lambda ^{n_1 k_2}\right){}^{l_1l_2}~\Lambda _{n_1}{}^{l_4}$, where $k_2$ can be contracted with $l_1,~ l_2,~ l_4$. If we contract $k_2$ with $l_4$ by multiplying $\eta _{k_2l_4}$, the result is 0. If we contract $k_2$ with $l_1$ or $l_2$, the result is $\left(\eta \sigma \Lambda \Lambda ^{n_1}\right){}^{l_2}~\Lambda _{n_1}{}^{l_4}$. At the end we obtain only one parameter with indices $k l$, that is $\left(\eta \sigma \Lambda \Lambda ^{n_1}\right){}^{l_1}~\Lambda _{n_1}{}^{l_2}$. Contract it with $\eta ^{k_1 k_2}$ we obtain parameter with indices $k~ l~ n ~o$, that is $\eta ^{k_1 k_2}~\left(\eta \sigma \Lambda \Lambda ^{n_1}\right){}^{l_1}~\Lambda _{n_1}{}^{l_2}$.
\end{enumerate}

\item 
Consider $\left(\sigma ^{k_1 k_2}\right){}^{\alpha _1\alpha _2}~\left(\sigma ^{k_3 k_4}\right){}^{\alpha _3\alpha _4}~\Lambda ^{l_1\beta _1}~\Lambda ^{l_2\beta _2}~\Lambda ^{l_3\beta _3}~\Lambda ^{l_4\beta _4}$, if we contract $\epsilon _{\alpha _1\beta _1}$~$\epsilon _{\alpha _2\beta _2}~\epsilon _{\alpha _3\beta _3}~\epsilon _{\alpha _4\beta _4}$ with it, the result is $\left(\sigma \Lambda \Lambda ^{k_1 k_2}\right){}^{l_1l_2}~\left(\sigma \Lambda \Lambda ^{k_3 k_4}\right){}^{l_3l_4}$. To construct parameters with indices $k~ l ~n ~o$, we should contract it with two $\eta$. $k_i$ can not be contracted. $l_i$ can not be contacted too. Further more, if we contract $k_i$ in $\left(\sigma \Lambda \Lambda ^{k_1 k_2}\right){}^{l_1l_2}$ with $l_i$ in $\left(\sigma \Lambda \Lambda ^{k_3 k_4}\right){}^{l_3l_4}$, we can transform it to contraction of $k_i$ with $l_j$, where $l_j$ is in $\left(\sigma \Lambda \Lambda ^{k_1 k_2}\right){}^{l_1l_2}$, by using Fierz identities,
\begin{eqnarray}
\label{}
&&\eta _{k_1l_4}~\left(\sigma \Lambda \Lambda ^{k_1 k_2}\right){}^{l_1l_2}~\left(\sigma \Lambda \Lambda ^{k_3 k_4}\right){}^{l_3l_4}\nonumber\\
&=&\eta _{k_1l_4}~\left(\sigma \Lambda ^{k_1 k_2}\right){}^{l_1\alpha }~\Lambda ^{l_2}{}_{\alpha }~\left(\sigma \Lambda ^{k_3 k_4}\right){}^{l_3\beta }~\Lambda ^{l_4}{}_{\beta }\nonumber\\
&=&-\eta _{k_1l_4}~\left(\sigma \Lambda ^{k_1 k_2}\right){}^{l_1\alpha }~\left(\sigma \Lambda ^{k_3 k_4}\right){}^{l_3\beta }~\Lambda ^{l_2}{}_{\alpha }~\Lambda ^{l_4}{}_{\beta }\nonumber\\
&=&-\eta _{k_1l_4}~\left(\sigma \Lambda ^{k_1 k_2}\right){}^{l_1\alpha }~\left(\sigma \Lambda ^{k_3 k_4}\right){}^{l_3\beta }~\left(\Lambda ^{l_2}{}_{\beta }~\Lambda ^{l_4}{}_{\alpha }+\epsilon _{\alpha  \beta }~\Lambda ^{l_2l_4}\right)\nonumber\\
&=&-\eta _{k_1l_4}~\left(\sigma \Lambda \Lambda ^{k_1 k_2}\right){}^{l_1l_4}~\left(\sigma \Lambda \Lambda ^{k_3 k_4}\right){}^{l_3l_2}-\eta _{k_1l_4}~\left(\sigma \Lambda ^{k_1 k_2}\right){}^{l_1\alpha }~\left(\sigma \Lambda ^{k_3 k_4}\right){}^{l_3\beta }~\epsilon _{\alpha  \beta }~\Lambda ^{l_2l_4}\nonumber\\
&=&\left(\eta \sigma \Lambda \Lambda ^{k_2}\right){}^{l_1}~\left(\sigma \Lambda \Lambda ^{k_3 k_4}\right){}^{l_3l_2}+X.\nonumber
\end{eqnarray}
where $X$ represents expression in which the number of $\sigma$ is reduced. So we only need to contract $\left(\sigma \Lambda \Lambda ^{k_1 k_2}\right){}^{l_1l_2}~\left(\sigma \Lambda \Lambda ^{k_3 k_4}\right){}^{l_3l_4}$ with $\eta _{k_2 l_2}$, the result is $\left(\eta \sigma \Lambda \Lambda ^{k_1}\right){}^{l_1}~\left(\sigma \Lambda \Lambda ^{k_3 k_4}\right){}^{l_3l_4}$. We need to contract it with a  $\eta$. There are 2 possibilities,
\begin{enumerate}
\label{}
\item 
Contract $\left(\eta \sigma \Lambda \Lambda ^{k_1}\right){}^{l_1}~\left(\sigma \Lambda \Lambda ^{k_3 k_4}\right){}^{l_3l_4}$ with $\eta _{k_1 l_1}$, we obtain $\sigma \Lambda \Lambda$. We do not consider this case.

\item 
Contract $\left(\eta \sigma \Lambda \Lambda ^{k_1}\right){}^{l_1}~\left(\sigma \Lambda \Lambda ^{k_3 k_4}\right){}^{l_3l_4}$ with $\eta _{k_3 l_3}$, the result is $\left(\eta \sigma \Lambda \Lambda ^{k_1}\right){}^{l_1}~\left(\eta \sigma \Lambda \Lambda ^{k_4}\right){}^{l_4}$, we can reduce the number of $\sigma$ by using identities.
\end{enumerate}
In summary, $\left(\sigma ^{k_1 k_2}\right){}^{\alpha _1\alpha _2}~\left(\sigma ^{k_3 k_4}\right){}^{\alpha _3\alpha _4}~\Lambda ^{l_1\beta _1}~\Lambda ^{l_2\beta _2}~\Lambda ^{l_3\beta _3}~\Lambda ^{l_4\beta _4}$ do not generate new parameters.

\item 
Consider $\left(\sigma ^{k_1 k_2}\right){}^{\alpha _1\alpha _2}~\left(\sigma ^{k_3 k_4}\right){}^{\alpha _3\alpha _4}~\left(\sigma ^{k_5 k_6}\right){}^{\alpha _5\alpha _6}~\Lambda ^{l_1\beta _1}~\Lambda ^{l_2\beta _2}~\Lambda ^{l_3\beta _3}~\Lambda ^{l_4\beta _4}$.
Because the number of $\alpha _i$ is 6, the number of $\beta _i$ is 4, at least 2 $\alpha _i$ have to be contracted with $\epsilon$. Then we can reduce the number of $\sigma$ and do not obtain new parameters.
\end{enumerate}
At the end, we obtain all parameters with indices $k~l$ and $k~l~n~o$, see Eq.~(\ref{Lambda4kl}), (\ref{Lambda4klno}).\\

In the following we will construct parameters with indices $k~ l ~\alpha ~\beta$ from parameters with indices $k~l$ or $k~l~n~o$. Some of them can be obtained by combine $\Lambda ^2$ or $\sigma \Lambda \Lambda$ and the parameters with indices $k~ l ~\alpha ~\beta$ at order $\Lambda ^2$. The others are new parameters. We list the them as follows,
\begin{enumerate}
\label{}
\item 
By combining $\epsilon ^{\alpha _1\beta _1}$ and parameters with indices $k~ l$, we obtain $\epsilon ^{\alpha _1\beta _1}~\left(\eta \sigma \Lambda \Lambda ^{n_1}\right){}^{l_1}~\Lambda _{n_1}{}^{l_2}$.

\item
Combining $\left(\sigma ^{k_1 k_2}\right){}^{\alpha _1\alpha _2}$ and parameters without indices do not give new parameters.

\item
By combining $\left(\sigma ^{k_1 k_2}\right){}^{\alpha _1\alpha _2}$ and parameters with indices $k~l$, we obtain $\left(\sigma ^{k_1 k_2}\right){}^{\alpha _1\alpha _2}~\left(\eta \sigma \Lambda \Lambda ^{n_1}\right){}^{l_1}~\Lambda _{n_1}{}^{l_2}$, it do not give new parameters.

\item
By combining $\left(\sigma ^{k_1 k_2}\right){}^{\alpha _1\alpha _2}$ and parameters with indices $k~l~n~o$, we have 3 possibilities,
\begin{enumerate}
\label{}
\item 
From $\left(\sigma ^{k_1 k_2}\right){}^{\alpha _1\alpha _2}~\Lambda ^{l_1l_2}~\Lambda ^{l_3l_4}$ we have $\left(\sigma ^{n_1 n_2}\right){}^{\alpha _1\alpha _2}~\Lambda _{n_1}{}^{l_2}~\Lambda _{n_2}{}^{l_4}$.

\item 
In $\left(\sigma ^{k_1 k_2}\right){}^{\alpha _1\alpha _2}~\epsilon ^{n_1n_2k_3k_4}~\Lambda _{n_1}{}^{l_1}~\Lambda _{n_2}{}^{l_2}$ we can remove $\epsilon ^{n_1n_2k_3k_4}$ and do not obtain new parameters.

\item 
For $\left(\sigma ^{k_1 k_2}\right){}^{\alpha _1\alpha _2}~\left(\eta \sigma \Lambda \Lambda ^{k_3}\right){}^{l_3}~\Lambda ^{l_1l_2}$, $\left(\sigma ^{k_1 k_2}\right){}^{\alpha _1\alpha _2}~\left(\sigma \Lambda \Lambda ^{n_1 k_3}\right){}^{l_1l_2}~\Lambda _{n_1}{}^{l_3}$ \\and $\left(\sigma ^{k_1 k_2}\right){}^{\alpha _1\alpha _2}~\eta ^{k_3 k_4}~\left(\eta \sigma \Lambda \Lambda ^{n_1}\right){}^{l_1}~\Lambda _{n_1}{}^{l_2}$, we can reduce the number of $\sigma$ in them and do not obtain new parameters.
\end{enumerate}
\end{enumerate}
In summary, we only need to use at most one $\left(\sigma ^{k_1 k_2}\right){}^{\alpha _1\alpha _2}$, one $\epsilon ^{k_1k_2k_3k_4}$ and $\Lambda ^{l_1\beta _1}~\Lambda ^{l_2\beta _2}~\Lambda ^{l_3\beta _3}~\Lambda ^{l_4\beta _4}$ to construct parameters with indices $k ~l ~\alpha  ~\beta$, we have 2 possibilities,
\begin{enumerate}
\label{}
\item 
Due to different contractions of $\left(\sigma ^{k_1 k_2}\right){}^{\alpha _1\alpha _2}~\Lambda ^{l_1\beta _1}~\Lambda ^{l_2\beta _2}~\Lambda ^{l_3\beta _3}~\Lambda ^{l_4\beta _4}$ with $\epsilon$, we can obtain 3 forms: $\left(\sigma ^{k_1 k_2}\right){}^{\alpha _1\alpha _2}~\Lambda ^{l_1l_2}~\Lambda ^{l_3l_4}$, $\left(\sigma \Lambda ^{k_1 k_2}\right){}^{l_1\alpha _2}~\Lambda ^{l_2\beta _2}~\Lambda ^{l_3l_4}$, $\left(\sigma \Lambda \Lambda ^{k_1 k_2}\right){}^{l_1l_2}~\Lambda ^{l_3\beta _3}~\Lambda ^{l_4\beta _4}$. 
By contracting them with $\eta$, we obtain 6 parameters,
\begin{eqnarray}
\label{}
&&\left(\sigma ^{n_1 n_2}\right){}^{\alpha _1\alpha _2}~\Lambda^{l_2} {}_{n_1}~\Lambda^{l_4} {}_{n_2},\quad
\left(\sigma \Lambda \Lambda ^{n_1 n_2}\right){}^{l_1l_2}~\Lambda _{n_1}{}^{\beta _3}~\Lambda _{n_2}{}^{\beta _4},\quad 
\left(\eta \sigma \Lambda \Lambda ^{n_1}\right){}^{l_2}~\Lambda _{n_1}{}^{\beta _3}~\Lambda ^{l_4\beta _4}.\nonumber\\
&&\left(\eta \sigma \Lambda ^{n_1}\right){}^{\alpha _2}~\Lambda ^{l_2\beta _2}~\Lambda^{l_4}{}_{n_1},\quad 
\left(\eta \sigma \Lambda ^{n_1}\right){}^{\alpha _2}~\Lambda _{n_1}{}^{\beta _2}~\Lambda ^{l_3l_4}, \quad
\left(\sigma \Lambda ^{n_1 n_2}\right){}^{l_3\alpha _2}~\Lambda^{l_4} {}_{n_2}~\Lambda _{n_1}{}^{\beta _2},\nonumber
\end{eqnarray}

\item
If we use $\epsilon ^{k_1k_2k_3k_4}$, the result is $\epsilon ^{k_1n_1n_2n_3}~\Lambda^{l_1}{}_{n_1}~\Lambda _{n_2}{}^{\beta _3}~\Lambda _{n_3}{}^{\beta _4}$.
\end{enumerate}
At the end we obtain all parameters with indices $k~ l~ \alpha ~\beta$, see Eq.~(\ref{Lambda4klab}).

\section{The 1/2 supersymmetry invariant action}
\subsection{$\Lambda ^2$}

We define,
{\small
\begin{eqnarray}
\label{}
X_i&=&\Lambda ^2~x_{i,1}+\sigma \Lambda \Lambda ~ x_{i,2},\label{X0}\\
X_i{}^{{k_1} {k_2}}&=&x_{i,1} ~\left(\eta \sigma \Lambda \Lambda ^{{k_1}}\right)^{{k_2}}
+x_{i,2} ~\eta ^{{k_1} {k_2}} ~\Lambda ^2
+x_{i,3}~ \Lambda ^{{k_1} {k_2}}
+x_{i,4}~\eta ^{{k_1} {k_2}}~ \sigma \Lambda \Lambda,\label{X2}\\
X_i{}^{{k_1} {k_2} {k_3} {k_4}}&=&x_{i,1} ~\eta ^{{k_1} {k_2}} ~\left(\eta \sigma \Lambda \Lambda ^{{k_3}}\right)^{{k_4}}
+x_{i,2} ~\eta ^{{k_1} {k_2}} ~\eta ^{{k_3} {k_4}} ~\Lambda ^2
+x_{i,3}~ \eta ^{{k_3} {k_4}} ~\Lambda ^{{k_1} {k_2}}\nonumber\\
&&+x_{i,4} ~\eta ^{{k_1} {k_2}} ~\eta ^{{k_3} {k_4}} ~\sigma \Lambda \Lambda,\label{X4-1}\\
\overline{X}_i{}^{{k_1} {k_2}, {k_3} {k_4}}&=&x_{i,1} ~\eta ^{{k_1} {k_2}} ~\left(\eta \sigma \Lambda \Lambda ^{{k_3}}\right)^{{k_4}}
+x_{i,2} ~\eta ^{{k_1} {k_3}} ~\left(\eta \sigma \Lambda \Lambda ^{{k_2}}\right)^{{k_4}}
+x_{i,3}~ \eta ^{{k_1} {k_3}}~ \eta ^{{k_2} {k_4}} ~\Lambda ^2\nonumber\\
&&+x_{i,4} ~\eta ^{{k_1} {k_2}} ~\eta ^{{k_3} {k_4}}~ \Lambda ^2
+x_{i,5}~ \eta ^{{k_1} {k_2}}~ \Lambda ^{{k_3} {k_4}}
+x_{i,6}~\eta ^{{k_1} {k_3}}~ \Lambda ^{{k_2} {k_4}}\nonumber\\
&&+x_{i,7}~\eta ^{{k_1} {k_3}}~ \eta ^{{k_2} {k_4}} ~\sigma \Lambda \Lambda 
+x_{i,8}~ \eta ^{{k_1} {k_2}} ~\eta ^{{k_3} {k_4}} \sigma \Lambda \Lambda 
+x_{i,9}~ \left(\sigma \Lambda \Lambda ^{{k_1} {k_3}}\right)^{{k_2} {k_4}},\label{X4-2}\\
X_i{}^{{k_1} \dot{\alpha }\beta }&=&x_{i,1} ~\eta _{{k_2} {k_3}} ~\left(\overline{\sigma }^{{k_3}}\right)^{\dot{\alpha } \beta } ~\left(\eta \sigma \Lambda \Lambda ^{{k_1}}\right)^{{k_2}}
+x_{i,2} ~\eta _{{k_2} {k_3}} ~\left(\overline{\sigma }^{{k_3}}\right)^{\dot{\alpha } \beta } ~\left(\eta \sigma \Lambda \Lambda ^{{k_2}}\right)^{{k_1}}\nonumber\\
&&+x_{i,3}~ \Lambda ^2~ \left(\overline{\sigma }^{{k_1}}\right)^{\dot{\alpha } \beta }
+x_{i,4}~ \sigma \Lambda \Lambda ~ \left(\overline{\sigma }^{{k_1}}\right)^{\dot{\alpha } \beta }
+x_{i,5} ~\left(\overline{\sigma }^{{k_2}}\right)^{\dot{\alpha } \beta } ~\Lambda ^{{k_1}}{}_{{k_2}},\label{X3}\\
X_i{}^{{k_1} {k_2} \alpha  \beta }&=&x_{i,1} ~\eta ^{{k_1} {k_2}} ~\epsilon ^{\alpha  \beta } ~\Lambda ^2
+x_{i,2} ~\eta ^{{k_1} {k_2}} ~\epsilon ^{\alpha  \beta } ~\sigma \Lambda \Lambda 
+x_{i,3} ~\epsilon ^{\alpha  \beta } ~\left(\eta \sigma \Lambda \Lambda ^{{k_1}}\right)^{{k_2}}
\nonumber\\&&
+x_{i,4} ~\epsilon ^{\alpha  \beta } ~\Lambda ^{{k_1} {k_2}}
+x_{i,5}~ \left(\sigma ^{{k_1} {k_3}}\right)^{\alpha  \beta }~ \Lambda ^{{k_2}}{}_{{k_3}}
+x_{i,6}~ \left(\sigma \Lambda ^{{k_1} {k_3}}\right)^{{k_2} \alpha }~ \Lambda _{{k_3}}{}^{\beta }
\nonumber\\&&
+x_{i,7} ~\left(\eta \sigma \Lambda ^{{k_1}}\right)^{\alpha }~ \Lambda ^{{k_2} \beta }
+x_{i,8} ~\eta ^{{k_1} {k_2}} ~\left(\eta \sigma \Lambda ^{{k_3}}\right)^{\alpha }~ \Lambda _{{k_3}}{}^{\beta },\label{X4-3}\\
\overline{X}_i{}^{{k_1} {k_2} \alpha  \beta }&=&x_{i,1} ~\eta ^{{k_1} {k_2}} ~\epsilon ^{\alpha  \beta } ~\Lambda ^2
+x_{i,2} ~\eta ^{{k_1} {k_2}}~ \epsilon ^{\alpha  \beta }~ \sigma \Lambda \Lambda 
+x_{i,3} ~\Lambda ^2 ~\left(\sigma ^{{k_1} {k_2}}\right)^{\alpha  \beta }
\nonumber\\&&
+x_{i,4} ~\sigma \Lambda \Lambda  ~\left(\sigma ^{{k_1} {k_2}}\right)^{\alpha  \beta }
+x_{i,5} ~\epsilon ^{\alpha  \beta } ~\left(\eta \sigma \Lambda \Lambda ^{{k_1}}\right)^{{k_2}}
+x_{i,6} ~\epsilon ^{\alpha  \beta } ~\Lambda ^{{k_1} {k_2}}
\nonumber\\&&
+x_{i,7} ~\Lambda ^{{k_1} \alpha } ~\Lambda ^{{k_2} \beta }
+x_{i,8}~ \epsilon ^{{k_1} {k_2} {k_3} {k_4}} ~\Lambda _{{k_3}}{}^{\alpha } ~\Lambda _{{k_4}}{}^{\beta }
+x_{i,9} ~\left(\sigma ^{{k_1} {k_3}}\right)^{\alpha  \beta } ~\Lambda ^{{k_2}}{}_{{k_3}}
\nonumber\\&&
+x_{i,10} ~\left(\sigma \Lambda ^{{k_1} {k_3}}\right)^{{k_2} \alpha } ~\Lambda _{{k_3}}{}^{\beta }
+x_{i,11}~ \left(\eta \sigma \Lambda ^{{k_1}}\right)^{\alpha } ~\Lambda ^{{k_2} \beta }.\label{X4-4}
\end{eqnarray}
}
where $x_{i, j}$ are free parameters.
They can combine with operators as follows,
{\small
\begin{eqnarray}
\label{}
&&X_i{}^{{k_1} {k_2}}~\Phi  ~\left(\partial _{{k_1}}\partial _{{k_2}}\Phi ^+\right),\quad 
X_i{}^{{k_1} {k_2} {k_3} {k_4}}~\Phi~  \left(\partial _{{k_1}}\partial _{{k_2}}\partial _{{k_3}}\partial _{{k_4}}\Phi ^+\right),\nonumber\\
&&\overline{X}_i{}^{{k_1} {k_2},{k_3} {k_4}} ~\left(\partial _{{k_1}}\partial _{{k_2}}\Phi ^+\right) \left(\partial _{{k_3}}\partial _{{k_4}}\Phi ^+\right) \Phi ^+,\quad
X_i{}^{{k_1} \dot{\alpha }\beta }~ \left(D_{\beta } \Phi \right) \left(\partial _{{k_1}}\Phi\right)  \left(\overline{D}_{\dot{\alpha }} \Phi ^+\right)\nonumber\\
&&X_i{}^{{k_1} {k_2} \alpha  \beta } ~\left(\partial _{{k_1}}\partial _{{k_2}}D_{\alpha } \Phi \right) \left(D_{\beta } \Phi \right) \Phi,\quad
\overline{X}_i{}^{{k_1} {k_2} \alpha  \beta } \left(\partial _{{k_1}}D_{\alpha } \Phi \right) \left(\partial _{{k_2}}D_{\beta } \Phi \right) \Phi.
\end{eqnarray}
}

\subsection{$\Lambda ^4$}
We define,
{\small
\begin{eqnarray}
\label{}
Y_i&=&y_{i,1}~ \Lambda ^2 ~\Lambda ^2
+y_{i,2}~ \Lambda ^2 ~\sigma \Lambda \Lambda 
+y_{i,3}~ \sigma \Lambda \Lambda ~ \sigma \Lambda \Lambda,\label{Y0}\\
Y_i{}^{{k_1} {k_2}}&=&y_{i,1}~ \eta ^{{k_1} {k_2}} ~\Lambda ^2 ~\Lambda ^2
+y_{i,2} ~\eta ^{{k_1} {k_2}} ~\Lambda ^2 ~\sigma \Lambda \Lambda 
+y_{i,3}~ \eta ^{{k_1} {k_2}} ~\sigma \Lambda \Lambda~  \sigma \Lambda \Lambda 
\nonumber\\&&
+y_{i,4} ~\Lambda ^2 ~\Lambda ^{{k_1} {k_2}}
+y_{i,5}~ \sigma \Lambda \Lambda  ~\Lambda ^{{k_1} {k_2}}
+y_{i,6} ~\Lambda ^2 \left(\eta \sigma \Lambda \Lambda ^{{k_1}}\right)^{{k_2}}
\nonumber\\&&
+y_{i,7} ~\sigma \Lambda \Lambda~  \left(\eta \sigma \Lambda \Lambda ^{{k_1}}\right)^{{k_2}},\\
\overline{Y}_i{}^{{k_1} {k_2}}&=&y_{i,1} ~\eta ^{{k_1} {k_2}}~ \Lambda ^2 ~\Lambda ^2
+y_{i,2} ~\eta ^{{k_1} {k_2}} ~\Lambda ^2 ~\sigma \Lambda \Lambda 
+y_{i,3} ~\eta ^{{k_1} {k_2}}~ \sigma \Lambda \Lambda  ~\sigma \Lambda \Lambda 
\nonumber\\&&
+y_{i,4} ~\Lambda ^2 ~\Lambda ^{{k_1} {k_2}}
+y_{i,5} ~\sigma \Lambda \Lambda ~ \Lambda ^{{k_1} {k_2}}
+y_{i,6} ~\Lambda ^2 \left(\eta \sigma \Lambda \Lambda ^{{k_1}}\right)^{{k_2}}
\nonumber\\&&
+y_{i,7}~ \Lambda ^2 \left(\eta \sigma \Lambda \Lambda ^{{k_2}}\right)^{{k_1}}
+y_{i,8} ~\sigma \Lambda \Lambda  \left(\eta \sigma \Lambda \Lambda ^{{k_1}}\right)^{{k_2}}
+y_{i,9} ~\sigma \Lambda \Lambda  \left(\eta \sigma \Lambda \Lambda ^{{k_2}}\right)^{{k_1}}
\nonumber\\&&
+y_{i,10} ~\left(\eta \sigma \Lambda \Lambda ^{{k_3}}\right)^{{k_1}} \Lambda ^{{k_2}}{}_{{k_3}}
+y_{i,11}~ \left(\eta \sigma \Lambda \Lambda ^{{k_3}}\right)^{{k_2}} \Lambda ^{{k_1}}{}_{{k_3}},\\
Y_i{}^{{k_1} {k_2} {k_3} {k_4}}&=&y_{i,1}~ \eta ^{{k_1} {k_2}} ~\eta ^{{k_3} {k_4}}~ \Lambda ^2 ~\Lambda ^2
+y_{i,2} ~\eta ^{{k_1} {k_2}} ~\eta ^{{k_3} {k_4}}~ \Lambda ^2~ \sigma \Lambda \Lambda 
+y_{i,3} ~\eta ^{{k_1} {k_2}} ~\eta ^{{k_3} {k_4}} ~\sigma \Lambda \Lambda ~ \sigma \Lambda \Lambda \nonumber\\&&
+y_{i,4} ~\eta ^{{k_1} {k_2}} ~\Lambda ^2 ~\Lambda ^{{k_3} {k_4}}
+y_{i,5} ~\eta ^{{k_1} {k_2}} ~\sigma \Lambda \Lambda  ~\Lambda ^{{k_3} {k_4}}
+y_{i,6} ~\eta ^{{k_1} {k_2}}~ \Lambda ^2 \left(\eta \sigma \Lambda \Lambda ^{{k_3}}\right)^{{k_4}}\nonumber\\&&
+y_{i,7} ~\eta ^{{k_1} {k_2}}~ \sigma \Lambda \Lambda  ~\left(\eta \sigma \Lambda \Lambda ^{{k_3}}\right)^{{k_4}},\\
\overline{Y}_i{}^{{k_1},{k_2} {k_3} {k_4}}&=&y_{i,1}~ \eta ^{{k_1} {k_2}} ~\eta ^{{k_3} {k_4}} ~\Lambda ^2 ~\Lambda ^2
+y_{i,2} ~\eta ^{{k_1} {k_2}} ~\eta ^{{k_3} {k_4}} ~\Lambda ^2 ~\sigma \Lambda \Lambda 
+y_{i,3}~ \eta ^{{k_1} {k_2}} ~\eta ^{{k_3} {k_4}} ~\sigma \Lambda \Lambda  ~\sigma \Lambda \Lambda \nonumber\\&&
+y_{i,4} ~\eta ^{{k_3} {k_4}} ~\Lambda ^2~ \Lambda ^{{k_1} {k_2}}
+y_{i,5} ~\eta ^{{k_1} {k_2}} ~\Lambda ^2 ~\Lambda ^{{k_3} {k_4}}
+y_{i,6}~ \eta ^{{k_2} {k_3}} ~\sigma \Lambda \Lambda  ~\Lambda ^{{k_1} {k_4}}
\nonumber\\&&
+y_{i,7} ~\eta ^{{k_1} {k_2}} ~\sigma \Lambda \Lambda  ~\Lambda ^{{k_3} {k_4}}
+y_{i,8} ~\eta ^{{k_2} {k_3}} ~\Lambda ^2 \left(\eta \sigma \Lambda \Lambda ^{{k_1}}\right)^{{k_4}}
+y_{i,9}~ \eta ^{{k_2} {k_3}} ~\Lambda ^2 \left(\eta \sigma \Lambda \Lambda ^{{k_4}}\right)^{{k_1}}
\nonumber\\&&
+y_{i,10}~ \eta ^{{k_1} {k_2}} ~\Lambda ^2 \left(\eta \sigma \Lambda \Lambda ^{{k_3}}\right)^{{k_4}}
+y_{i,11}~ \eta ^{{k_2} {k_3}} ~\sigma \Lambda \Lambda \left(\eta \sigma \Lambda \Lambda ^{{k_1}}\right)^{{k_4}}
\nonumber\\&&
+y_{i,12} ~\eta ^{{k_2} {k_3}} ~\sigma \Lambda \Lambda \left(\eta \sigma \Lambda \Lambda ^{{k_4}}\right)^{{k_1}}
+y_{i,13} ~\eta ^{{k_1} {k_2}}~ \sigma \Lambda \Lambda\left(\eta \sigma \Lambda \Lambda ^{{k_3}}\right)^{{k_4}}
\nonumber\\&&
+y_{i,14} ~\Lambda ^{{k_2} {k_3}} \left(\eta \sigma \Lambda \Lambda ^{{k_1}}\right)^{{k_4}}
+y_{i,15} ~\Lambda ^{{k_2} {k_3}} \left(\eta \sigma \Lambda \Lambda ^{{k_4}}\right)^{{k_1}}
\nonumber\\&&
+y_{i,16} ~\Lambda ^{{k_1} {k_2}} \left(\eta \sigma \Lambda \Lambda ^{{k_3}}\right)^{{k_4}}
+y_{i,17} ~\eta ^{{k_2} {k_3}} \left(\eta \sigma \Lambda \Lambda ^{{n_1}}\right)^{{k_4}} \Lambda ^{{k_1}}{}_{{n_1}}
\nonumber\\&&
+y_{i,18} ~\eta ^{{k_2} {k_3}} \left(\eta \sigma \Lambda \Lambda ^{{n_1}}\right)^{{k_1}} \Lambda ^{{k_4}}{}_{{n_1}},\\
\widetilde{Y}_i{}^{{k_1} {k_2},{k_3} {k_4}}&=&y_{i,1} ~\eta ^{{k_1} {k_3}}~ \eta ^{{k_2} {k_4}} ~\Lambda ^2 \Lambda ^2
+y_{i,2} ~\eta ^{{k_1} {k_2}} ~\eta ^{{k_3} {k_4}} ~\Lambda ^2 ~\Lambda ^2
+y_{i,3} ~\eta ^{{k_1} {k_3}} ~\eta ^{{k_2} {k_4}} ~\Lambda ^2 ~\sigma \Lambda \Lambda 
\nonumber\\&&
+y_{i,4} ~\eta ^{{k_1} {k_2}} ~\eta ^{{k_3} {k_4}} ~\Lambda ^2~ \sigma \Lambda \Lambda
+y_{i,5} ~\eta ^{{k_1} {k_3}} ~\eta ^{{k_2} {k_4}}~ \sigma \Lambda \Lambda  ~\sigma \Lambda \Lambda
\nonumber\\&&
+y_{i,6} ~\eta ^{{k_1} {k_2}} ~\eta ^{{k_3} {k_4}}~ \sigma \Lambda \Lambda  ~\sigma \Lambda \Lambda 
+y_{i,7} ~\eta ^{{k_1} {k_3}} ~\Lambda ^2 ~\Lambda ^{{k_2} {k_4}}
+y_{i,8} ~\eta ^{{k_1} {k_2}} ~\Lambda ^2 ~\Lambda ^{{k_3} {k_4}}
\nonumber\\&&
+y_{i,9}~ \eta ^{{k_1} {k_3}} ~\sigma \Lambda \Lambda ~ \Lambda ^{{k_2} {k_4}}
+y_{i,10} ~\eta ^{{k_1} {k_2}} ~\sigma \Lambda \Lambda ~ \Lambda ^{{k_3} {k_4}}
+y_{i,11} ~\eta ^{{k_1} {k_3}} ~\Lambda ^2 \left(\eta \sigma \Lambda \Lambda ^{{k_4}}\right)^{{k_2}}
\nonumber\\&&
+y_{i,12}~ \eta ^{{k_1} {k_2}} ~\Lambda ^2 \left(\eta \sigma \Lambda \Lambda ^{{k_3}}\right)^{{k_4}}
+y_{i,13} ~\eta ^{{k_1} {k_3}}~ \sigma \Lambda \Lambda  \left(\eta \sigma \Lambda \Lambda ^{{k_4}}\right)^{{k_2}}
\nonumber\\&&
+y_{i,14} ~\eta ^{{k_1} {k_2}} ~\sigma \Lambda \Lambda  \left(\eta \sigma \Lambda \Lambda ^{{k_3}}\right)^{{k_4}}
+y_{i,15} ~\Lambda ^2 \left(\sigma \Lambda \Lambda ^{{k_1} {k_3}}\right)^{{k_2} {k_4}}
\nonumber\\&&
+y_{i,16} ~\sigma \Lambda \Lambda  \left(\sigma \Lambda \Lambda ^{{k_1} {k_3}}\right)^{{k_2} {k_4}}
+y_{i,17} ~\Lambda ^{{k_1} {k_2}} ~\Lambda ^{{k_3} {k_4}}
+y_{i,18} ~\Lambda ^{{k_1} {k_2}} \left(\eta \sigma \Lambda \Lambda ^{{k_3}}\right)^{{k_4}}
\nonumber\\&&
+y_{i,19} ~\eta ^{{k_1} {k_3}} \left(\eta \sigma \Lambda \Lambda ^{{n_1}}\right)^{{k_4}} \Lambda ^{{k_2}}{}_{{n_1}}
+y_{i,20} ~\epsilon ^{{k_1} {k_3} {n_1} {n_2}}~ \Lambda ^{{k_2}}{}_{{n_1}} ~\Lambda ^{{k_4}}{}_{{n_2}},\\
\hat{Y}_i{}^{{k_1} {k_2},{k_3} {k_4}}&=&y_{i,1} ~\eta ^{{k_1} {k_3}} ~\eta ^{{k_2} {k_4}} ~\Lambda ^2 ~\Lambda ^2
+y_{i,2} ~\eta ^{{k_1} {k_2}} ~\eta ^{{k_3} {k_4}} ~\Lambda ^2 ~\Lambda ^2
+y_{i,3} ~\eta ^{{k_1} {k_3}} ~\eta ^{{k_2} {k_4}}~ \Lambda ^2 ~\sigma \Lambda \Lambda 
\nonumber\\&&
+y_{i,4} ~\eta ^{{k_1} {k_2}} \eta ^{{k_3} {k_4}} ~\Lambda ^2 ~\sigma \Lambda \Lambda
+y_{i,5} ~\eta ^{{k_1} {k_3}} ~\eta ^{{k_2} {k_4}} ~\sigma \Lambda \Lambda  ~\sigma \Lambda \Lambda
\nonumber\\&&
+y_{i,6} ~\eta ^{{k_1} {k_2}}~ \eta ^{{k_3} {k_4}} ~\sigma \Lambda \Lambda  ~\sigma \Lambda \Lambda 
+y_{i,7} ~\eta ^{{k_3} {k_4}} ~\Lambda ^2~ \Lambda ^{{k_1} {k_2}}
+y_{i,8}~ \eta ^{{k_1} {k_3}}~ \Lambda ^2~ \Lambda ^{{k_2} {k_4}}
\nonumber\\&&
+y_{i,9}~ \eta ^{{k_1} {k_2}} ~\Lambda ^2 ~\Lambda ^{{k_3} {k_4}}
+y_{i,10}~ \eta ^{{k_3} {k_4}} ~\sigma \Lambda \Lambda~  \Lambda ^{{k_1} {k_2}}
+y_{i,11}~ \eta ^{{k_1} {k_3}} ~\sigma \Lambda \Lambda ~ \Lambda ^{{k_2} {k_4}}
\nonumber\\&&
+y_{i,12} ~\eta ^{{k_1} {k_2}}~ \sigma \Lambda \Lambda ~ \Lambda ^{{k_3} {k_4}}
+y_{i,13} ~\eta ^{{k_1} {k_3}} ~\Lambda ^2 \left(\eta \sigma \Lambda \Lambda ^{{k_2}}\right)^{{k_4}}
\nonumber\\&&
+y_{i,14} ~\eta ^{{k_3} {k_4}} ~\Lambda ^2 \left(\eta \sigma \Lambda \Lambda ^{{k_1}}\right)^{{k_2}}
+y_{i,15} ~\eta ^{{k_1} {k_3}} ~\Lambda ^2 \left(\eta \sigma \Lambda \Lambda ^{{k_4}}\right)^{{k_2}}
\nonumber\\&&
+y_{i,16} ~\eta ^{{k_1} {k_2}}~ \Lambda ^2 \left(\eta \sigma \Lambda \Lambda ^{{k_3}}\right)^{{k_4}}
+y_{i,17}~ \eta ^{{k_1} {k_3}} ~\sigma \Lambda \Lambda  \left(\eta \sigma \Lambda \Lambda ^{{k_2}}\right)^{{k_4}}
\nonumber\\&&
+y_{i,18} ~\eta ^{{k_3} {k_4}} ~\sigma \Lambda \Lambda  \left(\eta \sigma \Lambda \Lambda ^{{k_1}}\right)^{{k_2}}
+y_{i,19} ~\eta ^{{k_1} {k_3}} ~\sigma \Lambda \Lambda  \left(\eta \sigma \Lambda \Lambda ^{{k_4}}\right)^{{k_2}}
\nonumber\\&&
+y_{i,20} ~\eta ^{{k_1} {k_2}} ~\sigma \Lambda \Lambda  \left(\eta \sigma \Lambda \Lambda ^{{k_3}}\right)^{{k_4}}
+y_{i,21}~ \Lambda ^2 \left(\sigma \Lambda \Lambda ^{{k_1} {k_3}}\right)^{{k_2} {k_4}}
\nonumber\\&&
+y_{i,22} ~\sigma \Lambda \Lambda  \left(\sigma \Lambda \Lambda ^{{k_1} {k_3}}\right)^{{k_2} {k_4}}
+y_{i,23}~ \Lambda ^{{k_1} {k_2}} ~\Lambda ^{{k_3} {k_4}}
+y_{i,24} ~\Lambda ^{{k_3} {k_4}} \left(\eta \sigma \Lambda \Lambda ^{{k_1}}\right)^{{k_2}}
\nonumber\\&&
+y_{i,25} ~\Lambda ^{{k_1} {k_2}} \left(\eta \sigma \Lambda \Lambda ^{{k_3}}\right)^{{k_4}}
+y_{i,26} ~\eta ^{{k_1} {k_3}} \left(\eta \sigma \Lambda \Lambda ^{{n_1}}\right)^{{k_4}} \Lambda ^{{k_2}}{}_{{n_1}}
\nonumber\\&&
+y_{i,27} ~\eta ^{{k_1} {k_3}} \left(\eta \sigma \Lambda \Lambda ^{{n_1}}\right)^{{k_2}} \Lambda ^{{k_4}}{}_{{n_1}}
+y_{i,28} ~\epsilon ^{{k_1} {k_3} {n_1} {n_2}} ~\Lambda ^{{k_2}}{}_{{n_1}} ~\Lambda ^{{k_4}}{}_{{n_2}},\\
\widetilde{\overline{Y}}_i{}^{{k_1},{k_2} {k_3},{k_4}}&=&y_{i,1} ~\eta ^{{k_1} {k_4}} ~\eta ^{{k_2} {k_3}} ~\Lambda ^2 ~\Lambda ^2
+y_{i,2}~\eta ^{{k_1} {k_3}} ~\eta ^{{k_2} {k_4}} ~\Lambda ^2 ~\Lambda ^2
+y_{i,3} ~\eta ^{{k_1} {k_4}} ~\eta ^{{k_2} {k_3}} ~\Lambda ^2~ \sigma \Lambda \Lambda 
\nonumber\\&&
+y_{i,4}~ \eta ^{{k_1} {k_3}} ~\eta ^{{k_2} {k_4}} ~\Lambda ^2 ~\sigma \Lambda \Lambda
+y_{i,5} ~\eta ^{{k_1} {k_4}}~ \eta ^{{k_2} {k_3}} ~\sigma \Lambda \Lambda  ~\sigma \Lambda \Lambda
\nonumber\\&&
+y_{i,6} ~\eta ^{{k_1} {k_3}}~ \eta ^{{k_2} {k_4}} ~\sigma \Lambda \Lambda  ~\sigma \Lambda \Lambda 
+y_{i,7} ~\eta ^{{k_2} {k_3}}~ \Lambda ^2~ \Lambda ^{{k_1} {k_4}}
+y_{i,8}~\eta ^{{k_2} {k_4}} ~\Lambda ^2~ \Lambda ^{{k_1} {k_3}}
\nonumber\\&&
+y_{i,9}~ \eta ^{{k_1} {k_2}} ~\Lambda ^2 ~\Lambda ^{{k_3} {k_4}}
+y_{i,10}~ \eta ^{{k_1} {k_4}}~ \Lambda ^2 ~\Lambda ^{{k_2} {k_3}}
+y_{i,11}~ \eta ^{{k_2} {k_3}} ~\sigma \Lambda \Lambda  ~\Lambda ^{{k_1} {k_4}}
\nonumber\\&&
+y_{i,12} ~\eta ^{{k_2} {k_4}} ~\sigma \Lambda \Lambda ~ \Lambda ^{{k_1} {k_3}}
+y_{i,13} ~\eta ^{{k_1} {k_2}} ~\sigma \Lambda \Lambda ~ \Lambda ^{{k_3} {k_4}}
+y_{i,14} ~\eta ^{{k_1} {k_4}} ~\sigma \Lambda \Lambda  ~\Lambda ^{{k_2} {k_3}}
\nonumber\\&&
+y_{i,15}~ \eta ^{{k_2} {k_3}} ~\Lambda ^2 \left(\eta \sigma \Lambda \Lambda ^{{k_1}}\right)^{{k_4}}
+y_{i,16} ~\eta ^{{k_2} {k_3}}~ \Lambda ^2 \left(\eta \sigma \Lambda \Lambda ^{{k_4}}\right)^{{k_1}}
\nonumber\\&&
+y_{i,17}~ \eta ^{{k_2} {k_4}} ~\Lambda ^2 \left(\eta \sigma \Lambda \Lambda ^{{k_1}}\right)^{{k_3}}
+y_{i,18} ~\eta ^{{k_1} {k_2}} ~\Lambda ^2 \left(\eta \sigma \Lambda \Lambda ^{{k_4}}\right)^{{k_3}}
\nonumber\\&&
+y_{i,19}~ \eta ^{{k_2} {k_4}} ~\Lambda ^2 \left(\eta \sigma \Lambda \Lambda ^{{k_3}}\right)^{{k_1}}
+y_{i,20} ~\eta ^{{k_1} {k_2}} ~\Lambda ^2 \left(\eta \sigma \Lambda \Lambda ^{{k_3}}\right)^{{k_4}}
\nonumber\\&&
+y_{i,21}~ \eta ^{{k_1} {k_4}} ~\Lambda ^2 \left(\eta \sigma \Lambda \Lambda ^{{k_2}}\right)^{{k_3}}
+y_{i,22}~ \eta ^{{k_2} {k_3}} ~\sigma \Lambda \Lambda  \left(\eta \sigma \Lambda \Lambda ^{{k_1}}\right)^{{k_4}}
\nonumber\\&&
+y_{i,23} ~\eta ^{{k_2} {k_3}}~\sigma \Lambda \Lambda  \left(\eta \sigma \Lambda \Lambda ^{{k_4}}\right)^{{k_1}}
+y_{i,24} ~\eta ^{{k_2} {k_4}} ~\sigma \Lambda \Lambda  \left(\eta \sigma \Lambda \Lambda ^{{k_1}}\right)^{{k_3}}
\nonumber\\&&
+y_{i,25}~ \eta ^{{k_1} {k_2}} ~\sigma \Lambda \Lambda  \left(\eta \sigma \Lambda \Lambda ^{{k_4}}\right)^{{k_3}}
+y_{i,26} ~\eta ^{{k_2} {k_4}} ~\sigma \Lambda \Lambda  \left(\eta \sigma \Lambda \Lambda ^{{k_3}}\right)^{{k_1}}
\nonumber\\&&
+y_{i,27} ~\eta ^{{k_1} {k_2}} ~\sigma \Lambda \Lambda  \left(\eta \sigma \Lambda \Lambda ^{{k_3}}\right)^{{k_4}}
+y_{i,28}~ \eta ^{{k_1} {k_4}} ~\sigma \Lambda \Lambda  \left(\eta \sigma \Lambda \Lambda ^{{k_2}}\right)^{{k_3}}
\nonumber\\&&
+y_{i,29}~ \epsilon ^{{k_1} {k_2} {k_4} {n_1}}~ \Lambda ^2~ \Lambda ^{{k_3}}{}_{{n_1}}
+y_{i,30}~ \Lambda ^2 \left(\sigma \Lambda \Lambda ^{{k_1} {k_2}}\right)^{{k_3} {k_4}}
+y_{i,31} ~\Lambda ^2 \left(\sigma \Lambda \Lambda ^{{k_2} {k_4}}\right)^{{k_1} {k_3}}
\nonumber\\&&
+y_{i,32} ~\sigma \Lambda \Lambda  \left(\sigma \Lambda \Lambda ^{{k_1} {k_2}}\right)^{{k_3} {k_4}}
+y_{i,33} ~\sigma \Lambda \Lambda  \left(\sigma \Lambda \Lambda ^{{k_2} {k_4}}\right)^{{k_1} {k_3}}
+y_{i,34}~ \Lambda ^{{k_1} {k_4}}~ \Lambda ^{{k_2} {k_3}}
\nonumber\\&&
+y_{i,35}~ \Lambda ^{{k_2} {k_3}} \left(\eta \sigma \Lambda \Lambda ^{{k_1}}\right)^{{k_4}}
+y_{i,36} ~\Lambda ^{{k_2} {k_3}} \left(\eta \sigma \Lambda \Lambda ^{{k_4}}\right)^{{k_1}}
+y_{i,37}~ \Lambda ^{{k_2} {k_4}} \left(\eta \sigma \Lambda \Lambda ^{{k_3}}\right)^{{k_1}}
\nonumber\\&&
+y_{i,38}~ \Lambda ^{{k_1} {k_2}} \left(\eta \sigma \Lambda \Lambda ^{{k_3}}\right)^{{k_4}}
+y_{i,39}~ \Lambda ^{{k_1} {k_4}} \left(\eta \sigma \Lambda \Lambda ^{{k_2}}\right)^{{k_3}}
\nonumber\\&&
+y_{i,40}~ \left(\sigma \Lambda \Lambda ^{{k_2} {n_1}}\right)^{{k_3} {k_4}} \Lambda ^{{k_1}}{}_{{n_1}}
+y_{i,41} ~\left(\sigma \Lambda \Lambda ^{{k_2} {n_1}}\right)^{{k_1} {k_3}} \Lambda ^{{k_4}}{}_{{n_1}}
\nonumber\\&&
+y_{i,42} ~\left(\sigma \Lambda \Lambda ^{{k_2} {n_1}}\right)^{{k_1} {k_4}} \Lambda ^{{k_3}}{}_{{n_1}}
+y_{i,43}~ \eta ^{{k_2} {k_3}} \left(\eta \sigma \Lambda \Lambda ^{{n_1}}\right)^{{k_4}} \Lambda ^{{k_1}}{}_{{n_1}}
\nonumber\\&&
+y_{i,44} ~\eta ^{{k_2} {k_3}} \left(\eta \sigma \Lambda \Lambda ^{{n_1}}\right)^{{k_1}} \Lambda ^{{k_4}}{}_{{n_1}}
+y_{i,45}~ \eta ^{{k_2} {k_4}} \left(\eta \sigma \Lambda \Lambda ^{{n_1}}\right)^{{k_3}} \Lambda ^{{k_1}}{}_{{n_1}}
\nonumber\\&&
+y_{i,46} ~\eta ^{{k_1} {k_2}} \left(\eta \sigma \Lambda \Lambda ^{{n_1}}\right)^{{k_3}} \Lambda ^{{k_4}}{}_{{n_1}}
+y_{i,47} ~\eta ^{{k_2} {k_4}} \left(\eta \sigma \Lambda \Lambda ^{{n_1}}\right)^{{k_1}} \Lambda ^{{k_3}}{}_{{n_1}}
\nonumber\\&&
+y_{i,48}~ \eta ^{{k_1} {k_2}} \left(\eta \sigma \Lambda \Lambda ^{{n_1}}\right)^{{k_4}} \Lambda ^{{k_3}}{}_{{n_1}}
+y_{i,49}~ \epsilon ^{{k_2} {k_4} {n_1} {n_2}}~ \Lambda ^{{k_1}}{}_{{n_2}} ~\Lambda ^{{k_3}}{}_{{n_1}}
\nonumber\\&&
+y_{i,50}~ \epsilon ^{{k_1} {k_2} {n_1} {n_2}} ~\Lambda ^{{k_3}}{}_{{n_2}} ~\Lambda ^{{k_4}}{}_{{n_1}},\\
Y_i{}^{{k_1} \dot{\alpha }\beta }&=&y_{i,1} ~\Lambda ^2 ~\Lambda ^2 \left(\overline{\sigma }^{{k_1}}\right)^{\dot{\alpha } \beta }
+y_{i,2} ~\Lambda ^2 ~\sigma \Lambda \Lambda  \left(\overline{\sigma }^{{k_1}}\right)^{\dot{\alpha } \beta }
+y_{i,3} ~\sigma \Lambda \Lambda  ~\sigma \Lambda \Lambda  \left(\overline{\sigma }^{{k_1}}\right)^{\dot{\alpha } \beta }
\nonumber\\&&
+y_{i,4}~ \Lambda ^2 \left(\overline{\sigma }^{{k_2}}\right)^{\dot{\alpha } \beta } \Lambda ^{{k_1}}{}_{{k_2}}
+y_{i,5}~ \sigma \Lambda \Lambda  \left(\overline{\sigma }^{{k_2}}\right)^{\dot{\alpha } \beta } \Lambda ^{{k_1}}{}_{{k_2}}
\nonumber\\&&
+y_{i,6} ~\Lambda ^2~ \eta _{{k_2} {k_3}} \left(\overline{\sigma }^{{k_3}}\right)^{\dot{\alpha } \beta } \left(\eta \sigma \Lambda \Lambda ^{{k_1}}\right)^{{k_2}}
+y_{i,7} ~\Lambda ^2 ~\eta _{{k_2} {k_3}} \left(\overline{\sigma }^{{k_3}}\right)^{\dot{\alpha } \beta } \left(\eta \sigma \Lambda \Lambda ^{{k_2}}\right)^{{k_1}}
\nonumber\\&&
+y_{i,8} ~\sigma \Lambda \Lambda  ~\eta _{{k_2} {k_3}} \left(\overline{\sigma }^{{k_3}}\right)^{\dot{\alpha } \beta } \left(\eta \sigma \Lambda \Lambda ^{{k_1}}\right)^{{k_2}}
+y_{i,9}~ \sigma \Lambda \Lambda ~ \eta _{{k_2} {k_3}} \left(\overline{\sigma }^{{k_3}}\right)^{\dot{\alpha } \beta } \left(\eta \sigma \Lambda \Lambda ^{{k_2}}\right)^{{k_1}}
\nonumber\\&&
+y_{i,10} ~\eta _{{k_3} {k_4}} \left(\overline{\sigma }^{{k_4}}\right)^{\dot{\alpha } \beta } \left(\eta \sigma \Lambda \Lambda ^{{k_2}}\right)^{{k_3}} \Lambda ^{{k_1}}{}_{{k_2}}
+y_{i,11}~ \Lambda _{{k_2} {k_3}} \left(\overline{\sigma }^{{k_3}}\right)^{\dot{\alpha } \beta } \left(\eta \sigma \Lambda \Lambda ^{{k_2}}\right)^{{k_1}},\\
Y_i{}^{{k_1} {k_2} \alpha  \beta }&=&y_{i,1}~ \eta ^{{k_1} {k_2}} ~\epsilon ^{\alpha  \beta } ~\Lambda ^2 \Lambda ^2
+y_{i,2}~ \eta ^{{k_1} {k_2}}~ \epsilon ^{\alpha  \beta }~ \Lambda ^2 ~\sigma \Lambda \Lambda
+y_{i,3} ~\eta ^{{k_1} {k_2}}~ \epsilon ^{\alpha  \beta }~ \sigma \Lambda \Lambda ~ \sigma \Lambda \Lambda
\nonumber\\&&
+y_{i,4} ~\epsilon ^{\alpha  \beta } ~\Lambda ^2 \left(\eta \sigma \Lambda \Lambda ^{{k_2}}\right)^{{k_1}}
+y_{i,5}~ \epsilon ^{\alpha  \beta } ~\sigma \Lambda \Lambda  \left(\eta \sigma \Lambda \Lambda ^{{k_2}}\right)^{{k_1}}
+y_{i,6} ~\epsilon ^{\alpha  \beta }~ \Lambda ^2 ~\Lambda ^{{k_1} {k_2}}
\nonumber\\&&
+y_{i,7}~\epsilon ^{\alpha  \beta } ~\sigma \Lambda \Lambda  ~\Lambda ^{{k_1} {k_2}}
+y_{i,8} ~\Lambda ^2 \left(\sigma ^{{k_1} {k_3}}\right)^{\alpha  \beta } \Lambda ^{{k_2}}{}_{{k_3}}
+y_{i,9} ~\sigma \Lambda \Lambda  \left(\sigma ^{{k_1} {k_3}}\right)^{\alpha  \beta } \Lambda ^{{k_2}}{}_{{k_3}}
\nonumber\\&&
+y_{i,10} ~\Lambda ^2 \left(\sigma \Lambda ^{{k_2} {k_3}}\right)^{{k_1} \alpha } \Lambda _{{k_3}}{}^{\beta }
+y_{i,11}~ \sigma \Lambda \Lambda  \left(\sigma \Lambda ^{{k_2} {k_3}}\right)^{{k_1} \alpha } \Lambda _{{k_3}}{}^{\beta }
\nonumber\\&&
+y_{i,12}~ \Lambda ^2 \left(\eta \sigma \Lambda ^{{k_2}}\right)^{\alpha } \Lambda ^{{k_1} \beta }
+y_{i,13} ~\sigma \Lambda \Lambda  \left(\eta \sigma \Lambda ^{{k_2}}\right)^{\alpha } \Lambda ^{{k_1} \beta }
\nonumber\\&&
+y_{i,14} ~\eta ^{{k_1} {k_2}}~ \Lambda ^2 \left(\eta \sigma \Lambda ^{{k_3}}\right)^{\alpha } \Lambda _{{k_3}}{}^{\beta }
+y_{i,15} ~\eta ^{{k_1} {k_2}} ~\sigma \Lambda \Lambda  \left(\eta \sigma \Lambda ^{{k_3}}\right)^{\alpha } \Lambda _{{k_3}}{}^{\beta }
\nonumber\\&&
+y_{i,16}~ \Lambda ^{{k_1} {k_2}} \left(\eta \sigma \Lambda ^{{k_3}}\right)^{\alpha } \Lambda _{{k_3}}{}^{\beta }
+y_{i,17}~\epsilon ^{{k_2} {k_3} {k_4} {n_1}} ~\Lambda ^{{k_1}}{}_{{n_1}} ~\Lambda _{{k_3}}{}^{\beta } ~\Lambda _{{k_4}}{}^{\alpha },\\
\overline{Y}_i{}^{{k_1} {k_2} \alpha  \beta }&=&y_{i,1} ~\eta ^{{k_1} {k_2}} ~\epsilon ^{\alpha  \beta } ~\Lambda ^2~ \Lambda ^2
+y_{i,2} ~\eta ^{{k_1} {k_2}} ~\epsilon ^{\alpha  \beta } ~\Lambda ^2 ~\sigma \Lambda \Lambda
+y_{i,3} ~\eta ^{{k_1} {k_2}}~ \epsilon ^{\alpha  \beta } ~\sigma \Lambda \Lambda ~ \sigma \Lambda \Lambda
\nonumber\\&&
+y_{i,4} ~\Lambda ^2 ~\Lambda ^2 \left(\sigma ^{{k_1} {k_2}}\right)^{\alpha  \beta }
+y_{i,5} ~\Lambda ^2 ~\sigma \Lambda \Lambda  \left(\sigma ^{{k_1} {k_2}}\right)^{\alpha  \beta }
+y_{i,6} ~\sigma \Lambda \Lambda~  \sigma \Lambda \Lambda  \left(\sigma ^{{k_1} {k_2}}\right)^{\alpha  \beta }
\nonumber\\&&
+y_{i,7}~ \epsilon ^{\alpha  \beta } ~\Lambda ^2 \left(\eta \sigma \Lambda \Lambda ^{{k_2}}\right)^{{k_1}}
+y_{i,8} ~\epsilon ^{\alpha  \beta } ~\sigma \Lambda \Lambda  \left(\eta \sigma \Lambda \Lambda ^{{k_2}}\right)^{{k_1}}
+y_{i,9} ~\epsilon ^{\alpha  \beta } \Lambda ^2 \Lambda ^{{k_1} {k_2}}
\nonumber\\&&
+y_{i,10}~ \epsilon ^{\alpha  \beta } ~\sigma \Lambda \Lambda  ~\Lambda ^{{k_1} {k_2}}
+y_{i,11} ~\Lambda ^2 ~\Lambda ^{{k_1} \beta } ~\Lambda ^{{k_2} \alpha }
+y_{i,12} ~\sigma \Lambda \Lambda  ~\Lambda ^{{k_1} \beta } ~\Lambda ^{{k_2} \alpha }
\nonumber\\&&
+y_{i,13} ~\epsilon ^{{k_1} {k_2} {k_3} {k_4}} ~\Lambda ^2 ~\Lambda _{{k_3}}{}^{\beta } ~\Lambda _{{k_4}}{}^{\alpha }
+y_{i,14}~\Lambda ^2 \left(\sigma ^{{k_2} {k_3}}\right)^{\alpha  \beta } \Lambda ^{{k_1}}{}_{{k_3}}
\nonumber\\&&
+y_{i,15}~ \sigma \Lambda \Lambda  \left(\sigma ^{{k_2} {k_3}}\right)^{\alpha  \beta } \Lambda ^{{k_1}}{}_{{k_3}}
+y_{i,16} ~\Lambda ^2 \left(\sigma \Lambda ^{{k_2} {k_3}}\right)^{{k_1} \alpha } \Lambda _{{k_3}}{}^{\beta }
\nonumber\\&&
+y_{i,17} ~\sigma \Lambda \Lambda  \left(\sigma \Lambda ^{{k_2} {k_3}}\right)^{{k_1} \alpha } \Lambda _{{k_3}}{}^{\beta }
+y_{i,18}~ \Lambda ^2 \left(\eta \sigma \Lambda ^{{k_2}}\right)^{\alpha } \Lambda ^{{k_1} \beta }
\nonumber\\&&
+y_{i,19} ~\sigma \Lambda \Lambda  \left(\eta \sigma \Lambda ^{{k_2}}\right)^{\alpha } \Lambda ^{{k_1} \beta }
+y_{i,20} ~\epsilon ^{\alpha  \beta } \left(\eta \sigma \Lambda \Lambda ^{{k_3}}\right)^{{k_2}} \Lambda ^{{k_1}}{}_{{k_3}}
\nonumber\\&&
+y_{i,21}~ \left(\sigma ^{{k_3} {k_4}}\right)^{\alpha  \beta } \Lambda ^{{k_1}}{}_{{k_3}} ~\Lambda ^{{k_2}}{}_{{k_4}}
+y_{i,22}~ \epsilon ^{{k_2} {k_3} {k_4} {n_1}} ~\Lambda ^{{k_1}}{}_{{n_1}} ~\Lambda _{{k_3}}{}^{\beta } ~\Lambda _{{k_4}}{}^{\alpha }
\nonumber\\&&
+y_{i,23}~ \left(\sigma \Lambda \Lambda ^{{k_3} {k_4}}\right)^{{k_1} {k_2}} \Lambda _{{k_3}}{}^{\beta } \Lambda _{{k_4}}{}^{\alpha }
+y_{i,24}~ \left(\eta \sigma \Lambda \Lambda ^{{k_3}}\right)^{{k_2}} \Lambda ^{{k_1} \alpha }~ \Lambda _{{k_3}}{}^{\beta }
\nonumber\\&&
+y_{i,25} ~\left(\sigma \Lambda ^{{k_3} {k_4}}\right)^{{k_2} \alpha } \Lambda ^{{k_1}}{}_{{k_4}} ~\Lambda _{{k_3}}{}^{\beta }
+y_{i,26} ~\left(\eta \sigma \Lambda ^{{k_3}}\right)^{\alpha } \Lambda ^{{k_2} \beta } ~\Lambda ^{{k_1}}{}_{{k_3}},\\
\widetilde{Y}_i{}^{{k_1} {k_2} \alpha  \beta }&=&y_{i,1} ~\eta ^{{k_1} {k_2}} ~\epsilon ^{\alpha  \beta } ~\Lambda ^2 \Lambda ^2
+y_{i,2} ~\eta ^{{k_1} {k_2}} ~\epsilon ^{\alpha  \beta } ~\Lambda ^2 ~\sigma \Lambda \Lambda
+y_{i,3} ~\eta ^{{k_1} {k_2}} ~\epsilon ^{\alpha  \beta }~ \sigma \Lambda \Lambda~ \sigma \Lambda \Lambda
\nonumber\\&&
+y_{i,4} ~\Lambda ^2 ~\Lambda ^2 \left(\sigma ^{{k_1} {k_2}}\right)^{\alpha  \beta }
+y_{i,5} ~\Lambda ^2 ~\sigma \Lambda \Lambda  \left(\sigma ^{{k_1} {k_2}}\right)^{\alpha  \beta }
\nonumber\\&&
+y_{i,6} ~\sigma \Lambda \Lambda ~ \sigma \Lambda \Lambda  \left(\sigma ^{{k_1} {k_2}}\right)^{\alpha  \beta }
+y_{i,7} ~\epsilon ^{\alpha  \beta } ~\Lambda ^2 \left(\eta \sigma \Lambda \Lambda ^{{k_1}}\right)^{{k_2}}
\nonumber\\&&
+y_{i,8}~ \epsilon ^{\alpha  \beta } ~\Lambda ^2 \left(\eta \sigma \Lambda \Lambda ^{{k_2}}\right)^{{k_1}}
+y_{i,9} ~\epsilon ^{\alpha  \beta }~ \sigma \Lambda \Lambda  \left(\eta \sigma \Lambda \Lambda ^{{k_1}}\right)^{{k_2}}
\nonumber\\&&
+y_{i,10} ~\epsilon ^{\alpha  \beta } ~\sigma \Lambda \Lambda  \left(\eta \sigma \Lambda \Lambda ^{{k_2}}\right)^{{k_1}}
+y_{i,11} ~\epsilon ^{\alpha  \beta }~ \Lambda ^2~ \Lambda ^{{k_1} {k_2}}
+y_{i,12} ~\epsilon ^{\alpha  \beta } ~\sigma \Lambda \Lambda  ~\Lambda ^{{k_1} {k_2}}
\nonumber\\&&
+y_{i,13}~ \Lambda ^2 ~\Lambda ^{{k_1} \alpha } ~\Lambda ^{{k_2} \beta }
+y_{i,14} ~\sigma \Lambda \Lambda ~ \Lambda ^{{k_1} \alpha } ~\Lambda ^{{k_2} \beta }
+y_{i,15} ~\epsilon ^{{k_1} {k_2} {k_3} {k_4}} ~\Lambda ^2~ \Lambda _{{k_3}}{}^{\beta } ~\Lambda _{{k_4}}{}^{\alpha }
\nonumber\\&&
+y_{i,16}~ \Lambda ^2 \left(\sigma ^{{k_1} {k_3}}\right)^{\alpha  \beta } \Lambda ^{{k_2}}{}_{{k_3}}
+y_{i,17}~ \Lambda ^2 \left(\sigma ^{{k_2} {k_3}}\right)^{\alpha  \beta } \Lambda ^{{k_1}}{}_{{k_3}}
\nonumber\\&&
+y_{i,18}~ \sigma \Lambda \Lambda  \left(\sigma ^{{k_1} {k_3}}\right)^{\alpha  \beta } \Lambda ^{{k_2}}{}_{{k_3}}
+y_{i,19} ~\sigma \Lambda \Lambda  \left(\sigma ^{{k_2} {k_3}}\right)^{\alpha  \beta } \Lambda ^{{k_1}}{}_{{k_3}}
\nonumber\\&&
+y_{i,20} ~\Lambda ^2 \left(\sigma \Lambda ^{{k_1} {k_3}}\right)^{{k_2} \alpha } \Lambda _{{k_3}}{}^{\beta }
+y_{i,21} ~\Lambda ^2 \left(\sigma \Lambda ^{{k_2} {k_3}}\right)^{{k_1} \alpha } \Lambda _{{k_3}}{}^{\beta }
\nonumber\\&&
+y_{i,22}~ \sigma \Lambda \Lambda  \left(\sigma \Lambda ^{{k_1} {k_3}}\right)^{{k_2} \alpha } \Lambda _{{k_3}}{}^{\beta }
+y_{i,23} ~\sigma \Lambda \Lambda  \left(\sigma \Lambda ^{{k_2} {k_3}}\right)^{{k_1} \alpha } \Lambda _{{k_3}}{}^{\beta }
\nonumber\\&&
+y_{i,24}~ \Lambda ^2 \left(\eta \sigma \Lambda ^{{k_1}}\right)^{\alpha } \Lambda ^{{k_2} \beta }
+y_{i,25}~ \Lambda ^2 \left(\eta \sigma \Lambda ^{{k_2}}\right)^{\alpha } \Lambda ^{{k_1} \beta }
\nonumber\\&&
+y_{i,26} ~\sigma \Lambda \Lambda  \left(\eta \sigma \Lambda ^{{k_1}}\right)^{\alpha } \Lambda ^{{k_2} \beta }
+y_{i,27}~ \sigma \Lambda \Lambda  \left(\eta \sigma \Lambda ^{{k_2}}\right)^{\alpha } \Lambda ^{{k_1} \beta }
\nonumber\\&&
+y_{i,28}~ \eta ^{{k_1} {k_2}} ~\Lambda ^2 \left(\eta \sigma \Lambda ^{{k_3}}\right)^{\alpha } \Lambda _{{k_3}}{}^{\beta }
+y_{i,29}~ \eta ^{{k_1} {k_2}}~ \sigma \Lambda \Lambda  \left(\eta \sigma \Lambda ^{{k_3}}\right)^{\alpha } \Lambda _{{k_3}}{}^{\beta }
\nonumber\\&&
+y_{i,30} ~\epsilon ^{\alpha  \beta } \left(\eta \sigma \Lambda \Lambda ^{{k_3}}\right)^{{k_1}} \Lambda ^{{k_2}}{}_{{k_3}}
+y_{i,31}~ \epsilon ^{\alpha  \beta } \left(\eta \sigma \Lambda \Lambda ^{{k_3}}\right)^{{k_2}} \Lambda ^{{k_1}}{}_{{k_3}}
\nonumber\\&&
+y_{i,32} ~\left(\sigma ^{{k_3} {k_4}}\right)^{\alpha  \beta } \Lambda ^{{k_1}}{}_{{k_3}} ~\Lambda ^{{k_2}}{}_{{k_4}}
+y_{i,33} ~\Lambda ^{{k_1} {k_2}} \left(\eta \sigma \Lambda ^{{k_3}}\right)^{\alpha } \Lambda _{{k_3}}{}^{\beta }
\nonumber\\&&
+y_{i,34}~ \epsilon ^{{k_2} {k_3} {k_4} {n_1}} ~\Lambda ^{{k_1}}{}_{{n_1}} ~\Lambda _{{k_3}}{}^{\beta } ~\Lambda _{{k_4}}{}^{\alpha }
+y_{i,35}~ \epsilon ^{{k_1} {k_3} {k_4} {n_1}} ~\Lambda ^{{k_2}}{}_{{n_1}} ~\Lambda _{{k_3}}{}^{\beta } ~\Lambda _{{k_4}}{}^{\alpha }
\nonumber\\&&
+y_{i,36}~ \left(\sigma \Lambda \Lambda ^{{k_3} {k_4}}\right)^{{k_1} {k_2}} \Lambda _{{k_3}}{}^{\beta } ~\Lambda _{{k_4}}{}^{\alpha }
+y_{i,37}~ \left(\eta \sigma \Lambda \Lambda ^{{k_3}}\right)^{{k_1}} \Lambda ^{{k_2} \alpha }~ \Lambda _{{k_3}}{}^{\beta }
\nonumber\\&&
+y_{i,38} ~\left(\eta \sigma \Lambda \Lambda ^{{k_3}}\right)^{{k_2}} \Lambda ^{{k_1} \alpha } ~\Lambda _{{k_3}}{}^{\beta }
+y_{i,39} ~\left(\sigma \Lambda ^{{k_3} {k_4}}\right)^{{k_1} \alpha } \Lambda ^{{k_2}}{}_{{k_4}}~ \Lambda _{{k_3}}{}^{\beta }
\nonumber\\&&
+y_{i,40} ~\left(\sigma \Lambda ^{{k_3} {k_4}}\right)^{{k_2} \alpha } \Lambda ^{{k_1}}{}_{{k_4}} ~\Lambda _{{k_3}}{}^{\beta }
+y_{i,41} ~\left(\eta \sigma \Lambda ^{{k_3}}\right)^{\alpha } \Lambda ^{{k_1} \beta } \Lambda ^{{k_2}}{}_{{k_3}}
\nonumber\\&&
+y_{i,42} ~\left(\eta \sigma \Lambda ^{{k_3}}\right)^{\alpha } \Lambda ^{{k_2} \beta } \Lambda ^{{k_1}}{}_{{k_3}}.
\label{Ylast}
\end{eqnarray}
}
where $y_{i, j}$ are free parameters.
They can combine with operators as follows,
{\small
\begin{eqnarray}
\label{}
&&Y_i{}^{{k_1} {k_2}}~\left(D^2 \Phi \right) \Phi  \left(\partial _{{k_1}}\partial _{{k_2}}\Phi ^+\right),\quad 
\overline{Y}_i{}^{{k_1} {k_2}} ~\left(\partial _{{k_1}}D^2 \Phi \right) \Phi  \left(\partial _{{k_2}}\Phi ^+\right) \Phi ^+\nonumber\\
&&Y_i{}^{{k_1} {k_2} {k_3} {k_4}} ~\Phi ~ \left(\partial _{{k_1}}\partial _{{k_2}}\partial _{{k_3}}\partial _{{k_4}}\Phi ^+ \right)\Phi ^+,\quad
\overline{Y}_i{}^{{k_1},{k_2} {k_3} {k_4}} ~\Phi  \left(\partial _{{k_2}}\partial _{{k_3}}\partial _{{k_4}}\Phi ^+ \right) \left(\partial _{{k_1}}\Phi ^+\right)\nonumber\\
&&\widetilde{Y}_i{}^{{k_1} {k_2},{k_3} {k_4}} ~\Phi  \left(\partial _{{k_1}}\partial _{{k_2}}\Phi ^+\right) \left(\partial _{{k_3}}\partial _{{k_4}}\Phi ^+\right),\quad 
\hat{Y}_i{}^{{k_1} {k_2},{k_3} {k_4}} \left(\partial _{{k_3}}\partial _{{k_4}}D^2 \Phi \right) \Phi  \left(\partial _{{k_1}}\partial _{{k_2}}\Phi ^+\right)
\nonumber\\&&
\widetilde{\overline{Y}}_i{}^{{k_1},{k_2} {k_3},{k_4}}\left(\partial _{{k_4}}D^2 \Phi \right) \Phi  \left(\partial _{{k_2}}\partial _{{k_3}}\Phi ^+\right) \left(\partial _{{k_1}}\Phi ^+\right),\quad 
Y_i{}^{{k_1} \dot{\alpha }\beta }\left(D_{\beta } \Phi \right) \left(\partial _{{k_1}}\Phi \right) \left(\overline{D}_{\dot{\alpha }} \Phi ^+\right) \Phi ^+ ~\Phi ^+
\nonumber\\&&
Y_i{}^{{k_1} {k_2} \alpha  \beta } \left(\partial _{{k_1}}\partial _{{k_2}}D_{\alpha } \Phi \right) \left(D_{\beta } \Phi \right) \Phi  ~\Phi ^+,\quad 
\overline{Y}_i{}^{{k_1} {k_2} \alpha  \beta } \left(\partial _{{k_1}}D_{\alpha } \Phi \right) \left(\partial _{{k_2}}D_{\beta } \Phi \right) \Phi  ~\Phi ^+
\nonumber\\&&
\widetilde{Y}_i{}^{{k_1} {k_2} \alpha  \beta } \left(\partial _{{k_1}}D_{\alpha } \Phi \right) \left(D_{\beta } \Phi \right) \left(\partial _{{k_2}}\Phi \right) \Phi ^+
\end{eqnarray}
}
By using Eq.~(\ref{Lambda4-type1}) we can construct the following action,
{\small
\begin{eqnarray}
\label{L4-1}
&&\int d^4x~d^4\theta~\Bigg\{
Y_1~\overline{\theta }^2 \left(D^2 \Phi \right) \left(D^2 \Phi \right) \Phi 
+Y_2~\overline{\theta }^2 \left(D^2 \Phi \right) \Phi  ~\left(\Phi ^+\right)^2
+Y_3~\overline{\theta }^2 \left(D^2 \Phi \right) \left(D^2 \Phi \right) \Phi  ~\Phi ^+
\nonumber\\&&
+Y_4~\overline{\theta }^2 \left(D^2 \Phi \right) \left(D^2 \Phi \right) \left(D^2 \Phi \right) \Phi 
+Y_5~\overline{\theta }^2~ \Phi ~ \left(\Phi ^+\right)^4
+Y_6~\overline{\theta }^2 \left(D^2 \Phi \right) \Phi  ~\left(\Phi ^+\right)^3
\nonumber\\&&
+Y_7~\overline{\theta }^2 \left(D^2 \Phi \right) \left(D^2 \Phi \right) \Phi  ~\left(\Phi ^+\right)^2
+Y_8~\overline{\theta }^2~ \Phi  ~\left(\Phi ^+\right)^5
+Y_9~\overline{\theta }^2 \left(D^2 \Phi \right) \Phi  ~\left(\Phi ^+\right)^4
\nonumber\\&&
+Y_{10}~\overline{\theta }^2~ \Phi  ~\left(\Phi ^+\right)^6
\Bigg\}.
\end{eqnarray}
}
By using Eq.~(\ref{Lambda4-type2}) we obtain,
{\small
\begin{eqnarray}
\label{L4-2}
&&\int d^4x~d^4\theta~\Bigg\{
Y_{11}{}^{{k_1} {k_2}}~\overline{\theta }^2 \left(D^2 \Phi \right) \Phi  \left(\partial _{{k_1}}\partial _{{k_2}}\Phi ^+\right)
+Y_{12}{}^{{k_1} {k_2}}~\overline{\theta }^2 \left(\partial _{{k_1}}\partial _{{k_2}}D^2 \Phi \right) \Phi  ~\Phi ^+
\nonumber\\&&
+\overline{Y}_{13}{}^{{k_1} {k_2}}~\overline{\theta }^2 \left(\partial _{{k_1}}D^2 \Phi \right) \Phi  \left(\partial _{{k_2}}\Phi ^+\right)
+Y_{14}{}^{{k_1} {k_2}}~\overline{\theta }^2 \left(\partial _{{k_1}}D^2 \Phi \right) \left(\partial _{{k_2}}D^2 \Phi \right) \Phi
\nonumber\\&&
+Y_{15}{}^{{k_1} {k_2}}~\overline{\theta }^2 \left(\partial _{{k_1}}\partial _{{k_2}}D^2 \Phi  \right)\left(D^2 \Phi \right) \Phi 
+Y_{16}{}^{{k_1} {k_2}}~\overline{\theta }^2~ \Phi  \left(\partial _{{k_1}}\Phi ^+\right) \left(\partial _{{k_2}}\Phi ^+\right) \Phi ^+
\nonumber\\&&
+Y_{17}{}^{{k_1} {k_2}}~\overline{\theta }^2 ~\Phi  \left(\partial _{{k_1}}\partial _{{k_2}}\Phi ^+\right) \left(\Phi ^+\right)^2
+Y_{18}{}^{{k_1} {k_2}}~\overline{\theta }^2 \left(D^2 \Phi \right) \Phi  \left(\partial _{{k_1}}\Phi ^+\right) \left(\partial _{{k_2}}\Phi ^+\right)
\nonumber\\&&
+Y_{19}{}^{{k_1} {k_2}}~\overline{\theta }^2 \left(D^2 \Phi \right) \Phi  \left(\partial _{{k_1}}\partial _{{k_2}}\Phi ^+\right) \Phi ^+
+Y_{20}{}^{{k_1} {k_2}}~\overline{\theta }^2 \left(\partial _{{k_1}}\partial _{{k_2}}D^2\right) \Phi^2 \left(\Phi ^+\right)^2
\nonumber\\&&
+\overline{Y}_{21}{}^{{k_1} {k_2}}~\overline{\theta }^2 \left(\partial _{{k_1}}D^2 \Phi \right) \Phi  \left(\partial _{{k_2}}\Phi ^+\right) \Phi ^+
+Y_{22}{}^{{k_1} {k_2}}~\overline{\theta }^2 \left(D^2 \Phi \right) \left(D^2 \Phi \right) \Phi  \left(\partial _{{k_1}}\partial _{{k_2}}\Phi ^+\right)
\nonumber\\&&
+Y_{23}{}^{{k_1} {k_2}}~\overline{\theta }^2 \left(\partial _{{k_1}}D^2 \Phi \right) \left(\partial _{{k_2}}D^2 \Phi \right) \Phi ~ \Phi ^+
+Y_{24}{}^{{k_1} {k_2}}~\overline{\theta }^2 \left(\partial _{{k_1}}\partial _{{k_2}}D^2 \Phi \right) \left(D^2 \Phi \right) \Phi  ~\Phi ^+
\nonumber\\&&
+\overline{Y}_{25}{}^{{k_1} {k_2}}~\overline{\theta }^2 \left(\partial _{{k_1}}D^2 \Phi \right) \left(D^2 \Phi \right) \Phi  \left(\partial _{{k_2}}\Phi ^+\right)
+Y_{26}{}^{{k_1} {k_2}}~\overline{\theta }^2~ \Phi  \left(\partial _{{k_1}}\Phi ^+\right) \left(\partial _{{k_2}}\Phi ^+\right) \left(\Phi ^+\right)^2
\nonumber\\&&
+Y_{27}{}^{{k_1} {k_2}}~\overline{\theta }^2~ \Phi  \left(\partial _{{k_1}}\partial _{{k_2}}\Phi ^+\right) \left(\Phi ^+\right)^3
+Y_{28}{}^{{k_1} {k_2}}~\theta ^4 \left(\partial _{{k_1}}\partial _{{k_2}}\Phi ^+\right) \left(\Phi ^+\right)^4
\nonumber\\&&
+Y_{29}{}^{{k_1} {k_2}}~\overline{\theta }^2 \left(D^2 \Phi \right) \Phi  \left(\partial _{{k_1}}\Phi ^+\right) \left(\partial _{{k_2}}\Phi ^+\right) \Phi ^+
+Y_{30}{}^{{k_1} {k_2}}~\overline{\theta }^2 \left(D^2 \Phi \right) \Phi  \left(\partial _{{k_1}}\partial _{{k_2}}\Phi ^+\right) \left(\Phi ^+\right)^2
\nonumber\\&&
+Y_{31}{}^{{k_1} {k_2}}~\overline{\theta }^2 \left(\partial _{{k_1}}\partial _{{k_2}}D^2 \Phi \right) \Phi  ~\left(\Phi ^+\right)^3
+\overline{Y}_{32}{}^{{k_1} {k_2}}~\overline{\theta }^2 \left(\partial _{{k_2}}D^2 \Phi \right) \Phi  \left(\partial _{{k_1}}\Phi ^+\right) \left(\Phi ^+\right)^2
\nonumber\\&&
+Y_{33}{}^{{k_1} {k_2}}~\overline{\theta }^2 ~\Phi \left(\partial _{{k_1}}\Phi ^+\right) \left(\partial _{{k_2}}\Phi ^+\right) \left(\Phi ^+\right)^3
+Y_{34}{}^{{k_1} {k_2}}~\overline{\theta }^2 ~\Phi  \left(\partial _{{k_1}}\partial _{{k_2}}\Phi ^+\right) \left(\Phi ^+\right)^4
\nonumber\\&&
+Y_{35}{}^{{k_1} {k_2}}~\theta ^4 \left(\partial _{{k_1}}\partial _{{k_2}}\Phi ^+\right) \left(\Phi ^+\right)^5
+Y_{36}{}^{{k_1} {k_2}}~\theta ^4 \left(\partial _{{k_1}}\partial _{{k_2}}\Phi ^+\right)\left(\Phi ^+\right)^6
\Bigg\}.
\end{eqnarray}
}
By using Eq.~(\ref{Lambda4-type3}) we obtain,
{\small
\begin{eqnarray}
\label{L4-3}
&&\int d^4x~d^4\theta~\Bigg\{
Y_{37}{}^{{k_1} {k_2} {k_3} {k_4}}~\overline{\theta }^2 \left(\partial _{{k_1}}\partial _{{k_2}}\partial _{{k_3}}\partial _{{k_4}}D^2 \Phi \right) \Phi 
+Y_{38}{}^{{k_1} {k_2} {k_3} {k_4}}~\overline{\theta }^2~ \Phi  \left(\partial _{{k_1}}\partial _{{k_2}}\partial _{{k_3}}\partial _{{k_4}}\Phi ^+\right) \Phi ^+
\nonumber\\&&
+\overline{Y}_{39}{}^{{k_1},{k_2} {k_3} {k_4}}~\overline{\theta }^2 ~\Phi  \left(\partial _{{k_2}}\partial _{{k_3}}\partial _{{k_4}}\Phi ^+\right) \left(\partial _{{k_1}}\Phi ^+\right)
+\widetilde{Y}_{40}{}^{{k_1} {k_2},{k_3} {k_4}}~\overline{\theta }^2 ~\Phi  \left(\partial _{{k_1}}\partial _{{k_2}}\Phi ^+\right) \left(\partial _{{k_3}}\partial _{{k_4}}\Phi ^+\right)
\nonumber\\&&
+Y_{41}{}^{{k_1} {k_2} {k_3} {k_4}}~\overline{\theta }^2\left(\partial _{{k_1}}\partial _{{k_2}}\partial _{{k_3}}\partial _{{k_4}}D^2 \Phi \right) \Phi ~ \Phi ^+
+\overline{Y}_{42}{}^{{k_1},{k_2} {k_3} {k_4}}~\overline{\theta }^2 \left(\partial _{{k_2}}\partial _{{k_3}}\partial _{{k_4}}D^2 \Phi \right) \Phi  \left(\partial _{{k_1}}\Phi ^+\right)
\nonumber\\&&
+\hat{Y}_{43}{}^{{k_1} {k_2},{k_3} {k_4}}~\overline{\theta }^2 \left(\partial _{{k_3}}\partial _{{k_4}}D^2 \Phi \right) \Phi  \left(\partial _{{k_1}}\partial _{{k_2}}\Phi ^+\right)
+\overline{Y}_{44}{}^{{k_1},{k_2} {k_3} {k_4}}~\overline{\theta }^2 \left(\partial _{{k_1}}D^2 \Phi \right) \Phi  \left(\partial _{{k_2}}\partial _{{k_3}}\partial _{{k_4}}\Phi ^+\right)
\nonumber\\&&
+Y_{45}{}^{{k_1} {k_2} {k_3} {k_4}}~\overline{\theta }^2 \left(D^2 \Phi \right) \Phi  \left(\partial _{{k_1}}\partial _{{k_2}}\partial _{{k_3}}\partial _{{k_4}}\Phi ^+\right)
\nonumber\\&&
+Y_{46}{}^{{k_1} {k_2} {k_3} {k_4}}~\overline{\theta }^2 \left(\partial _{{k_1}}\partial _{{k_2}}\partial _{{k_3}}\partial _{{k_4}}D^2 \Phi \right) \left(D^2 \Phi \right) \Phi
\nonumber\\&&
+\overline{Y}_{47}{}^{{k_1},{k_2} {k_3} {k_4}}~\overline{\theta }^2 \left(\partial _{{k_2}}\partial _{{k_3}}\partial _{{k_4}}D^2 \Phi \right) \left(\partial _{{k_1}}D^2 \Phi \right) \Phi 
\nonumber\\&&
+\widetilde{Y}_{48}{}^{{k_1} {k_2},{k_3} {k_4}}~\overline{\theta }^2 \left(\partial _{{k_1}}\partial _{{k_2}}D^2 \Phi \right) \left(\partial _{{k_3}}\partial _{{k_4}}D^2 \Phi \right) \Phi 
\nonumber\\&&
+Y_{49}{}^{{k_1} {k_2} {k_3} {k_4}}~\theta ^4 \left(\partial _{{k_1}}\partial _{{k_2}}\partial _{{k_3}}\partial _{{k_4}}\Phi ^+\right)\left(\Phi ^+\right)^3
\nonumber\\&&
+\widetilde{Y}_{50}{}^{{k_1} {k_2},{k_3} {k_4}}~\theta ^4 \left(\partial _{{k_1}}\partial _{{k_2}}\Phi ^+\right) \left(\partial _{{k_3}}\partial _{{k_4}}\Phi ^+\right) \left(\Phi ^+\right)^2
\nonumber\\&&
+Y_{51}{}^{{k_1} {k_2} {k_3} {k_4}}~\overline{\theta }^2 ~\Phi  \left(\partial _{{k_1}}\partial _{{k_2}}\partial _{{k_3}}\partial _{{k_4}}\Phi ^+\right) \left(\Phi ^+\right)^2
\nonumber\\&&
+\overline{Y}_{52}{}^{{k_1},{k_2} {k_3} {k_4}}~\overline{\theta }^2 ~\Phi  \left(\partial _{{k_2}}\partial _{{k_3}}\partial _{{k_4}}\Phi ^+\right) \left(\partial _{{k_1}}\Phi ^+\right) \Phi ^+
\nonumber\\&&
+\hat{Y}_{53}{}^{{k_1} {k_2},{k_3} {k_4}}~\overline{\theta }^2 ~\Phi  \left(\partial _{{k_3}}\partial _{{k_4}}\Phi ^+\right) \left(\partial _{{k_1}}\Phi ^+\right) \left(\partial _{{k_2}}\Phi ^+\right)
\nonumber\\&&
+\widetilde{Y}_{54}{}^{{k_1} {k_2},{k_3} {k_4}}~\overline{\theta }^2~ \Phi  \left(\partial _{{k_1}}\partial _{{k_2}}\Phi ^+\right) \left(\partial _{{k_3}}\partial _{{k_4}}\Phi ^+\right) \Phi ^+
\nonumber\\&&
+Y_{55}{}^{{k_1} {k_2} {k_3} {k_4}}~\overline{\theta }^2 \left(\partial _{{k_1}}\partial _{{k_2}}\partial _{{k_3}}\partial _{{k_4}}D^2 \Phi \right) \Phi  ~\left(\Phi ^+\right)^2
\nonumber\\&&
+\overline{Y}_{56}{}^{{k_1},{k_2} {k_3} {k_4}}~\overline{\theta }^2\left(\partial _{{k_2}}\partial _{{k_3}}\partial _{{k_4}}D^2 \Phi \right) \Phi  \left(\partial _{{k_1}}\Phi ^+\right) \Phi ^+
\nonumber\\&&
+\hat{Y}_{57}{}^{{k_1} {k_2},{k_3} {k_4}}~\overline{\theta }^2 \left(\partial _{{k_3}}\partial _{{k_4}}D^2 \Phi \right) \Phi  \left(\partial _{{k_1}}\partial _{{k_2}}\Phi ^+\right) \Phi ^+
\nonumber\\&&
+\hat{Y}_{58}{}^{{k_1} {k_2},{k_3} {k_4}}~\overline{\theta }^2 \left(\partial _{{k_3}}\partial _{{k_4}}D^2 \Phi \right) \Phi  \left(\partial _{{k_1}}\Phi ^+\right) \left(\partial _{{k_2}}\Phi ^+\right)
\nonumber\\&&
+\overline{Y}_{59}{}^{{k_1},{k_2} {k_3} {k_4}}~\overline{\theta }^2 \left(\partial _{{k_1}}D^2 \Phi \right) \Phi  \left(\partial _{{k_2}}\partial _{{k_3}}\partial _{{k_4}}\Phi ^+\right) \Phi ^+
\nonumber\\&&
+\widetilde{\overline{Y}}_{60}{}^{{k_1},{k_2} {k_3},{k_4}}~\overline{\theta }^2 \left(\partial _{{k_4}}D^2 \Phi \right) \Phi  \left(\partial _{{k_2}}\partial _{{k_3}}\Phi ^+\right) \left(\partial _{{k_1}}\Phi ^+\right)\nonumber\\&&
+Y_{61}{}^{{k_1} {k_2} {k_3} {k_4}}~\overline{\theta }^2 \left(D^2 \Phi \right) \Phi  \left(\partial _{{k_1}}\partial _{{k_2}}\partial _{{k_3}}\partial _{{k_4}}\Phi ^+\right) \Phi ^+
\nonumber\\&&
+\overline{Y}_{62}{}^{{k_1},{k_2} {k_3} {k_4}}~\overline{\theta }^2\left(D^2 \Phi \right) \Phi  \left(\partial _{{k_2}}\partial _{{k_3}}\partial _{{k_4}}\Phi ^+\right) \left(\partial _{{k_1}}\Phi ^+\right)
\nonumber\\&&
+\widetilde{Y}_{63}{}^{{k_1} {k_2},{k_3} {k_4}}~\overline{\theta }^2 \left(D^2 \Phi \right) \Phi  \left(\partial _{{k_1}}\partial _{{k_2}}\Phi ^+\right) \left(\partial _{{k_3}}\partial _{{k_4}}\Phi ^+\right)\nonumber\\&&
+Y_{64}{}^{{k_1} {k_2} {k_3} {k_4}}~\theta ^4 \left(\partial _{{k_1}}\partial _{{k_2}}\partial _{{k_3}}\partial _{{k_4}}\Phi ^+\right) \left(\Phi ^+\right)^4
\nonumber\\&&
+\widetilde{Y}_{65}{}^{{k_1} {k_2},{k_3} {k_4}}~\theta ^4 \left(\partial _{{k_1}}\partial _{{k_2}}\Phi ^+\right) \left(\partial _{{k_3}}\partial _{{k_4}}\Phi ^+\right) \left(\Phi ^+\right)^3
\nonumber\\&&
+Y_{66}{}^{{k_1} {k_2} {k_3} {k_4}}~\overline{\theta }^2 ~\Phi  \left(\partial _{{k_1}}\partial _{{k_2}}\partial _{{k_3}}\partial _{{k_4}}\Phi ^+\right) \left(\Phi ^+\right)^3
\nonumber\\&&
+\overline{Y}_{67}{}^{{k_1},{k_2} {k_3} {k_4}}~\overline{\theta }^2~ \Phi  \left(\partial _{{k_2}}\partial _{{k_3}}\partial _{{k_4}}\Phi ^+\right) \left(\partial _{{k_1}}\Phi ^+\right) \left(\Phi ^+\right)^2
\nonumber\\&&
+\widetilde{Y}_{68}{}^{{k_1} {k_2},{k_3} {k_4}}~\overline{\theta }^2 ~\Phi  \left(\partial _{{k_1}}\partial _{{k_2}}\Phi ^+\right) \left(\partial _{{k_3}}\partial _{{k_4}}\Phi ^+\right)\left(\Phi ^+\right)^2
\nonumber\\&&
+\hat{Y}_{69}{}^{{k_1} {k_2},{k_3} {k_4}}~\overline{\theta }^2 ~\Phi  \left(\partial _{{k_3}}\partial _{{k_4}}\Phi ^+\right)\left( \partial _{{k_1}}\Phi ^+\right) \left(\partial _{{k_2}}\Phi ^+\right) \Phi ^+
\nonumber\\&&
+Y_{70}{}^{{k_1} {k_2} {k_3} {k_4}}~\overline{\theta }^2 ~\Phi  \left(\partial _{{k_1}}\Phi ^+\right)\left(\partial _{{k_2}}\Phi ^+\right)\left(\partial _{{k_3}}\Phi ^+\right)\left( \partial _{{k_4}}\Phi ^+\right)\nonumber\\&&+Y_{71}{}^{{k_1} {k_2} {k_3} {k_4}}~\theta ^4 \left(\partial _{{k_1}}\partial _{{k_2}}\partial _{{k_3}}\partial _{{k_4}}\Phi ^+\right) \left(\Phi ^+\right)^5
\nonumber\\&&
+\widetilde{Y}_{72}{}^{{k_1} {k_2},{k_3} {k_4}}~\theta ^4 \left(\partial _{{k_1}}\partial _{{k_2}}\Phi ^+\right) \left(\partial _{{k_3}}\partial _{{k_4}}\Phi ^+\right) \left(\Phi ^+\right)^4
\Bigg\}.
\end{eqnarray}
}
By using Eq.~(\ref{Lambda4-type4}) we obtain,
{\small
\begin{eqnarray}
\label{L4-4}
&&\int d^4x~d^4\theta~\Bigg\{
Y_{73}~\overline{\theta }^2 \left(D^2 \Phi \right) \left(D^2 \Phi \right) \Phi^2
+Y_{74}~\overline{\theta }^2 \left(D^2 \Phi \right) \Phi^2 \left(\Phi ^+\right)^2
+Y_{75}~\overline{\theta }^2 \left(D^2 \Phi \right) \left(D^2 \Phi \right) \Phi^3 
\nonumber\\&&
+Y_{76}~\overline{\theta }^2 \left(D^2 \Phi \right) \left(D^2 \Phi \right) \Phi^2~\Phi ^+
+Y_{77}~\overline{\theta }^2 \left(D^2 \Phi \right) \left(D^2 \Phi \right) \left(D^2 \Phi \right) \Phi^2 
+Y_{78}~\overline{\theta }^2 ~\Phi^2 \left(\Phi ^+\right)^4
\nonumber\\&&
+Y_{79}~\overline{\theta }^2 \left(D^2 \Phi \right) \Phi^3 \left(\Phi ^+\right)^2
+Y_{80}~\overline{\theta }^2 \left(D^2 \Phi \right) \Phi^2 \left(\Phi ^+\right)^3
+Y_{81}~\overline{\theta }^2 \left(D^2 \Phi \right) \left(D^2 \Phi \right) \Phi^3~ \Phi ^+
\nonumber\\&&
+Y_{82}~\overline{\theta }^2 \left(D^2 \Phi \right) \left(D^2 \Phi \right) \Phi^2 \left(\Phi ^+\right)^2
+Y_{83}~\overline{\theta }^2 \left(D^2 \Phi \right) \left(D^2 \Phi \right) \left(D^2 \Phi \right) \Phi^3
+Y_{84}~\overline{\theta }^2~ \Phi^3 \left(\Phi ^+\right)^4
\nonumber\\&&
+Y_{85}~\overline{\theta }^2~ \Phi^2 \left(\Phi ^+\right)^5
+Y_{86}~\overline{\theta }^2 \left(D^2 \Phi \right) \Phi^3 \left(\Phi ^+\right)^3
+Y_{87}~\overline{\theta }^2 \left(D^2 \Phi \right) \Phi^2 \left(\Phi ^+\right)^4
\nonumber\\&&
+Y_{88}~\overline{\theta }^2 \left(D^2 \Phi \right) \left(D^2 \Phi \right) \Phi^3 \left(\Phi ^+\right)^2
+Y_{89}~\overline{\theta }^2~ \Phi^3 \left(\Phi ^+\right)^5
+Y_{90}~\overline{\theta }^2~ \Phi^2 \left(\Phi ^+\right)^6
\nonumber\\&&
+Y_{91}~\overline{\theta }^2 \left(D^2 \Phi \right) \Phi^3 \left(\Phi ^+\right)^4
+Y_{92}~\overline{\theta }^2 ~\Phi^3 \left(\Phi ^+\right)^6
\Bigg\}.
\end{eqnarray}
}
By using Eq.~(\ref{Lambda4-type5}) we obtain,
{\small
\begin{eqnarray}
\label{L4-5}
&&\int d^4x~d^4\theta~\Bigg\{
Y_{93}{}^{{k_1} {k_2}}~\overline{\theta }^2~ \left(\partial _{{k_2}}\partial _{{k_1}}D^2 \Phi \right) \left(D^2 \Phi \right) \Phi^2
+Y_{94}{}^{{k_1} {k_2}}~\overline{\theta }^2~ \left(\partial _{{k_1}}D^2 \Phi \right) \left(\partial _{{k_2}}D^2 \Phi \right) \Phi^2
\nonumber\\&&
+Y_{95}{}^{{k_1} {k_2}}~\overline{\theta }^2 \left(D^2 \Phi \right) \left(D^2 \Phi \right) \left(\partial _{{k_1}}\partial _{{k_2}}\Phi \right) \Phi
+Y_{96}{}^{{k_1} {k_2} \alpha  \beta }~\overline{\theta }^2~ \left(\partial _{{k_2}}\partial _{{k_1}}D_{\alpha } \Phi \right) \left(D_{\beta } \Phi \right) \left(D^2 \Phi \right) \Phi
\nonumber\\&&
+\overline{Y}_{97}{}^{{k_1} {k_2} \alpha  \beta }~\overline{\theta }^2~ \left(\partial _{{k_1}}D_{\alpha } \Phi \right) \left(\partial _{{k_2}}D_{\beta } \Phi \right) \left(D^2 \Phi \right) \Phi
+\widetilde{Y}_{98}{}^{{k_1} {k_2} \alpha  \beta }~\overline{\theta }^2~ \left(\partial _{{k_2}}D_{\beta } \Phi \right) \left(D_{\alpha } \Phi \right) \left(\partial _{{k_1}}D^2 \Phi \right) \Phi
\nonumber\\&&
+Y_{99}{}^{{k_1} {k_2}}~\epsilon ^{\alpha  \beta }~\overline{\theta }^2 \left(D_{\alpha } \Phi \right) \left(D_{\beta } \Phi \right) \left(\partial _{{k_2}}\partial _{{k_1}}D^2 \Phi \right) \Phi
+Y_{100}~\overline{\theta }^2 \left(D^2 \Phi \right) \left(D^2 \Phi \right) \Phi  \left(\overline{D}^2 \Phi ^+\right)
\nonumber\\&&
+Y_{101}~\epsilon ^{\alpha  \beta }~\overline{\theta }^2 \left(D_{\alpha } \Phi \right) \left(D_{\beta } \Phi \right) \left(D^2 \Phi \right) \left(\overline{D}^2 \Phi ^+\right)
+Y_{102}{}^{{k_1} \dot{\alpha }\beta }~\overline{\theta }^2~ \left(\partial _{{k_1}}D_{\beta } \Phi \right) \left(D^2 \Phi \right) \Phi  \left(\overline{D}_{\dot{\alpha }} \Phi ^+\right)
\nonumber\\&&
+Y_{103}{}^{{k_1} \dot{\alpha }\beta }~\overline{\theta }^2 \left(D_{\beta } \Phi \right) \left(\partial _{{k_1}}D^2 \Phi \right) \Phi  \left(\overline{D}_{\dot{\alpha }} \Phi ^+\right)
+Y_{104}{}^{{k_1} \dot{\alpha }\beta }~\overline{\theta }^2 \left(D_{\beta } \Phi \right) \left(D^2 \Phi \right) \left(\partial _{{k_1}}\Phi\right)  \left(\overline{D}_{\dot{\alpha }} \Phi ^+\right)
\nonumber\\&&
+Y_{105}{}^{{k_1} {k_2}}~\overline{\theta }^2~ \left(\partial _{{k_2}}\partial _{{k_1}}D^2 \Phi \right) \Phi^2  ~\Phi ^+
+\overline{Y}_{106}{}^{{k_1} {k_2}}~\overline{\theta }^2~ \left(\partial _{{k_1}}D^2 \Phi \right) \left(\partial _{{k_2}}\Phi\right)  \Phi ~ \Phi ^+
\nonumber\\&&
+Y_{107}{}^{{k_1} {k_2}}~\overline{\theta }^2 \left(D^2 \Phi \right) \left(\partial _{{k_1}}\partial _{{k_2}}\Phi\right)  \Phi  ~\Phi ^+
+Y_{108}{}^{{k_1} {k_2}}~\overline{\theta }^2 \left(D^2 \Phi \right) \left(\partial _{{k_1}}\Phi\right)  \left(\partial _{{k_2}}\Phi\right)  \Phi ^+
\nonumber\\&&
+Y_{109}{}^{{k_1} {k_2} \alpha  \beta }~\overline{\theta }^2~ \left(\partial _{{k_2}}\partial _{{k_1}}D_{\alpha } \Phi \right) \left(D_{\beta } \Phi \right) \Phi ~ \Phi ^+
+\overline{Y}_{110}{}^{{k_1} {k_2} \alpha  \beta }~\overline{\theta }^2~ \left(\partial _{{k_1}}D_{\alpha } \Phi \right) \left(\partial _{{k_2}}D_{\beta } \Phi \right) \Phi  ~\Phi ^+
\nonumber\\&&
+\widetilde{Y}_{111}{}^{{k_1} {k_2} \alpha  \beta }~\overline{\theta }^2~ \left(\partial _{{k_2}}D_{\beta } \Phi \right) \left(D_{\alpha } \Phi \right) \left(\partial _{{k_1}}\Phi\right)  \Phi ^+
+Y_{112}{}^{{k_1} {k_2}}~\epsilon ^{\alpha  \beta }~\overline{\theta }^2 \left(D_{\alpha } \Phi \right) \left(D_{\beta } \Phi \right) \left(\partial _{{k_1}}\partial _{{k_2}}\Phi\right)  \Phi ^+
\nonumber\\&&
+Y_{113}~\overline{\theta }^2 \left(D^2 \Phi \right) \Phi  \left(\overline{D}^2 \Phi ^+\right) \left(\Phi ^+\right)^2
+Y_{114}~\epsilon ^{\alpha  \beta }~\overline{\theta }^2 \left(D_{\alpha } \Phi \right) \left(D_{\beta } \Phi \right) \left(\overline{D}^2 \Phi ^+\right) \left(\Phi ^+\right)^2
\nonumber\\&&
+Y_{115}~\epsilon ^{\dot{\alpha } \dot{\beta }}~\overline{\theta }^2 \left(D^2 \Phi \right) \Phi  \left(\overline{D}_{\dot{\alpha }} \Phi ^+\right) \left(\overline{D}_{\dot{\beta }} \Phi ^+\right) \Phi ^+
+Y_{116}{}^{{k_1} \dot{\alpha }\beta }~\overline{\theta }^2~ \left(\partial _{{k_1}}D_{\beta } \Phi \right) \Phi  \left(\overline{D}_{\dot{\alpha }} \Phi ^+\right) \left(\Phi ^+\right)^2
\nonumber\\&&
+Y_{117}{}^{{k_1} \dot{\alpha }\beta }~\overline{\theta }^2 \left(D_{\beta } \Phi \right) \left(\partial _{{k_1}}\Phi\right)  \left(\overline{D}_{\dot{\alpha }} \Phi ^+\right) \left(\Phi ^+\right)^2
+Y_{118}{}^{{k_1} \dot{\alpha }\beta }~\overline{\theta }^2 \left(D_{\beta } \Phi \right) \Phi  \left(\partial _{{k_1}}\overline{D}_{\dot{\alpha }} \Phi ^+\right) \left(\Phi ^+\right)^2
\nonumber\\&&
+Y_{119}{}^{{k_1} {k_2}}~\overline{\theta }^2~ \left(\partial _{{k_1}}\partial _{{k_2}}\Phi\right)  \Phi  \left(\Phi ^+\right)^3
+Y_{120}{}^{{k_1} {k_2}}~\overline{\theta }^2~ \left(\partial _{{k_1}}\Phi\right)  \left(\partial _{{k_2}}\Phi  \right) \left(\Phi ^+\right)^3
\nonumber\\&&
+Y_{121}{}^{{k_1} {k_2}}~\overline{\theta }^2 ~\Phi^2 \left(\partial _{{k_1}}\partial _{{k_2}}\Phi ^+\right) \left(\Phi ^+\right)^2
+Y_{122}~\overline{\theta }^2 \left(D^2 \Phi \right) \left(D^2 \Phi \right) \Phi  \left(\overline{D}^2 \Phi ^+\right) \Phi ^+
\nonumber\\&&
+Y_{123}~\epsilon ^{\alpha  \beta }~\overline{\theta }^2 \left(D_{\alpha } \Phi \right) \left(D_{\beta } \Phi \right) \left(D^2 \Phi \right) \left(\overline{D}^2 \Phi ^+\right) \Phi ^+
\nonumber\\&&
+Y_{124}~\epsilon ^{\dot{\alpha } \dot{\beta }}~\overline{\theta }^2 \left(D^2 \Phi \right) \left(D^2 \Phi \right) \Phi  \left(\overline{D}_{\dot{\alpha }} \Phi ^+\right) \left(\overline{D}_{\dot{\beta }} \Phi ^+\right)
\nonumber\\&&
+Y_{125}{}^{{k_1} \dot{\alpha }\beta }~\overline{\theta }^2~ \left(\partial _{{k_1}}D_{\beta } \Phi \right) \left(D^2 \Phi \right) \Phi  \left(\overline{D}_{\dot{\alpha }} \Phi ^+\right) \Phi ^+
+Y_{126}{}^{{k_1} \dot{\alpha }\beta }~\overline{\theta }^2 \left(D_{\beta } \Phi \right) \left(\partial _{{k_1}}D^2 \Phi \right) \Phi  \left(\overline{D}_{\dot{\alpha }} \Phi ^+\right) \Phi ^+
\nonumber\\&&
+Y_{127}{}^{{k_1} \dot{\alpha }\beta }~\overline{\theta }^2 \left(D_{\beta } \Phi \right) \left(D^2 \Phi \right) \left(\partial _{{k_1}}\Phi\right)  \left(\overline{D}_{\dot{\alpha }} \Phi ^+\right) \Phi ^+
+Y_{128}{}^{{k_1} \dot{\alpha }\beta }~\overline{\theta }^2 \left(D_{\beta } \Phi \right) \left(D^2 \Phi \right) \Phi  \left(\partial _{{k_1}}\overline{D}_{\dot{\alpha }} \Phi ^+\right) \Phi ^+
\nonumber\\&&
+Y_{129}{}^{{k_1} {k_2}}~\overline{\theta }^2~\left(\partial _{{k_2}}\partial _{{k_1}}D^2 \Phi \right) \Phi^2\left(\Phi ^+\right)^2
+\overline{Y}_{130}{}^{{k_1} {k_2}}~\overline{\theta }^2~ \left(\partial _{{k_1}}D^2 \Phi \right) \left(\partial _{{k_2}}\Phi\right)  \Phi  \left(\Phi ^+\right)^2
\nonumber\\&&
+Y_{131}{}^{{k_1} {k_2}}~\overline{\theta }^2 \left(D^2 \Phi \right) \left(\partial _{{k_1}}\partial _{{k_2}}\Phi\right)  \Phi  \left(\Phi ^+\right)^2
+Y_{132}{}^{{k_1} {k_2}}~\overline{\theta }^2 \left(D^2 \Phi \right) \left(\partial _{{k_1}}\Phi\right)  \left(\partial _{{k_2}}\Phi\right) \left(\Phi ^+\right)^2
\nonumber\\&&
+Y_{133}{}^{{k_1} {k_2}}~\overline{\theta }^2 \left(D^2 \Phi \right) \Phi^2 \left(\partial _{{k_1}}\partial _{{k_2}}\Phi ^+\right) \Phi ^+
+Y_{134}{}^{{k_1} {k_2} \alpha  \beta }~\overline{\theta }^2~ \left(\partial _{{k_2}}\partial _{{k_1}}D_{\alpha } \Phi \right) \left(D_{\beta } \Phi \right) \Phi  \left(\Phi ^+\right)^2
\nonumber\\&&
+\overline{Y}_{135}{}^{{k_1} {k_2} \alpha  \beta }~\overline{\theta }^2~ \left(\partial _{{k_1}}D_{\alpha } \Phi \right) \left(\partial _{{k_2}}D_{\beta } \Phi \right) \Phi \left(\Phi ^+\right)^2
+\widetilde{Y}_{136}{}^{{k_1} {k_2} \alpha  \beta }~\overline{\theta }^2~ \left(\partial _{{k_2}}D_{\beta } \Phi \right) \left(D_{\alpha } \Phi \right) \left(\partial _{{k_1}}\Phi\right)  \left(\Phi ^+\right)^2
\nonumber\\&&
+Y_{137}{}^{{k_1} {k_2}}~\epsilon ^{\alpha  \beta }~\overline{\theta }^2 \left(D_{\alpha } \Phi \right) \left(D_{\beta } \Phi \right) \left(\partial _{{k_1}}\partial _{{k_2}}\Phi\right)  \left(\Phi ^+\right)^2
\nonumber\\&&
+Y_{138}{}^{{k_1} {k_2}}~\epsilon ^{\alpha  \beta }~\overline{\theta }^2 \left(D_{\alpha } \Phi \right) \left(D_{\beta } \Phi \right) \Phi  \left(\partial _{{k_1}}\partial _{{k_2}}\Phi ^+\right) \Phi ^+
+Y_{139}~\overline{\theta }^2 \left(D^2 \Phi \right) \left(D^2 \Phi \right) \left(D^2 \Phi \right) \Phi  \left(\overline{D}^2 \Phi ^+\right)
\nonumber\\&&
+Y_{140}~\epsilon ^{\alpha  \beta }~\overline{\theta }^2 \left(D_{\alpha } \Phi \right) \left(D_{\beta } \Phi \right) \left(D^2 \Phi \right) \left(D^2 \Phi \right) \left(\overline{D}^2 \Phi ^+\right)
\nonumber\\&&
+Y_{141}{}^{{k_1} \dot{\alpha }\beta }~\overline{\theta }^2~ \left(\partial _{{k_1}}D_{\beta } \Phi \right) \left(D^2 \Phi \right) \left(D^2 \Phi \right) \Phi  \left(\overline{D}_{\dot{\alpha }} \Phi ^+\right)
\nonumber\\&&
+Y_{142}{}^{{k_1} \dot{\alpha }\beta }~\overline{\theta }^2 \left(D_{\beta } \Phi \right) \left(\partial _{{k_1}}D^2 \Phi \right) \left(D^2 \Phi \right) \Phi  \left(\overline{D}_{\dot{\alpha }} \Phi ^+\right)
\nonumber\\&&
+Y_{143}{}^{{k_1} \dot{\alpha }\beta }~\overline{\theta }^2 \left(D_{\beta } \Phi \right) \left(D^2 \Phi \right) \left(D^2 \Phi \right) \left(\partial _{{k_1}}\Phi\right)  \left(\overline{D}_{\dot{\alpha }} \Phi ^+\right)
+Y_{144}{}^{{k_1} {k_2}}~\overline{\theta }^2~ \left(\partial _{{k_2}}\partial _{{k_1}}D^2 \Phi \right) \left(D^2 \Phi \right) \Phi^2 ~\Phi ^+
\nonumber\\&&
+Y_{145}{}^{{k_1} {k_2}}~\overline{\theta }^2~ \left(\partial _{{k_1}}D^2 \Phi \right) \left(\partial _{{k_2}}D^2 \Phi \right) \Phi^2 ~\Phi ^+
+\overline{Y}_{146}{}^{{k_1} {k_2}}~\overline{\theta }^2~ \left(\partial _{{k_1}}D^2 \Phi \right) \left(D^2 \Phi \right) \left(\partial _{{k_2}}\Phi\right)  \Phi  ~\Phi ^+
\nonumber\\&&
+Y_{147}{}^{{k_1} {k_2}}~\overline{\theta }^2 \left(D^2 \Phi \right) \left(D^2 \Phi \right) \left(\partial _{{k_1}}\partial _{{k_2}}\Phi\right)  \Phi  ~\Phi ^+
+Y_{148}{}^{{k_1} {k_2}}~\overline{\theta }^2 \left(D^2 \Phi \right) \left(D^2 \Phi \right) \left(\partial _{{k_1}}\Phi\right)  \left(\partial _{{k_2}}\Phi\right)  \Phi ^+
\nonumber\\&&
+Y_{149}{}^{{k_1} {k_2} \alpha  \beta }~\overline{\theta }^2~ \left(\partial _{{k_2}}\partial _{{k_1}}D_{\alpha } \Phi \right) \left(D_{\beta } \Phi \right) \left(D^2 \Phi \right) \Phi  ~\Phi ^+
\nonumber\\&&
+\overline{Y}_{150}{}^{{k_1} {k_2} \alpha  \beta }~\overline{\theta }^2~ \left(\partial _{{k_1}}D_{\alpha } \Phi \right) \left(\partial _{{k_2}}D_{\beta } \Phi \right) \left(D^2 \Phi \right) \Phi ~ \Phi ^+
\nonumber\\&&
+\widetilde{Y}_{151}{}^{{k_1} {k_2} \alpha  \beta }~\overline{\theta }^2~ \left(\partial _{{k_2}}D_{\beta } \Phi \right) \left(D_{\alpha } \Phi \right) \left(D^2 \Phi \right) \left(\partial _{{k_1}}\Phi\right)  \Phi ^+
\nonumber\\&&
+\widetilde{Y}_{152}{}^{{k_1} {k_2} \alpha  \beta }~\overline{\theta }^2~ \left(\partial _{{k_2}}D_{\beta } \Phi \right) \left(D_{\alpha } \Phi \right) \left(\partial _{{k_1}}D^2 \Phi \right) \Phi ~ \Phi ^+
\nonumber\\&&
+Y_{153}{}^{{k_1} {k_2}}~\epsilon ^{\alpha  \beta }~\overline{\theta }^2 \left(D_{\alpha } \Phi \right) \left(D_{\beta } \Phi \right) \left(\partial _{{k_2}}\partial _{{k_1}}D^2 \Phi \right) \Phi  ~\Phi ^+
\nonumber\\&&
+\overline{Y}_{154}{}^{{k_1} {k_2}}~\epsilon ^{\alpha  \beta }~\overline{\theta }^2 \left(D_{\alpha } \Phi \right) \left(D_{\beta } \Phi \right) \left(\partial _{{k_1}}D^2 \Phi \right) \left(\partial _{{k_2}}\Phi\right)  \Phi ^+
\nonumber\\&&
+Y_{155}{}^{{k_1} {k_2}}~\epsilon ^{\alpha  \beta }~\overline{\theta }^2 \left(D_{\alpha } \Phi \right) \left(D_{\beta } \Phi \right) \left(D^2 \Phi \right) \left(\partial _{{k_1}}\partial _{{k_2}}\Phi\right)  \Phi ^+
+Y_{156}~\overline{\theta }^2 ~\Phi  \left(\overline{D}^2 \Phi ^+\right) \left(\Phi ^+\right)^4
\nonumber\\&&
+Y_{157}~\epsilon ^{\dot{\alpha } \dot{\beta }}~\overline{\theta }^2 ~\Phi  \left(\overline{D}_{\dot{\alpha }} \Phi ^+\right) \left(\overline{D}_{\dot{\beta }} \Phi ^+\right) \left(\Phi ^+\right)^3
+Y_{158}~\overline{\theta }^2 \left(D^2 \Phi \right) \Phi  \left(\overline{D}^2 \Phi ^+\right) \left(\Phi ^+\right)^3
\nonumber\\&&
+Y_{159}~\epsilon ^{\alpha  \beta }~\overline{\theta }^2 \left(D_{\alpha } \Phi \right) \left(D_{\beta } \Phi \right) \left(\overline{D}^2 \Phi ^+\right) \left(\Phi ^+\right)^3
\nonumber\\&&
+Y_{160}~\epsilon ^{\dot{\alpha } \dot{\beta }}~\overline{\theta }^2 \left(D^2 \Phi \right) \Phi  \left(\overline{D}_{\dot{\alpha }} \Phi ^+\right) \left(\overline{D}_{\dot{\beta }} \Phi ^+\right) \left(\Phi ^+\right)^2
+Y_{161}{}^{{k_1} \dot{\alpha }\beta }~\overline{\theta }^2~ \left(\partial _{{k_1}}D_{\beta } \Phi \right) \Phi  \left(\overline{D}_{\dot{\alpha }} \Phi ^+\right) \left(\Phi ^+\right)^3
\nonumber\\&&
+Y_{162}{}^{{k_1} \dot{\alpha }\beta }~\overline{\theta }^2 \left(D_{\beta } \Phi \right) \left(\partial _{{k_1}}\Phi\right)  \left(\overline{D}_{\dot{\alpha }} \Phi ^+\right) \left(\Phi ^+\right)^3
\nonumber\\&&
+Y_{163}{}^{{k_1} \dot{\alpha }\beta }~\overline{\theta }^2 \left(D_{\beta } \Phi \right) \Phi  \left(\partial _{{k_1}}\overline{D}_{\dot{\alpha }} \Phi ^+\right) \left(\Phi ^+\right)^3
+Y_{164}{}^{{k_1} {k_2}}~\overline{\theta }^2~ \left(\partial _{{k_1}}\partial _{{k_2}}\Phi\right)  \Phi  \left(\Phi ^+\right)^4
\nonumber\\&&
+Y_{165}{}^{{k_1} {k_2}}~\overline{\theta }^2~ \left(\partial _{{k_1}}\Phi\right)  \left(\partial _{{k_2}}\Phi  \right) \left(\Phi ^+\right)^4
+Y_{166}{}^{{k_1} {k_2}}~\overline{\theta }^2 ~\Phi^2  \left(\partial _{{k_1}}\partial _{{k_2}}\Phi ^+\right) \left(\Phi ^+\right)^3
\nonumber\\&&
+Y_{167}~\overline{\theta }^2 \left(D^2 \Phi \right) \left(D^2 \Phi \right) \Phi  \left(\overline{D}^2 \Phi ^+\right) \left(\Phi ^+\right)^2
+Y_{168}~\epsilon ^{\alpha  \beta }~\overline{\theta }^2 \left(D_{\alpha } \Phi \right) \left(D_{\beta } \Phi \right) \left(D^2 \Phi \right) \left(\overline{D}^2 \Phi ^+\right) \left(\Phi ^+\right)^2
\nonumber\\&&
+Y_{169}~\epsilon ^{\dot{\alpha } \dot{\beta }}~\overline{\theta }^2 \left(D^2 \Phi \right) \left(D^2 \Phi \right) \Phi  \left(\overline{D}_{\dot{\alpha }} \Phi ^+\right) \left(\overline{D}_{\dot{\beta }} \Phi ^+\right) \Phi ^+
\nonumber\\&&
+Y_{170}{}^{{k_1} \dot{\alpha }\beta }~\overline{\theta }^2~ \left(\partial _{{k_1}}D_{\beta } \Phi \right) \left(D^2 \Phi \right) \Phi  \left(\overline{D}_{\dot{\alpha }} \Phi ^+\right) \left(\Phi ^+\right)^2
\nonumber\\&&
+Y_{171}{}^{{k_1} \dot{\alpha }\beta }~\overline{\theta }^2 \left(D_{\beta } \Phi \right) \left(\partial _{{k_1}}D^2 \Phi \right) \Phi  \left(\overline{D}_{\dot{\alpha }} \Phi ^+\right) \left(\Phi ^+\right)^2
\nonumber\\&&
+Y_{172}{}^{{k_1} \dot{\alpha }\beta }~\overline{\theta }^2 \left(D_{\beta } \Phi \right) \left(D^2 \Phi \right) \left(\partial _{{k_1}}\Phi\right)  \left(\overline{D}_{\dot{\alpha }} \Phi ^+\right) \left(\Phi ^+\right)^2
\nonumber\\&&
+Y_{173}{}^{{k_1} \dot{\alpha }\beta }~\overline{\theta }^2 \left(D_{\beta } \Phi \right) \left(D^2 \Phi \right) \Phi  \left(\partial _{{k_1}}\overline{D}_{\dot{\alpha }} \Phi ^+\right) \left(\Phi ^+\right)^2
+Y_{174}{}^{{k_1} {k_2}}~\overline{\theta }^2~ \left(\partial _{{k_2}}\partial _{{k_1}}D^2 \Phi \right) \Phi^2  \left(\Phi ^+\right)^3
\nonumber\\&&
+\overline{Y}_{175}{}^{{k_1} {k_2}}~\overline{\theta }^2~ \left(\partial _{{k_2}}D^2 \Phi \right) \left(\partial _{{k_1}}\Phi\right)  \Phi  \left(\Phi ^+\right)^3
+Y_{176}{}^{{k_1} {k_2}}~\overline{\theta }^2 \left(D^2 \Phi \right) \left(\partial _{{k_1}}\partial _{{k_2}}\Phi\right)  \Phi  \left(\Phi ^+\right)^3
\nonumber\\&&
+Y_{177}{}^{{k_1} {k_2}}~\overline{\theta }^2 \left(D^2 \Phi \right) \left(\partial _{{k_1}}\Phi\right)  \left(\partial _{{k_2}}\Phi\right)  \left(\Phi ^+\right)^3
+Y_{178}{}^{{k_1} {k_2}}~\overline{\theta }^2 \left(D^2 \Phi \right) \Phi  \Phi  \left(\partial _{{k_1}}\partial _{{k_2}}\Phi ^+ \right) \left(\Phi ^+\right)^2
\nonumber\\&&
+Y_{179}{}^{{k_1} {k_2} \alpha  \beta }~\overline{\theta }^2~ \left(\partial _{{k_2}}\partial _{{k_1}}D_{\alpha } \Phi \right) \left(D_{\beta } \Phi \right) \Phi \left(\Phi ^+\right)^3
\nonumber\\&&
+\overline{Y}_{180}{}^{{k_1} {k_2} \alpha  \beta }~\overline{\theta }^2~ \left(\partial _{{k_1}}D_{\alpha } \Phi \right) \left(\partial _{{k_2}}D_{\beta } \Phi \right) \Phi  \left(\Phi ^+\right)^3
+\widetilde{Y}_{181}{}^{{k_1} {k_2} \alpha  \beta }~\overline{\theta }^2~ \left(\partial _{{k_2}}D_{\beta } \Phi \right) \left(D_{\alpha } \Phi \right) \left(\partial _{{k_1}}\Phi\right)  \left(\Phi ^+\right)^3
\nonumber\\&&
+Y_{182}{}^{{k_1} {k_2}}~\epsilon ^{\alpha  \beta }~\overline{\theta }^2 \left(D_{\alpha } \Phi \right) \left(D_{\beta } \Phi \right) \left(\partial _{{k_1}}\partial _{{k_2}}\Phi\right)  \left(\Phi ^+\right)^3
\nonumber\\&&
+Y_{183}{}^{{k_1} {k_2}}~\epsilon ^{\alpha  \beta }~\overline{\theta }^2 \left(D_{\alpha } \Phi \right) \left(D_{\beta } \Phi \right) \Phi  \left(\partial _{{k_1}}\partial _{{k_2}}\Phi ^+\right) \left(\Phi ^+\right)^2
\nonumber\\&&
+Y_{184}~\overline{\theta }^2 ~\Phi  \left(\overline{D}^2 \Phi ^+\right) \left(\Phi ^+\right)^5
+Y_{185}~\epsilon ^{\dot{\alpha } \dot{\beta }}~\overline{\theta }^2~ \Phi  \left(\overline{D}_{\dot{\alpha }} \Phi ^+\right) \left(\overline{D}_{\dot{\beta }} \Phi ^+\right) \left(\Phi ^+\right)^4
\nonumber\\&&
+Y_{186}~\overline{\theta }^2 \left(D^2 \Phi \right) \Phi  \left(\overline{D}^2 \Phi ^+\right) \left(\Phi ^+\right)^4
+Y_{187}~\epsilon ^{\alpha  \beta }~\overline{\theta }^2 \left(D_{\alpha } \Phi \right) \left(D_{\beta } \Phi \right) \left(\overline{D}^2 \Phi ^+\right) \left(\Phi ^+\right)^4
\nonumber\\&&
+Y_{188}~\epsilon ^{\dot{\alpha } \dot{\beta }}~\overline{\theta }^2 \left(D^2 \Phi \right) \Phi  \left(\overline{D}_{\dot{\alpha }} \Phi ^+\right) \left(\overline{D}_{\dot{\beta }} \Phi ^+\right) \left(\Phi ^+\right)^3
+Y_{189}{}^{{k_1} \dot{\alpha }\beta }~\overline{\theta }^2~ \left(\partial _{{k_1}}D_{\beta } \Phi \right) \Phi  \left(\overline{D}_{\dot{\alpha }} \Phi ^+\right) \left(\Phi ^+\right)^4
\nonumber\\&&
+Y_{190}{}^{{k_1} \dot{\alpha }\beta }~\overline{\theta }^2 \left(D_{\beta } \Phi \right) \left(\partial _{{k_1}}\Phi\right)  \left(\overline{D}_{\dot{\alpha }} \Phi ^+\right) \left(\Phi ^+\right)^4
+Y_{191}{}^{{k_1} \dot{\alpha }\beta }~\overline{\theta }^2 \left(D_{\beta } \Phi \right) \Phi  \left(\partial _{{k_1}}\overline{D}_{\dot{\alpha }} \Phi ^+\right) \left(\Phi ^+\right)^4
\nonumber\\&&
+Y_{192}{}^{{k_1} {k_2}}~\overline{\theta }^2~ \left(\partial _{{k_1}}\partial _{{k_2}}\Phi\right)  \Phi  \left(\Phi ^+\right)^5
+Y_{193}{}^{{k_1} {k_2}}~\overline{\theta }^2~ \left(\partial _{{k_1}}\Phi\right)  \left(\partial _{{k_2}}\Phi  \right) \left(\Phi ^+\right)^5
\nonumber\\&&
+Y_{194}{}^{{k_1} {k_2}}~\overline{\theta }^2 ~\Phi^2  \left(\partial _{{k_1}}\partial _{{k_2}}\Phi ^+\right) \left(\Phi ^+\right)^4
+Y_{195}~\overline{\theta }^2 ~\Phi  \left(\overline{D}^2 \Phi ^+\right) \left(\Phi ^+\right)^6
\nonumber\\&&
+Y_{196}~\epsilon ^{\dot{\alpha } \dot{\beta }}~\overline{\theta }^2~ \Phi  \left(\overline{D}_{\dot{\alpha }} \Phi ^+\right) \left(\overline{D}_{\dot{\beta }} \Phi ^+\right) \left(\Phi ^+\right)^5
\Bigg\}.
\end{eqnarray}
}

\end{CJK}
\end{document}